\documentclass[pre,letterpaper,superscriptaddress,reprint]{revtex4-1}

\usepackage{graphicx}
\usepackage{pslatex}
\usepackage{amsmath}
\usepackage{booktabs, caption}
\usepackage[flushleft]{threeparttable}
\usepackage{color}   
\usepackage[dvipsnames]{xcolor} 
\usepackage{hyperref}
\hypersetup{
    colorlinks=true, 
    linktoc=all,     
    linkcolor=blue,  
    citecolor=violet
}

\newcommand{\atUH}{Dept. of Physics and Astronomy, Univ. of Hawaii, Manoa, HI 96822.}
\newcommand{\atSLAC}{SLAC National Accelerator Laboratory, Menlo Park, CA, 94025.}

\begin{document}

\title{Picosecond Timing-Planes for Future Collider Detectors}

\author{P.~W.~Gorham}
\affiliation{\atUH}
\author{C.~Campos}
\affiliation{\atUH}
\author{C.~Hast}
\affiliation{\atSLAC}
\author{K.~Jobe}
\affiliation{\atSLAC}
\author{C.~Miki}
\affiliation{\atUH}
\author{L.~Nguyen}
\affiliation{\atUH}
\author{M.~Olmedo}
\affiliation{\atUH}
\author{R.~Prechelt}
\affiliation{\atUH}
\author{C.~Sonoda}
\affiliation{\atUH}
\author{G.~S.~Varner}
\affiliation{\atUH}

\begin{abstract}

We report experimental test-beam results on dielectric-loaded waveguide detectors that utilize microwave Cherenkov signals to time and characterize high energy particle showers. These results are used to validate models and produce high-fidelity simulations of timing plane systems which yield picosecond time tags and millimeter spatial coordinates for the shower centroid. These timing planes, based on the Askaryan effect in solid dielectrics, are most effective at the high center-of-momentum energies planned for the Future Circular Collider (FCC-hh), and are of particular interest in the forward region due to their high radiation immunity. We use our beam test results and GEANT4 simulations to validate a hybrid microwave detector model, which is used to simulate a reference timing plane design for the FCC forward calorimeters. Our results indicate that 0.5-3 ps particle timing is possible for a wide range of collision products in the reference FCC hadron collider detector, even with current technology. 

\end{abstract}
\maketitle
\setcounter{tocdepth}{1}
\tableofcontents

\section{Introduction}

In current high energy particle colliders,
time resolutions of the arrival of particles or showers in a detector
are usually determined using the risetime of some type of photonics-based instrument. In the past the fastest risetimes available were from photomultipliers and avalanche photodiodes, which could yield timing precisions of several hundred picoseconds at best, for nanosecond-scale risetimes. Recent advances in the development of silicon photomultipliers (SIPM) have improved the device risetime, now giving time resolutions below 100 ps in devices with small areas, typically 1~mm$^2$ or so. However, intrinsic jitter in SIPM is likely to limit the improvement in time resolutons of these devices to no better than several tens of picoseconds, as recent measurements have shown~\cite{SIPMtiming1, SIPMtiming2, SIPMtiming3}. For the high luminosity Large Hadron Collider (HL-LHC), the Compact Muon Solenoid (CMS) detector has plans to implement a new timing layer design with 30~ps resolution for all minimum ionizing particles~\cite{CMS2019,PhotonTiming2015,CMScalorTiming} up to a pseudorapidity $|\eta|=3$, representing the state of the art for a large scale detector using SIPMs and avalanche photodetectors. Even for strong signals of tens of photoelectrons or more, timing improvement eventually saturates due to crosstalk or other effects. 

Recent developments with low-gain avalanche detectors (LGADs), tailored specifically to improve their timing characteristics, show promise that such technology may achieve the 10~ps level in the future~\cite{Cartiglia14}. These detectors, termed ultra-fast silicon detectors (UFSDs), use relatively thin silicon layers with doping implants that create a compact high-field region several microns deep in the material, producing an avalanche with a gain of 10-100, compared to $\sim10^4$ or more for SIPMs. Timing precisions of $\sim16$~ps~\cite{Cartiglia17} have been recently demonstrated through combining signals from a three-element detector ensemble. These devices allow tracking of minimum-ionizing particles with tens-of-micron precision. The major challenge for UFSDs in a collider detector is their loss of gain at high radiation fluences; recent work has shown substantial gain damage at neutron equivalent fluences of $\sim10^{15}$~neq/cm$^2$, more than an order of magnitude below the exposure expected at the HL-LHC or the Future Circular Hadron Collider~\cite{FCChh}. While work to mitigate these problems is very active, and some types of dopants show promise for improving the radiation hardness~\cite{LGAD21}, there is still no current technology identified that can achieve stable operation at the expected ambient radiation fluences for future collider detectors.

While $\Delta t \simeq 15-30$~ps timing represents a major advance over prior timing resolutions, the corresponding spatial resolution $\Delta x = c\Delta t$ is still an order-of-magnitude mismatched compared to tracker spatial resolution. The speed of light $c$ in relevant terms is about 3.33 mm per picosecond, and thus the current state-of-the-art UFSD devices can provide spatial-equivalent constraints of $\sim 5$ mm, 2 orders of magnitude worse than the 10-100 micron resolution possible with silicon strip or pixel detectors commonly used in tracker instruments. In addition, such detectors will not provide useful information on photons or other neutrals that are straightforward to detect in calorimeters.  There is therefore a need for methodologies that can extend picosecond timing to the calorimeter layers of large collider detectors, regardless of whether tracker-like technologies can also achieve such timing precision. 

We report here on tests of a timing-plane technology that is based on $Al_2O_3$ dielectric-loaded microwave waveguides which intersect short longitudinal portions of electromagnetic showers. We have previously reported a beam test which yielded timing precisions of a few picoseconds per detector element~\cite{ACE16}. These timing elements are inherently radiation-hard, and provide mm-or-better spatial resolution in addition to picosecond timing. Because they operate on particle showers rather than individual ionizing particles, responding to the electromagnetic energy of the shower, they are more naturally described as timing calorimeters, although the calorimetric function is not currently our focus. We initially termed these devices Askaryan Calorimeter Elements (ACE), because they rely in part on the Askaryan effect in high-energy electromagnetic showers: coherent microwave pulses due to the negative charge asymmetry in the shower. The effect was initially proposed by G. Askaryan~\cite{Ask62} and experimentally confirmed in measurements at SLAC in 2000~\cite{SLAC01} . We retain the ACE acronym designation in this report which focuses on the timing characteristics, providing better calibration, and more detailed validation compared to our earlier work.

\subsection{Physics drivers for picosecond timing.}

In addition to the general goal of achieving timing precision that is commensurate with the spatial precision of trackers, picosecond timing bears directly on physics applications in a future collider. We summarize some of these here, and address them in more detail in a later section.

\subsubsection{Event vertex pile-up.}

In the luminous region of a collision of two opposing particle bunches, events may occur at the same location at different times, or at different locations with the same time, leading to ambiguities in uniquely associating tracks and showers with a specific vertex in the collision, an effect known as {\it pile up}. For the Future Circular hadron-hadron Collider (FCC-hh)~\cite{FCChh}  $\sim 1000$ interactions per bunch crossing will occur, with very significant pile up expected lacking any time discrimination; timing resolution of below 5-10 ps appears necessary to reduce it to an acceptable level.  Pileup at the HL-LHC, which is far lower, is already a background concern for some physics goals, for example, dark matter candidate searches~\cite{DMtiming2019}, and $t\bar{t}$ pair production measurements~\cite{ttbar2020}. Currently there is no photonics- or silicon-based technology yet identified that has achieved even 10~ps precision, although UFSDs are approaching that goal as noted above.

\subsubsection{Time-of-flight (TOF) measurements.}
    
TOF results at the picosecond level for timing planes in locations such as in front of, or between electromagnetic and hadronic calorimeters (EMCAL \& HCAL) would provide a fractional resolution in Lorentz $\beta$ of order 
$$\frac{\Delta\beta}{ \beta } \simeq \frac{0.3~{\rm mm}}{L}~ \frac{\sigma_{\tau}}{1~{\rm ps}}$$
where $L$ is the distance from the interaction point in mm, assuming that a high-precision time tag for the centroid of the interaction region exists. For timing layers in the barrel region (2-3m radius) picosecond resolution  corresponds to  $\Delta\beta / \beta \simeq 10^{-4}$, and in forward detectors this can be an order of magnitude tighter. 
      
This can allow identification of heavy quasi-stable beyond-standard-model (BSM) particles. For example, a highly boosted $M=100$~GeV BSM particle with $p \simeq 3000$~GeV/c can be identified with $\sim 5\sigma$ confidence for $\sim 1$~ps timing at 3~m from the interaction point. For neutral heavy particles, such timing would be critical to identification, assuming lifetimes (or interaction cross sections) that are commensurate with the length scales involved. 

\subsubsection{Jet physics.}

Jets may be accompanied by one or more forward protons which must be correlated to the jet vertices in the presence of soft proton production processes with large cross sections which produce unwanted correlations of jets with background protons. To identify a desired event signature such as a dijet with two diffractive forward protons requires rejection of these pileup events with high confidence. Studies of these effects in the presence of high precision timing indicate that picosecond precision can lead to an order of magnitude improvement in background rejection for such processes~\cite{JetTiming2019, Cerny21}.

Timing of massive jet component particles may yield time-of-flight momentum values that are useful for constraining the jet kinematics, or even more exotic phenomena such as emerging jets~\cite{EmergingJetsTriggering, EmergingJets, Chekanov20}. At the energies of future colliders, boosted jets with much higher total energies become more prevalent, and methods which are effective for jet physics in the barrel region will require revision to operate under the extreme conditions -- both in the radiation environment, and in the high dynamic range required -- that will obtain there.

\subsection{Our prior work.}

In prior work we have reported picosecond-level timing precision of high-energy particle showers in a beam test using dielectric-loaded microwave rectangular waveguides, which generate band-limited microwave impulses when excited by a charge passing transversely through the waveguide~\cite{ACE16}. The fast-impulse excitation of the lowest order $TE_{10}$ mode of the waveguide occurs because the charged bunch induces guided microwave Cherenkov emission in the dielectric, in our case $Al_2O_3$ or alumina, a very radiation-hard insulating material which is also extremely transparent to microwaves, with a dielectric constant of $\epsilon \simeq 10$ and a loss tangent $\tan \delta \leq 10^{-4}$. We use standard WR51 waveguide, with an inside cross section of $6.35 \times 12.7$~mm.

Without dielectric loading, WR51 is normally used in the 15-22~GHz range, but after accounting for the dielectric loading, the single mode $TE_{10}$ center frequency is 6.5 GHz, with a bandwidth of $\sim \pm 1.5$ GHz. The resulting signals have risetimes below 40 ps, nearly an order of magnitude better than the fastest SIPM or LGAD risetime. Since these microwave impulses also couple directly to the current of the transiting particle bunch, there is no induced timing jitter due to any intermediate avalanche process. 

The significant limitation of this microwave timing method is the relatively high turn-on threshold due to thermal noise limitations. Unlike optical systems where photon or shot noise presents the detection noise floor, in radio and microwave systems the effective system temperature $T_{sys}$ is well above the broadband quantum limit $T_q = h\nu/k_B$ for Planck's constant $h$, frequency $\nu$, and Boltzmann's constant $k_B$. In our case at 6.5~GHz, $T_q = 0.25$K, and thus unless cooled well below 1K, microwave systems cannot achieve single-photon sensitivity. Rather, each mode of the electromagnetic field has a relatively high occupancy of thermal microwave photons. As a result, the induced field from a single transiting charge is undetectable above noise, but since the intensity grows with $(Ze)^2$ for a bunch charge number of $Z$, once this level is exceeded, a signal-to-noise ratio (SNR) adequate for picosecond timing precision is soon achieved. 

For these reasons, our method is currently applied to electromagnetic showers, which rapidly convert single-particle momentum to a compact, many-particle shower bunch, dominated by $e^+e^-$ pairs, with a total number proportional to the primary particle energy. The Askaryan effect~\cite{Ask62}, which selectively enhances the electrons over the positrons, then guarantees a charge excess of typically $\sim 25$\% of the electron+positron number. 

Our previous work~\cite{ACE16}  established the methodology and achieved timing precision approaching 1 ps, but due to limitations in our calibration of the number of electrons per bunch, the charge per bunch was not well-established.
In the beam tests we report here, we have used a calibrated silicon pixel detector to precisely measure the transiting charge per bunch, establishing both the number of electrons and the bunch spatial distribution, an important parameter for determining the sensitivity of the method. We have cooled our system to liquid helium (LHe) temperatures to reduce thermal noise and better establish limitations imposed by it. We report also on the waveguide response to off-center charge transits. Combining various measures from these data with GEANT4~\cite{GEANT} simulations, we then update our simulations of picosecond timing plane configurations that may be appropriate to the FCC-hh or other future colliders. We find that timing planes based on these methods are most appropriate for use in the high-rapidity regions of the detector, where radiation levels are most challenging, and a large fraction of the collision-induced particles retain very high energies.

In this report we first review the background theory and methods employed. We then detail our experimental setup for the beam test, and present the results of that test. These are analyzed to yield scaling relations which then anchor detailed modeling, including both GEANT results and microwave simulations. We then adopt a working design for timing planes for electromagnetic and hadronic showers, focusing our study on the forward region with pseudorapidity $|\eta|\geq 2.5$. We conclude our report by explore via simulation the timing capability and signal efficiency for several physics cases appropriate to the FCC-hh.

\section{Background theory}

The study of high energy charged bunches interacting with waveguides, both dielectric-filled and in vacuum, has been developed extensively for the purpose of investigation of wakefield acceleration~\cite{WF1,WF2,WF3}. In these cases the charged particle bunch is usually propagating along the longitudinal axis of the beam pipe or some other structure that acts as the waveguide. Cherenkov (or Vavilov-Cherenkov, VC) radiation is produced whether or not there is a dielectric present in the waveguide or beam pipe, since even for vacuum conditions the group velocity of radiation in the waveguide is below the free-space speed of light. Such radiation propagates out in a radial cone until it reflects from the side walls of the waveguide, and reconverges toward the axis, where it forms a series of nodes with high field amplitudes, that may be used for accelerating a trailing particle bunch. 

In our application, the charged bunch moves orthogonal to the waveguide longitudinal axis, entering and exiting the waveguide across its shortest dimension, which thus creates an apparent step-like current element which is parallel to, and couples to, the transverse electric ($TE_{mn}$) modes of the waveguide. To an observer within the waveguide, the charge appears suddenly on one wall, transits the short dimension of the waveguide at the bunch velocity $\beta c$, and then disappears suddenly as it exits the other wall.  

In our previous study~\cite{ACE16} we analyzed this current element in terms of finite track-length VC emission, under the framework developed by I. Tamm as a special case of the general infinite-track theory given by the well-known Frank-Tamm methodology of estimating Cherenkov emission. Tamm's solution~\cite{Tamm39} includes effects that account for the transition radiation in the forward and backward directions due to the sudden appearance and disappearance of the charge, and the transit of the dielectric leads to an additional term capturing the induced polarization wave in the material, the condition for Cherenkov radiation. However, Tamm's method cannot account directly for the waveguide boundary conditions, which act selectively to restrict the propagation modes of the resulting radiation.

In our prior work, we also developed a finite-difference time domain (FDTD)~\cite{FDTD, XF7} computational model for the charged particle transit of the loaded waveguide, and confirmed agreement with Tamm's theory to first order, but this required an ansatz in the application of Tamm's solution to account for the lack of a way of computing the coupling coefficient of the radiation to the waveguide modes. 

Here we take a different and complementary approach, computing the theoretical time-domain electric field of the injected pulse using the vector potential of a current element from first principles, and again comparing it to our FDTD model.


\subsection{Vector potential approach.}

The cutoff frequency of rectangular waveguide is given by $$f_c = (2a\sqrt{\mu\epsilon})^{-1/2} = c'/2a$$ where $a$ is the width of the waveguide, $\mu,~\epsilon$ are the permeability and permittivity of whatever fills the waveguide, and $c' = c/\sqrt{\epsilon}$ is the speed of light in a pure dielectric where the relative permeability $\mu = \mu_0$.  In our application, the length of the current element through the waveguide is thus $\lambda/4$ at the cutoff, and about $\lambda/2$ an octave higher in frequency where the next order modes begin to develop. This means the Cherenkov radiation always occurs in the limit of subwavelength track length, and the usual considerations of angular dependence for VC radiation are largely suppressed.

Under these conditions we can gain insight into the electric field strengths and time-domain response with analysis based on vector potential methods which have been used with success to describe radio Cherenkov emission from particle cascades~\cite{FORTE,PracticalAccurate}. 

Following reference~\cite{FORTE}, we start by modeling the transiting charge bunch as a point charge moving at the speed of light along the $z$-axis, with the $x$-axis along the longitudinal waveguide axis and $y$ the transverse axis. The current density is thus 
$$J_z(\mathbf{r},t) = cq(z)\delta(\mathbf{r} - ct\hat{z}) $$
with $J_x = J_y = 0$. We express the current density in the frequency domain via a Fourier over frequency $\omega$:
\begin{equation}
    \Tilde{J}_z(\mathbf{r},\omega) = 2 \int_{-\infty}^{\infty} J_z(\mathbf{r},t) e^{i\omega t} dt = 2q(z) \delta(x)\delta(y) e^{i\omega z/c}
\end{equation}
where the tilde indicates the Fourier dual, and factor of two is required for consistency with the definition of the frequency domain electric field $\Tilde{E}(\omega)$. The vector potential in the frequency domain can be determined from the Helmholtz equation
\begin{equation}
    \nabla^2 \Tilde{A}_z + k^2 \Tilde{A} = \mu_0 \Tilde{J}_z
\end{equation}
where we have assumed the relative permeability is unity. Since the current has only $z$ components, $\Tilde{A}_{x}=\Tilde{A}_y = 0$; the wavenumber $k = n\omega/c$. 

In the waveguide only plane waves propagate, with propagation constant $\beta = k\sqrt{1 - (f_c^2/f^2)}$, where we ignore waveguide attenuation here.
A solution for observation point $\mathbf{x}$ along the waveguide longitudinal axis is
\begin{equation}
    \Tilde{A}_z(\mathbf{x},\omega) = \mu_0 \int_V e^{i\beta|\mathbf{x}-\mathbf{r}|} \Tilde{J}(\mathbf{r},\omega) d^3\mathbf{r}
\end{equation}
Assuming the observation point is some distance down the waveguide so that $x\gg b$, where $b=a/2$ is the height of standard rectangular waveguide, is satisfied, $$e^{i\beta|\mathbf{x}-\mathbf{r}|} \simeq e^{i\beta x}~$$
and the resulting vector potential for the current traversing the waveguide from $z=-b/2$ to $z=b/2$ is

\begin{equation}
    \Tilde{A}_z(\mathbf{x},\omega) = \mu_0 e^{i\beta x}  \int_{-b/2}^{b/2} q(z) e^{iz\omega/c} dz
\end{equation}
We assume that the charge profile is constant during the waveguide transit, thus $q(z) = q_0$. 
The result of the integral is:
\begin{equation}
    \Tilde{A}_z({x},\omega) = 2\mu_0 q_0 c ~e^{i\beta x} ~ \frac{\sin \left ( \frac{b\omega}{2c} \right )}{ \omega} 
\end{equation}
From the frequency domain vector potential, we perform an inverse Fourier transform to recover the time domain vector potential $A(t)$:
\begin{equation}
    {A}_z(x,t) = 2\mu_0 q_0 c ~e^{i\beta x} ~ \int_{-\infty}^{\infty} \frac{\sin \left ( \frac{b\omega}{2c} \right )}{ \omega}~ e^{i\omega t} d\omega
\end{equation}
This integral does not have a closed analytic solution, although it can be expressed in terms of a linear combination of exponential integrals. We thus integrate it numerically  via a discrete Fourier transform, and the resulting electric field is then determined from a numerical derivative: $$\mathbf{E}(t) = \frac{\partial{\mathbf{A}}}{ \partial{t}}$$.

\begin{figure}[htb!]
 \includegraphics[width=3.5in]{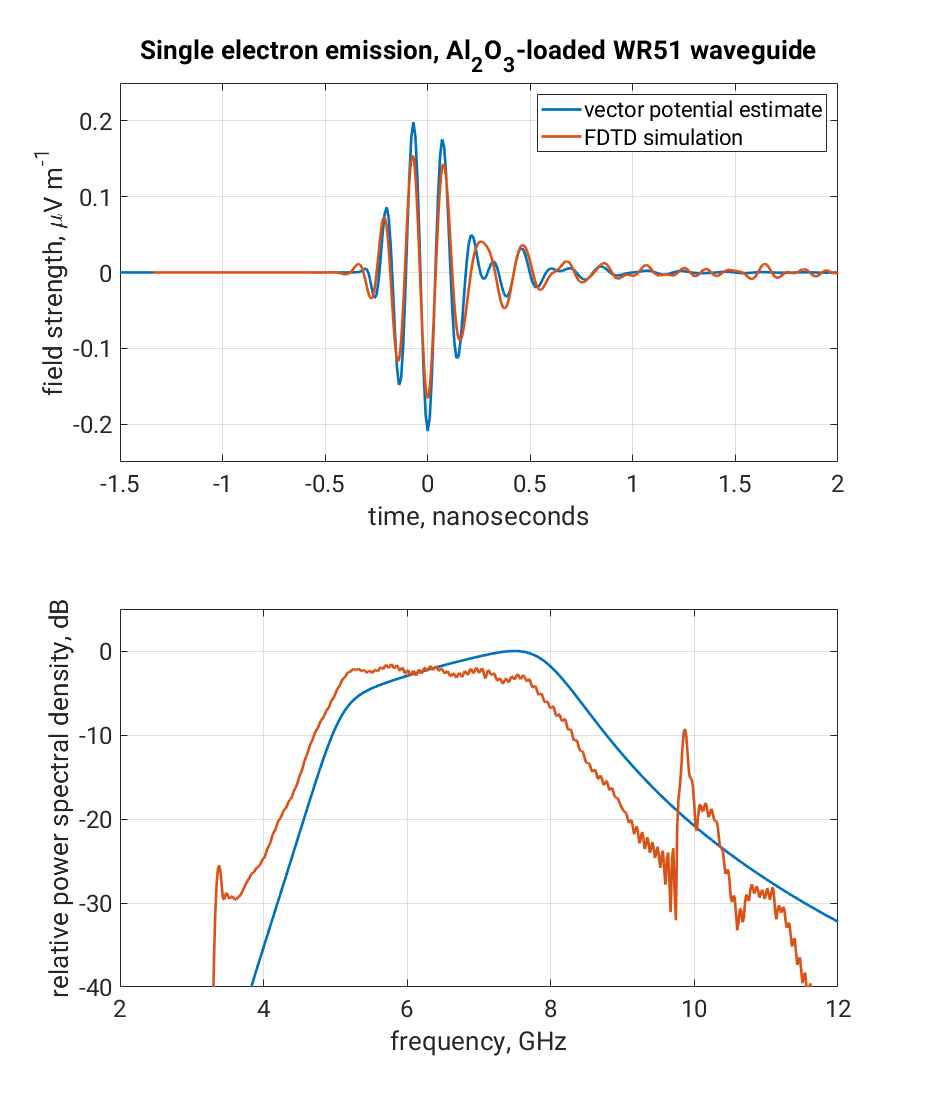}
 \caption{ Top: waveforms for a single electron transit of a WR51 waveguide loaded with alumina. The estimate from vector potential theory is shown in blue, and data from an XF simulation is shown in red. Bottom: power spectral density for the two waveforms in the top pane. A causal passband filter has been applied to emphasize the $TE_{10}$ mode response for the two cases.
 \label{vecpot}}
 \end{figure}

\subsection{Comparison with FDTD simulations.}

Fig.~\ref{vecpot} shows the results of a comparison between these analytic results for the time-domain electric field, and a finite-difference time domain (FDTD) simulation using Remcom's XF7.9~\cite{XF7}, for a single relativistic electron transiting alumina-loaded WR51 waveguide. In order to create the simulation, the relativistic electron is introduced as a series of short discrete current elements, typically 0.5 mm or less in length for this case, and arranged end-to-end transverse to the waveguide, with sequential delays corresponding to the average Lorentz beta for electrons at the critical energy in Alumina, $\sim 50$~MeV. These current elements are individually excited by a short current pulse with a peak current equal to the electron's apparent current $-ec$. Since the results of the theory are only valid for the lowest-order $TE_{10}$ mode, we have applied a causal infinite-impulse-response filter (8th order) to isolate the signal in the 5-8~GHz portion of the band.
These results have only a single degree of freedom in the comparison, and this is not arbitrary as we discuss below. The observed agreement in absolute amplitude is therefore also not arbitrary.

The FDTD results show a flat $TE_{10}$ passband signal, but also strong resonant response near the waveguide cutoff, and also at one of the higher-order mode cutoff frequencies. The vector potential theory, which assumes only propagating modes along the waveguide axis (eg., the $TE_{10}$ mode in the passband) does not show these effects. In the simulation, although the length of the current element due to the transiting electron is subwavelength, there is still power injected preferentially at the Cherenkov angle, about $70^{\circ}$ for alumina, and this power enters the waveguide about $20^{\circ}$ off-axis, where it can excite non-propagating modes. In addition, the TR at entry and exit, while exciting the $TE_{10}$ mode in part, also has forward and backward components that are far from the axis of propagation, thus also exciting these non-propagating modes near the cutoff frequencies. The additional low frequency power observed in the FDTD simulation but not in the theory is likely from these effects.

The vector potential theory predicts a rising spectrum within the $TE_{10}$ mode, and an extended high-frequency component (even after the filter) which is not observed in the FDTD simulation results. A spectrum that rises with frequency is characteristic of coherent Cherenkov emission in unbounded dielectrics~\cite{SLAC01, Tak2000}; thus we tentatively conclude that the flat spectrum observed in the simulation is a limitation of the FDTD application here. Requirements to maintain the validity of the FDTD simulation introduce one external ansatz: the shape of the current pulse. It can be a relatively short Gaussian, approximating a delta function, but computational gridding and meshing requirements require a minimum width to avoid artifacts. Thus the width of the Gaussian is determined by the limitations of the simulation space, and leads directly to an attenuation of high frequencies in the results. It is unclear whether this explains the differences; further work on higher fidelity simulations may elicit an answer.

\begin{figure*}[htb!]
\centerline{ \includegraphics[width=3.5in]{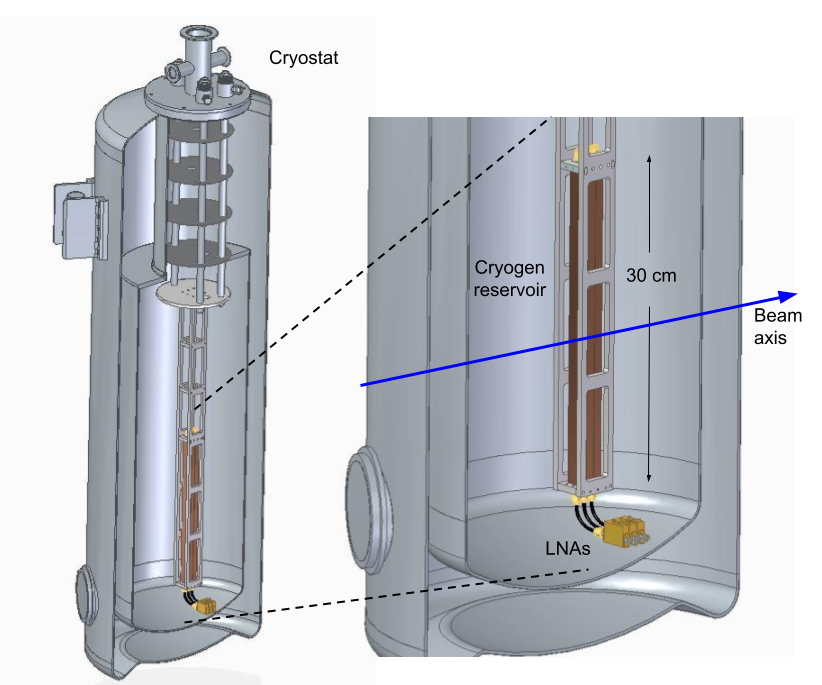}
\includegraphics[width=3in]{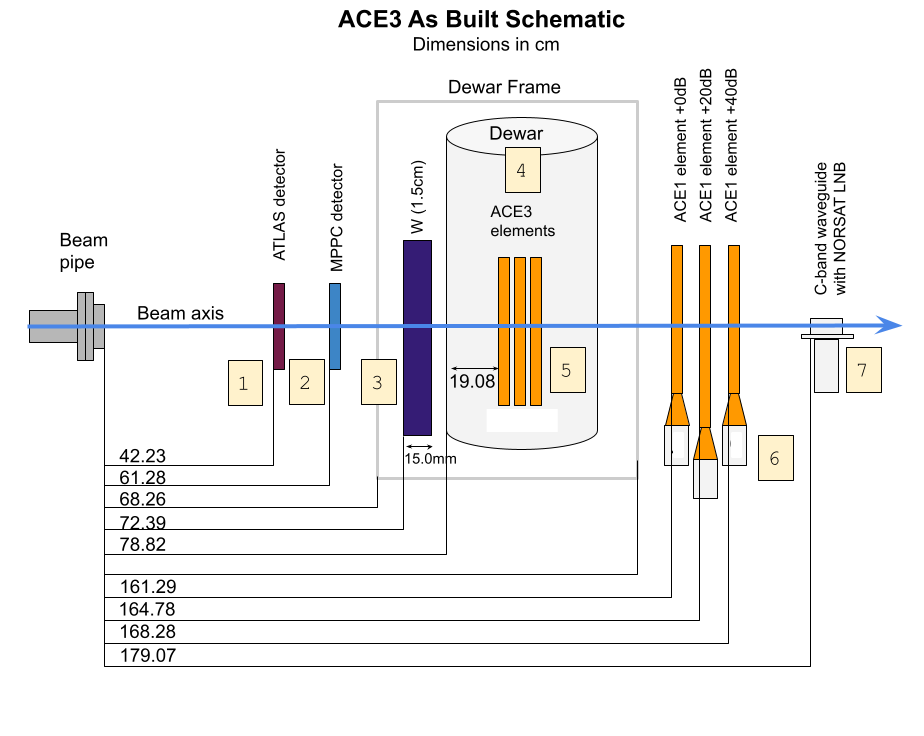}}
\centerline{ \includegraphics[width=6.5in]{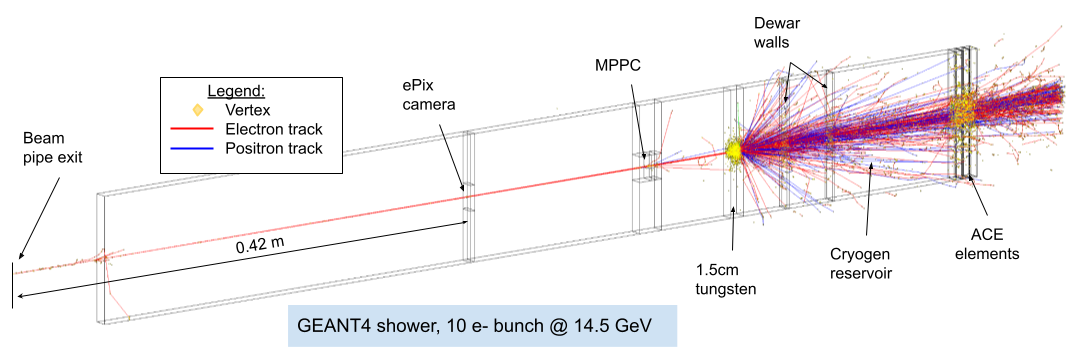}}
 \caption{  Upper Left: a rendering of the dewar containing the ACE elements, and an expanded view of the 60 cm long active section, which includes three WR51 waveguide elements. Upper Right: Dimensions in cm of the different portions of the experimental setup. Bottom: a GEANT4 simulation slice through the system from the beam pipe exit to the ACE elements within the dewar. This shows the charged portion of the  electromagnetic shower for a bunch with 10 electrons of 14.5 GeV energy each. Electrons are in red, positrons in blue, and each interaction vertex is marked with a yellow dot.
\label{setup}}
\end{figure*}

Despite these differences, the agreement with the theory is remarkably good given the simplicity of the vector potential approach, and it does provide confidence that we understand the physics of the relativistic current coupling to the waveguide modes to first order. Varying the width of the current-element Gaussian within a moderate range, consistent with the gridding,  results in variations of the FDTD simulation amplitude by typically 1-2 dB; we have used a mean value for the Gaussian width here. While it may be possible to refine the theory and simulations to get better agreement, the timing properties are largely unaffected by precise knowledge of the amplitude, as we will see later on.

It is interesting that, contrary to the usual analysis of Cherenkov detectors, the Cherenkov emission angle in this case plays only a secondary role, since only modes consistent with axial propagation are allowed. The dominant coupling to the waveguide can be understood completely in terms of the vector potential of a simple transverse current, truncated at the waveguide edges. In this case, the dielectric performs the role of creating a more compact geometry due to the increased refractive index, and the corresponding lowering of the waveguide frequency range. In the 5-8 GHz range where excellent low-noise amplifiers are available and attenuation losses are relatively low, an unloaded waveguide would be three times larger in its transverse dimensions, making a timing plane rather unwieldy, with correspondingly lower spatial resolution.

\section{Beam test at SLAC, 2018}

In 2018, we performed a beam test in the SLAC National Accelerator Laboratory, in End Station A, using beam energies of 14.5 GeV, and electron bunches with bunch charges that were modulated by using both screens and momentum selection so that we could go from unattenuated bunch charges of $10^{10}$ electrons per bunch down to $\leq 10$ electrons per bunch as needed. 

\subsection{Experimental design.}

In our 2018 beam test, we had two primary goals: (1) to measure with better accuracy the number of electrons per bunch, to calibrate the response of ACE at low beam currents, and (2) to use ultra-low noise amplifiers ({\it Low Noise Factory} model number LNF\_LNC4\_8C with of order 2 K noise figure at LHe ambient) to determine the practical limits of ACE sensitivity. This latter goal required the use of liquid helium as a cryogen to achieve a 4-5K ambient temperature in which the entire detector was immersed, but in fact we expect that improvements in cold-head technology in cryogenics would allow for point-cooling of the LNA if the sensitivity afforded by this approach is deemed necessary. In practice we believe that even with liquid nitrogen or liquid argon cooling, ACE's sensitivity is useful and effective for a range of future collider applications.

Fig.~\ref{setup} shows a rendering of the cryostat and ACE elements, along with dimensions for the experiment. Included are the locations of the relevant components of the experimental setup (indicated in yellow boxes on the schematic): 
\begin{enumerate}
\item A silicon integrating pixel detector, the ePix 100~\cite{ePix}; 
\item A custom multi-pixel photon counting system (MPPC) coupled to a Cherenkov radiator; 
\item A 15mm thick tungsten-alloy block (alloy EF-17, 90\% W, 7\% Ni, 3\% Fe, radiation length about 4.7mm) to pre-shower the bunches entering the ACE-3 elements; 
\item The cryostat containing the elements; 
\item The 3 sequential WR51 waveguide elements used in this experiment; 
\item The three original waveguide elements used in our first ACE experiment, which were placed to the rear of the dewar and used at room temperature only at high beam currents; 
\item A closed air-filled waveguide-short centered on the beam axis, coupled to a C-band (3.4-4.2 GHz) low-noise amplifier block.  
\end{enumerate}
Each of these was measured carefully for materials and thicknesses so that we could create an accurate GEANT4 model of showers that would develop. 

Fig.~\ref{setup}(bottom) shows an example of a typical simulated shower: in this case a compact bunch containing ten 14.5~GeV electrons propagate through air from the beampipe exit to the tungsten pre-shower block and the resulting shower passes through the ACE elements within and external to the dewar. The tungsten-alloy block was about 3.2 radiation lengths thick resulting in efficient conversion of most of the electron energy into showers with minimal losses.

All of the detectors which appear on the downstream side of the ACE elements within the dewar were used primarily for triggering and diagnostics and their signals and waveforms are similar to our previous studies. They are not part of the scope of this report.

\begin{figure}[htb!]
 \includegraphics[width=3.5in]{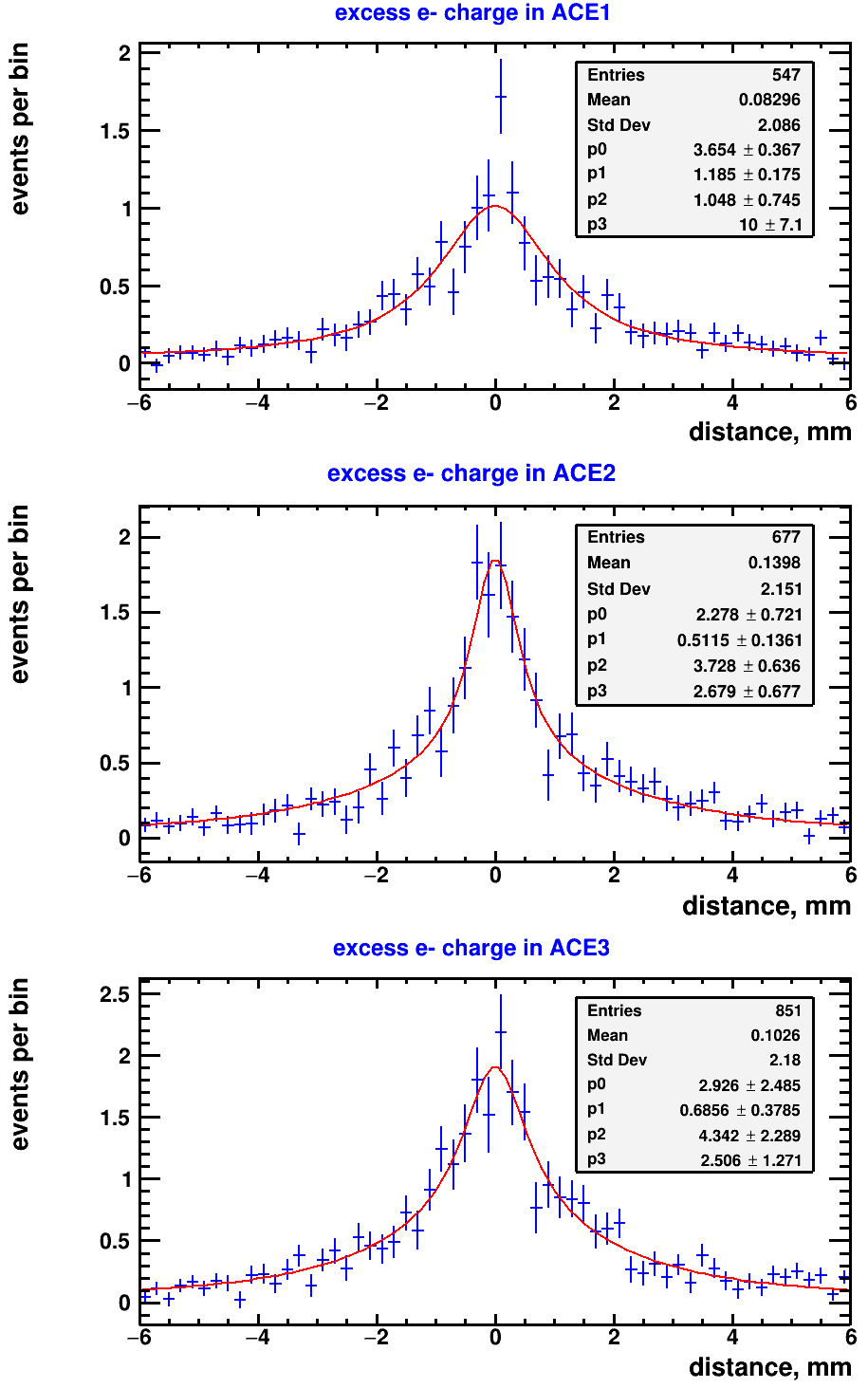}
 \caption{ Distributions of excess charge in each of the three ACE elements, along with fitted
 parameters for the double Lorentzian profiles.
 \label{ACELfit1}}
 \end{figure}

\subsection{GEANT4 simulations}

Originally we intended to submerge the tungsten pre-shower block in the cryogen to avoid the 20 cm gap between it and the ACE elements, but this was deemed impractical because of the large loss of liquid helium that would have been necessary to cool such a large mass. The showers that develop from the tungsten acquire large emittance, and are thus are not fully contained in the ACE elements, and we rely on GEANT4 simulations to estimate the effects of the shower development in this case. 
Fig.~\ref{ACELfit1} shows the results of the average of many such showers in the configuration shown. Here we have plotted a slice through the radially symmetric distribution of excess electrons passing through each element, and they are fitted to a double-Lorentzian function, which was found to provide a reasonable empirical match, providing a type of ``core + halo'' combination that represented the data well:
\begin{equation}
  N(r) = \frac{p_0}{\pi p_1(1 + (r/p_1)^2)} + \frac{p_2}{\pi p_1(1 + (r/p_3)^2)}
\end{equation}
where $N(r)$ is the radial density at distance $r$ from the beam center, here centered on the waveguide center axis, and the $p_i$ are the double-Lorentz parameters. 

\begin{figure}[htb!]
 \includegraphics[width=3.5in]{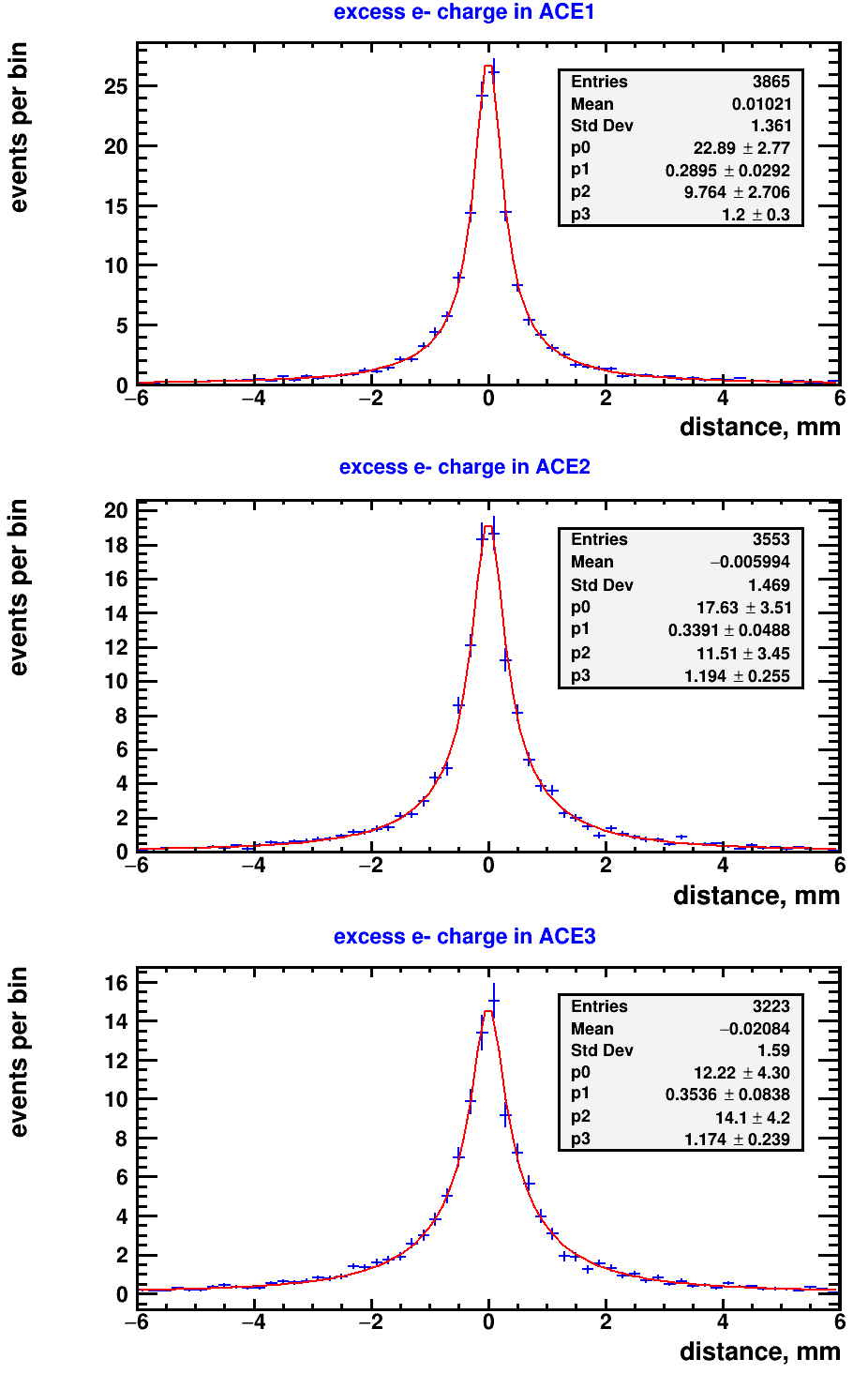}
 \caption{ Similar to Fig.~\ref{ACELfit1}, but now without any gaps between the pre-shower tungsten block and detector elements.
 \label{noGaps}}
 \end{figure}

This shape will affect the emission properties of the radiation that is coupled to the waveguide, since it is clearly not consistent with our original assumption of a line-current on the center axis of the waveguide. These effects can be accounted for by a convolution of this form factor with the fields of a single charged particle (or line current) on axis; the coherently summed field, including phase factors for the offset positions of the tracks, is given by
\begin{equation}
\mathbf{E}_{tot} ~=~ \sum_{j=1}^N \mathbf{E}_j \exp \left ( \frac{2\pi i f}{c} \hat{r}\cdot \mathbf{x}_j \right )~.
\label{formfactor-eq}
\end{equation}
Here $\mathbf{E}_j$ is the field from the $j$th particle,  
$\mathbf{x}_j$ is its position, and $\hat{r}$ the unit vector in the direction of observation.

\begin{figure*}[htb!]
 \centerline{\includegraphics[width=3.5in]{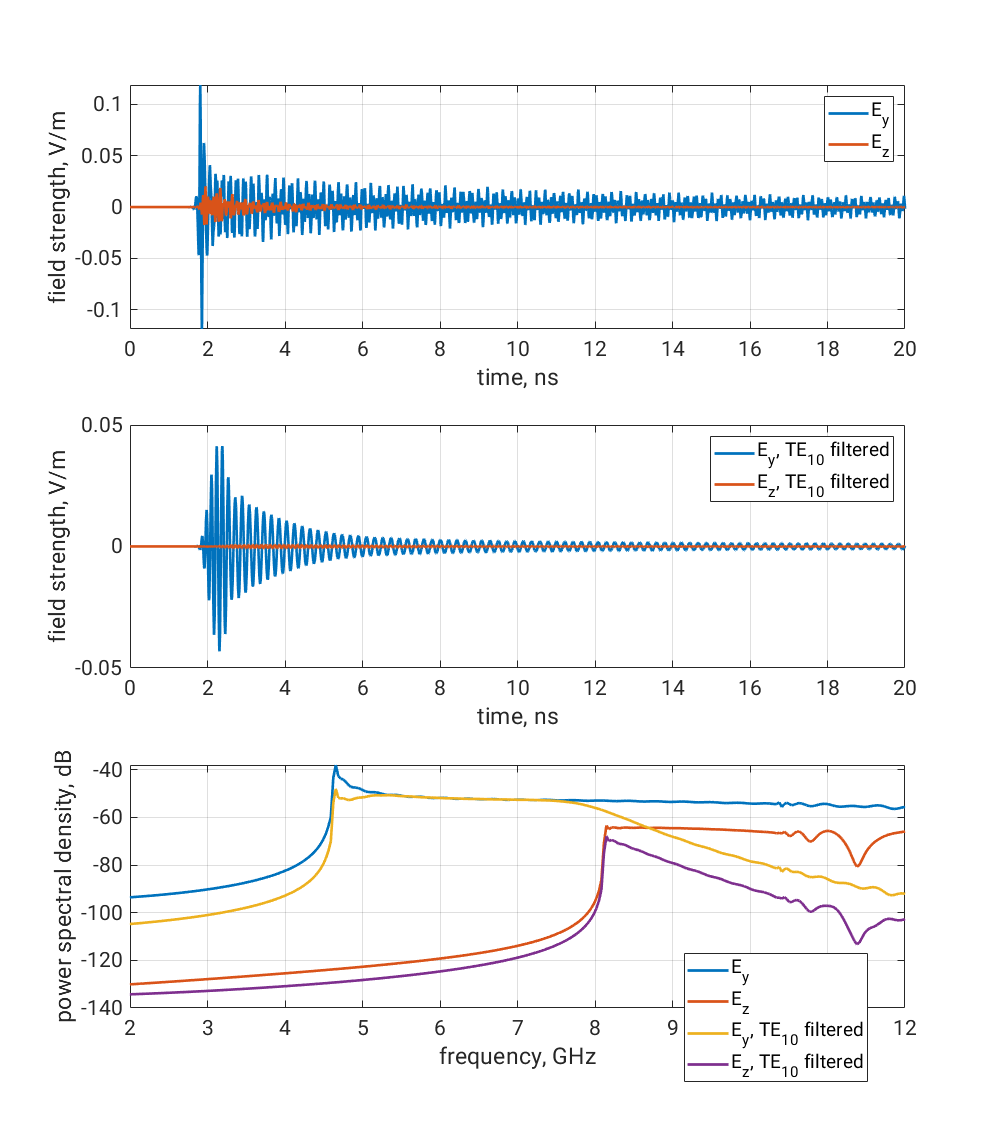}
 \includegraphics[width=3.5in]{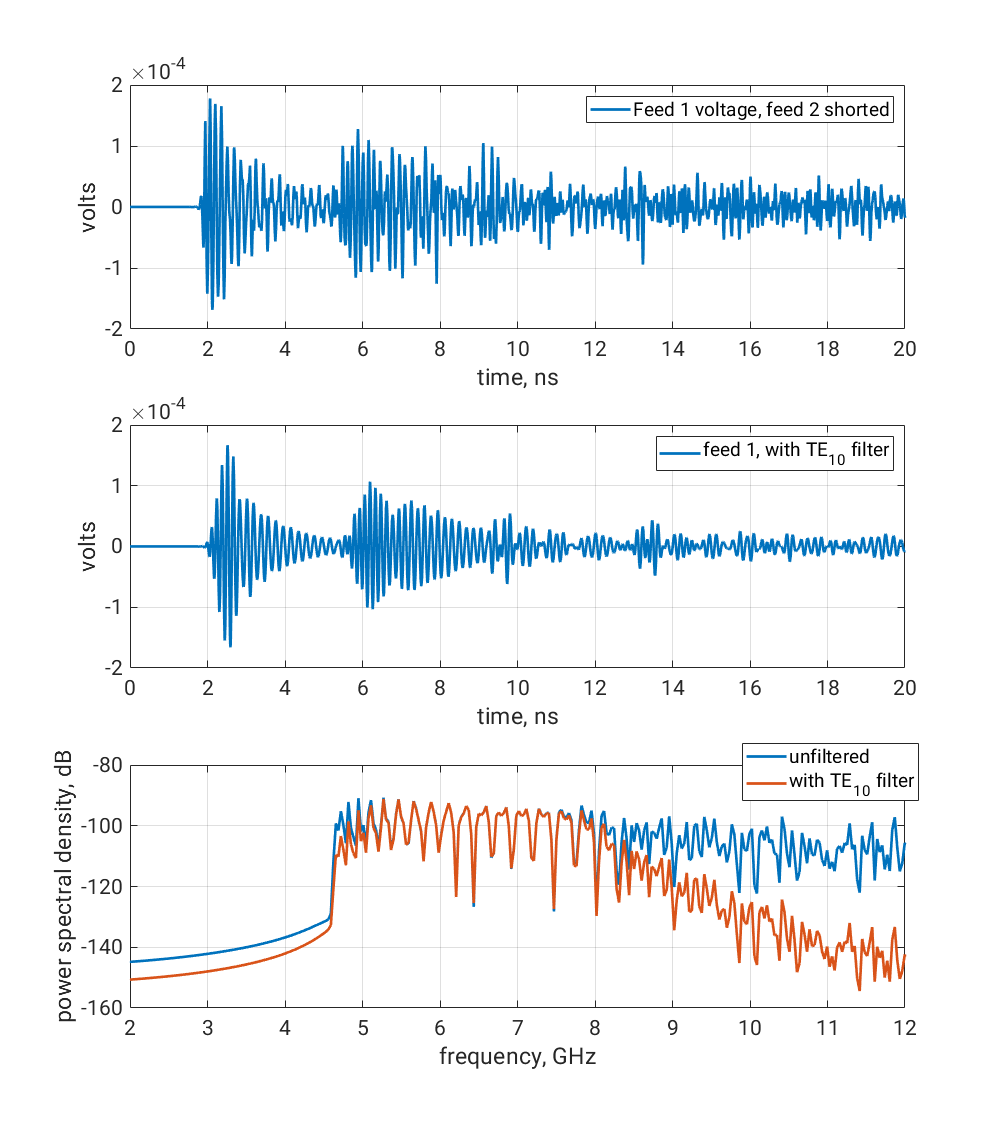}}
 \caption{Left column: FDTD-simulated electric fields for the as-built ACE element at a point just prior to the end
 of the waveguide. For these, the reflection from the shorted far end has been suppressed. Left top: the raw induced waveform with no $TE_{10}$ passband filtering imposed; blue is the $TE_{mn}$ transverse mode, and red the parasitic longitudinal mode. Left middle: same as left top, with a 5-8~GHz filter applied. Left bottom: power spectral density of each of the signals shown above. Right: the same sequence of signals is shown after coupling through the waveguide-to-coaxial adapter, and now including the reflection from the shorted end. Imperfections in the impedance matching are included at each end to produce a realistic signal. The power spectra (right bottom) show a high degree of modulation due to the effects of the Fourier transform of the double-pulse.
 \label{fdtd-wfms}}
 \end{figure*}
 
The double Lorentzian function estimated here is specific to our setup, which included the large gap and additional cryogen between the pre-shower block and the ACE elements. We are interested in establishing the shower energy threshold for the case where the pre-shower block location is optimized, and to do this we also simulated a setup where the pre-shower block appears just upstream of the ACE elements, allowing for a much larger fraction of the shower to be captured, and for lower divergence of the shower itself.
Fig.~\ref{noGaps} shows the results of this GEANT4 simulation and the fitted functions, and it is very evident that the shower containment is much better, and that the compactness of the shower is and order of magnitude higher than previously. We will use these results to scale from our measured data to the more ideal case. 

The choice of three sequential ACE elements rather than a more populous array was dictated by constraints on our procurement of
ultra-low-noise microwave amplifiers; those acquired represent the current state of the art for such LNAs in an immersed LHe bath with noise temperatures of $\sim 2$~K over the 4-8~GHz band. These same LNAs have equivalent noise figures of around 8K at 77K liquid nitrogen (LN$_2$) temperatures; we utilized LN$_2$ for a portion of our tests. Since our detector elements have outputs on either end, to capture both propagating modes we shorted one end to produce a reflection of one of the propagating modes to the opposite output, as we had done successfully in prior studies. This has the effect of combining the two opposite propagating signals into a single waveform 

\subsection{Tailored electromagnetic simulations.}

Our previous studies~\cite{ACE16} had showed good agreement between FDTD simulations and data, especially when all details of the as-built detectors were modeled carefully in the simulation. Thus we created a precise XF model of the 30~cm waveguides, with waveguide-to-coaxial adapters on either end, and a short at the output of one of the waveguides to produce a reflection from one end of the element, which is measured at the other end, and contributes to the timing measurements by adding the phase-inverted reflection to the detected waveform with a propagation delay proportional to the length of the element.

Any realization of a waveguide-to-coaxial adapter for a dielectric loaded waveguide must account for the much lower waveguide impedance due to the presence of the dielectric, and we thus developed a custom adapter to provide the necessary impedance matching. The coupler had a design efficiency of about 90\% in power over the $TE_{10}$ band, but this does mean that the amplitude of in-band reflections from either end is of order 30\%, which leads to some ringing of the system as the modes damp out. The coaxial coupling for higher-order modes is even less efficient, and thus such modes tend to ring more than the primary mode, although they represent out-of-band power that can be filtered to retain only the $TE_{10}$ mode as needed.

These effects are evident in the results of our FDTD simulation, as shown in Fig.~\ref{fdtd-wfms}. In the left column of the figure panes, we show the intrinsic field strength profiles as measured by a sensor probe placed in the simulation on the waveguide axis just prior to the coaxial adapter. This shows that the induced signal has a very strong initial impulsive broadband component, which in fact extends far above the nominal single-mode portion of the $TE_{10}$ spectrum. The trailing signals are due to strongly dispersive frequencies near the turn-on of two or more modes, including $TE_{10}$ at around 4.7~GHz. Very close to the waveguide turn-on, these modes are effectively non-propagating as their delays extend out far past the region of interest. The blue curve shows the primary transverse electric signal $E_y$, and red curve a subdominant $E_z$ mode that is excited by the field component of the Cherenkov emission that lies in the longitudinal direction. 

For comparison, we also show the $E_y, E_z$ waveforms after filtering to isolate the nominal $5-8$~GHz $TE_{10}$ operational passband. The sharp high-frequency impulse is now significantly reduced by the low-pass filter edge, and the high-pass edge has suppressed the non-propagating modes near the waveguide cutoff. In addition, the parasitic longitudinal modes are also suppressed. In Fig.~\ref{fdtd-wfms} left bottom, the power spectral density of each of the waveforms above them is shown for both the unfiltered and filtered waveforms. The flat, broadband power of the initial spike is evident, continuing above 12 GHz, and the parasitic mode turn-on is also evident at about 8.2 GHz. With the 8th order time-domain causal filter used, both the parasitic and non-propagating modes are seen to be removed.

On the right-hand column of Fig.~\ref{fdtd-wfms} we show simulated signals which are representative of what can be measured in our experiment, as the model includes a realistic coaxial adapter, necessary to couple the waveguide modes to a practical system using coaxial cables. We also model the shorted end, producing a reflection which must propagate about 3 times the distance of the prompt signal, and which thus acquires additional frequency dependent group-delay, changing the shape of the reflected waveform compared to the prompt pulse, which is itself somewhat dispersed as well. Because of impedance-matching imperfections, the signals now pick up some structure due to reflections at the couplings. The sequence of panes for the right column is the same as the left, with the exception now that there is no measurement of the parasitic $E_z$ mode, since the coaxial coupler can only respond to transverse modes. The strong modulation of the frequency domain power spectral density is due to the Fourier beating effects of the two time-domain peaks. As we will see in the next section, these results closely match what we have measured.


\subsection{Bunch shape and charge distribution}

Our experiments were performed during the period from 14-24 October 2018 in the End Station A hall of the SLAC National Accelerator laboratory. Beam parameters were determined by the needs of primary users of SLAC at the time, the Linac Coherent Light Source (LCLS), for which bunches were supplied at bunch charges of around 0.25 nC ($N_e = 1.6 \times 10^9$), at a rate of about 100 Hz and beam energies of 14.5 GeV. End Station A's test beam experiments received about 5 Hz of these same bunches, and beam tuning requirements are dictated by the primary users, leading to periods when we were unable to operate with stable conditions; data was not taken during these conditions to avoid uncertainties in the analysis.

 \begin{figure}[htb!]
 \includegraphics[width=3.5in]{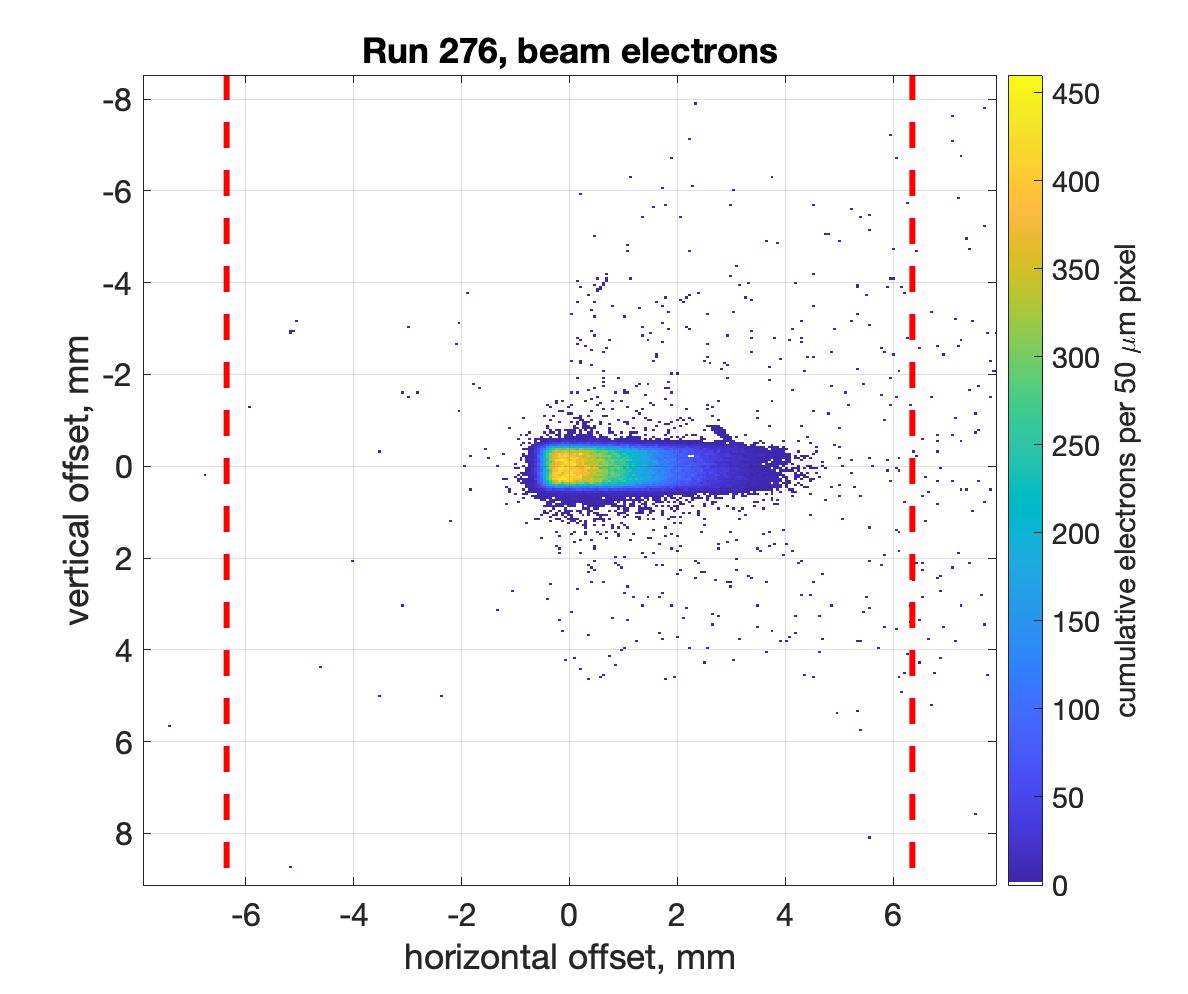}
 \caption{ Composite image of the beam primary electron distribution arriving at the ePix 100 silicon detector, showing the pattern due to imperfect collimation of the beam arriving at End Station A. The vertical dashed lines show the location of the waveguide side wall inner edges.
 \label{ePix1}}
 \end{figure}

As noted above, a key goal of these tests was to establish the absolute scale of the microwave emission, tied to the total energy of the transiting shower, which is the composite energy of the electron bunch. To do this we used the ePix camera~\cite{ePix} to determine $N_e$. The ePix 100, with a read noise of order $50$~e$^-$ rms, gives an SNR of $>40$ for $\sim 8~$KeV photons, and detects single high-energy electrons with effectively no background apart from ambient cosmic rays or radiation. In practice its accuracy in the electron beam count measurement far exceeded our requirement for about 3\% accuracy, given other experimental limitations. 

Fig.~\ref{ePix1} shows a composite camera frame for one of our runs during the test, approximately 2500 beam shots. Due to limitations in the collimation of the beam upstream of End Station A, the beam arrived with a non-uniform rectangular pattern as seen in the figure, with a Full-Width at Half-Maximum (FWHM) in each dimension of about $1 \times 1.75$~mm. This shape was therefore encoded for use in estimating the effect of phase factors in reducing the waveguide response; the orientation of the waveguide side walls is shown in the figure with dashed red lines.

\begin{figure}[htb!]
 \includegraphics[width=3.5in]{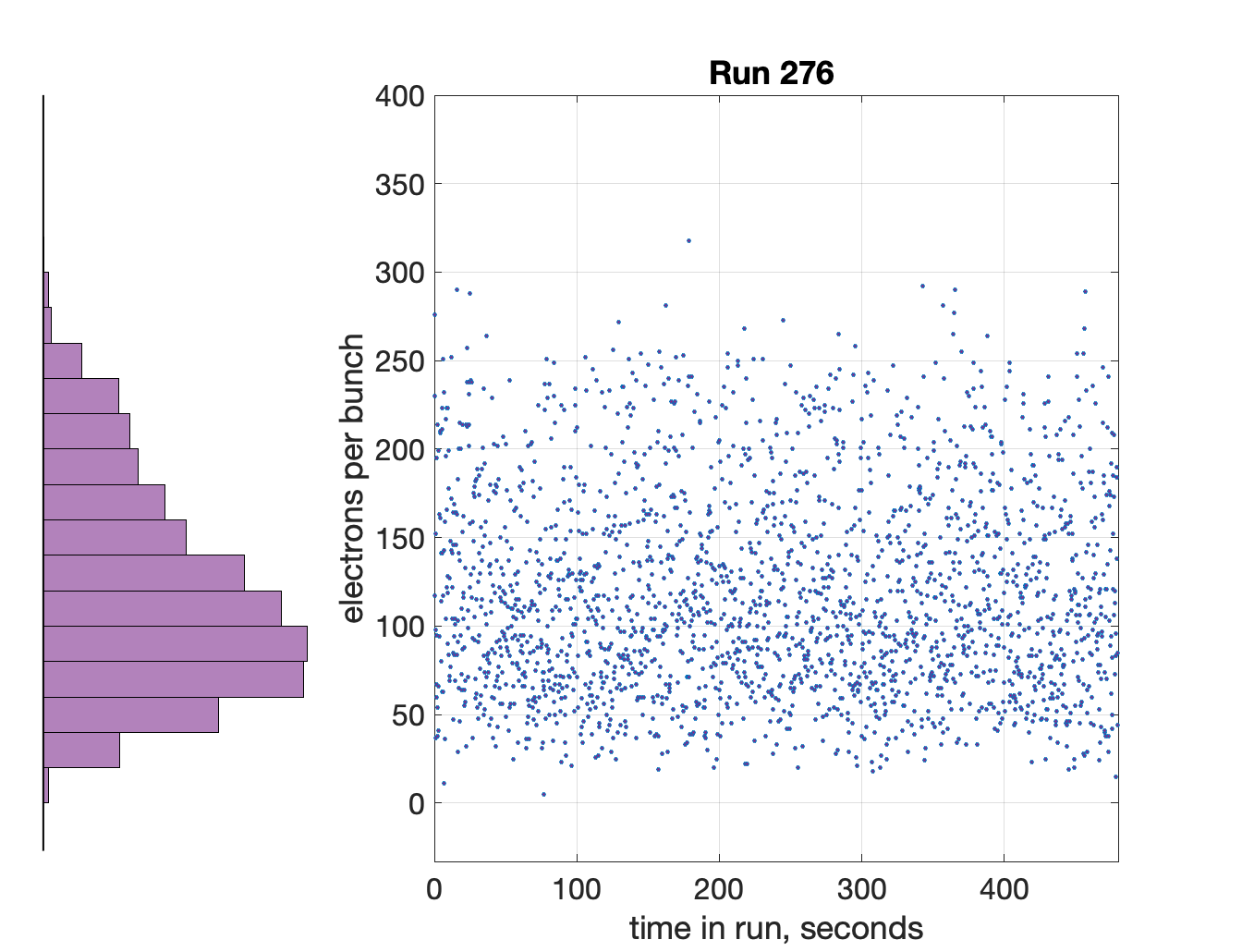}
 \caption{ For the same run shown in Fig.~\ref{ePix1}, we show the individual frame electron counts per bunch, and a marginal histogram on the left, showing the distribution. 
 \label{ePix2}}
 \end{figure}
 
Fig.~\ref{ePix2} shows the individual frame counts of electrons per bunch integrated over the detector frame, along with the distribution, for which the mean and standard deviation for this run is $N_e = 121 \pm 62.6$ electrons per bunch. The electron energy was 14.5~GeV, leading to a total shower energy of $1754\pm 907$~GeV. 
 
Limitations of our data acquisition system prevented us from keeping up with the 5 Hz rate of bunch crossings in the detectors, and while the ePix detector was able to record every bunch, we were unable to synchronize our data to provide a precise measurement of the bunch energy for each of our detected events in ACE.  The ePix data are thus used to estimate statistical quantities related to shower energy during the runs, thus achieving calibration of the response for the statistical ensemble of events rather than individual events.

\subsection{ACE element response.}

\begin{figure}[htb!]
 \includegraphics[width=3.5in]{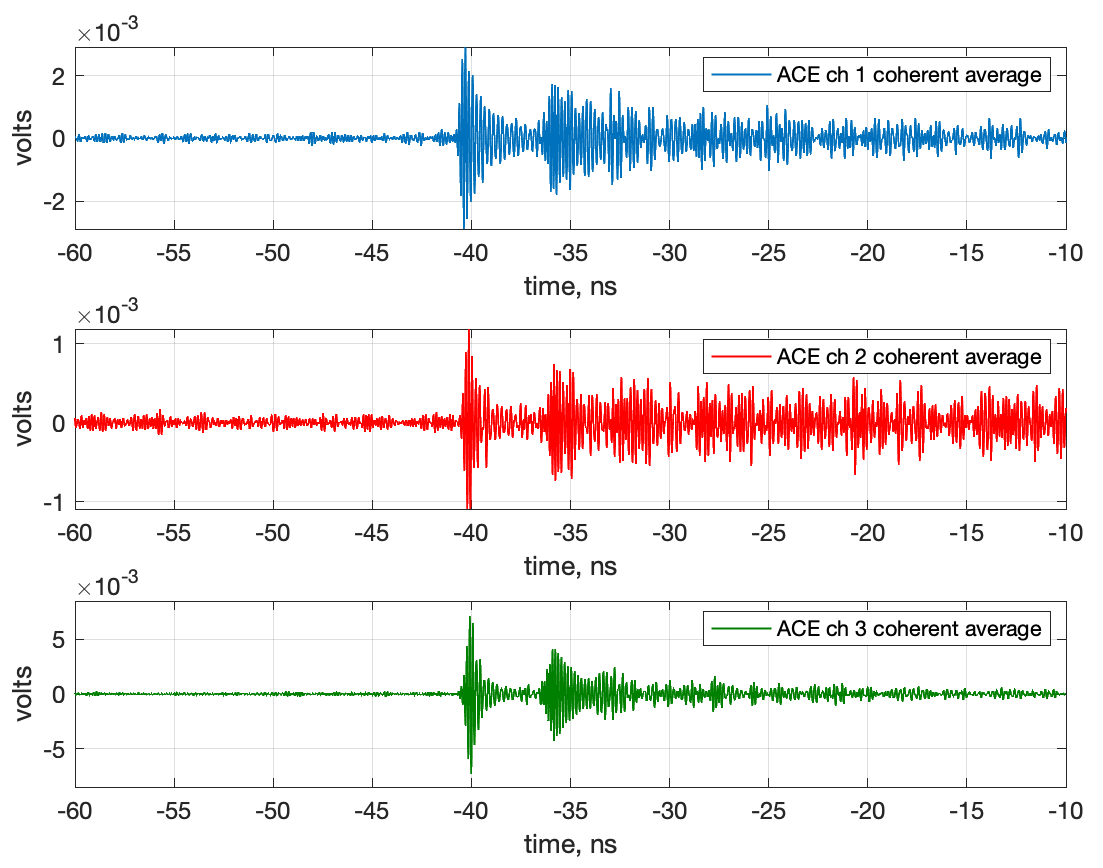}
 \caption{Coherent average of runs 268-292 (2400 ACE events), showing all three ACE waveform responses, including the reflection from the shorted end about 4~ns after the initial pulse, to showers generated by electron bunches with parameters similar to those shown in Figures~\ref{ePix1} and \ref{ePix2}. The first two elements (in blue and red) had lower efficiency due to degradation in the coupling at liquid helium temperatures. The third element (green) retained its full design efficiency and is used for all amplitude-dependent estimates of absolute response.
 \label{ace_avg}}
 \end{figure}

The three ACE elements used for shower detection were fabricated with a newly designed waveguide-to-coaxial adapter, which was fully tested at 77K using liquid nitrogen prior to the SLAC experiment, but could not be tested to liquid helium (LHe) temperatures prior to the beam test due to constraints on LHe availability in Hawaii. The adapters were designed to accommodate to the thermal strains expected, and all survived at 4.2K, but two of the three elements experience a loss of transmission efficiency at the adapter interfaces, leading to reduced signal in the main pulse and larger degrees of ringing in the tails of the measured pulses. 
Fig.~\ref{ace_avg} shows the coherently  averaged profile of the received signals for a range of runs similar to that shown in Figures~\ref{ePix1} and \ref{ePix2}. 

Tests of the impedance matching of the third ACE element before and after LHe immersion showed that its adapters on either end remained within the acceptable range of at least -10 dB return loss ($>90$\% power transmission efficiency) within the passband. The first two elements show factors of $\sim 2$ and $\sim 5$ lower amplitude coupling efficiency however, and the larger reflection amplitudes in the tails are also correlated to this effect. While these elements still preserve pulse shape and timing information, we do not use them for absolute scaling, and all amplitude-dependent quantities are referred only to the third element which retained its design efficiency.

 The shape of these pulses, which include a 5-8GHz bandpass filter, compare well to predictions from the FDTD simulations in Fig.~\ref{fdtd-wfms} above, reproducing the primary and reflected pulse as well as their dispersion, and the ringing in the tail which is due both to residual non-propagating modes and multiple reflections. These coherently averaged measurements provide effective template waveforms for estimating the location of the pulses for individual waveforms using matched filter or cross-correlation methods, under conditions where the signal may be otherwise buried in the noise. (In a practical timing plane application each detector would come with a calibration archive covering its entire active area; such archives would be developed using both direct measurements and validated time-domain models.)
 
 \subsection{Estimates of shower energy threshold.}
\label{SNRest}

We can approximate the observed SNR $S_{obs}$ in a single ACE element as a combination of the intrinsic SNR $S_0$ for the ideal case, and loss factors from two sources:
$$S_{obs} =F_{\phi } G_{gap} ~S_0$$
where $F_{\phi }$ is the loss factor due to partially destructive interference from the spatially dispersed form factor of the shower as it enters the ACE element, and $G_{gap}$ is the amplitude loss factor due to the dilution of the shower peak current density, arising mainly from the gap between the tungsten and the detector, but also affected by the additional intermediate material. The ideal case assumes that the shower is produced in a 15mm tungsten absorber just upstream of the ACE elements.

To determine these factors, we utilize the results from GEANT4 simulations of the ACE3 as-built system, shown in Fig.~\ref{ACELfit1} above, compared to the simulations, shown in Fig.~\ref{noGaps}, where we remove the gaps and intermediate materials. These results are then also be convolved with the shape of the electron distribution as observed by the ePix camera, although it turns out that this convolution leads to only a few percent loss in amplitude.

To estimate the phase factors, we use a Monte Carlo (MC) numerical integral which distributes the electrons across the convolution of the GEANT4 shower distribution and the input electron distribution. The MC computes the electric field phase factor for the Alumina-loaded WR51 $TE_{10}$ mid-band frequency 6.5 GHz using:
 $$E_{net} =E_0 \sum_{j=1}^N \cos \left(\frac{\pi }{2}\frac{|\mathbf{x_j} \cdot \hat{y} |}{w}\right)\exp (-i~\mathbf{k}\cdot \mathbf{x_j} ))$$
where $E_0$ is the electric field per single electron centered in the guide,  $k=(\omega n/c)\hat{z}$ is the wave vector along the waveguide axis, for frequency $\omega =2\pi f$ and index of refraction $n$, and $x_j$ is the position vector of the $j^{th}$ transiting electron at its internal trajectory midpoint as it passes through the waveguide and normal to it.   The cosine term applies the half-cosine factor to the amplitude depending on the transverse location (taken as the $\hat{y}$ direction) in the guide, which has half-width $w$.
 
For the ACE CH3 element, which we use as the reference for all of these calculations because it had the best SNR and efficiency, the fitted values are shown in Figures~\ref{ACELfit1} and \ref{noGaps} above.
With the fitted parameters, the MC can then generate the relevant distributions of excess electrons. When this is done, we find that the amplitude factor for no gaps is $E_{net}^{(n)} =0.442$ and with the gaps and intervening dewar $E_{net}^{(g)} =0.232$, each with negligible statistical error. The ratio of the two gives the relative loss factor for the observed SNR:

$$F_{\phi } =\frac{E_{net}^{(g)} }{E_{net}^{(n)} }=0.525$$

The amplitude factor $G_{gap}$ is estimated using the ratio of GEANT4 estimate of the total excess charge for the two cases: using the slices above (which have adequate statistics), the ratio gives $G_{gap} =851/3223=0.264\pm 0.01$. (For later use, we note that our GEANT4 results for ACE CH1 and ACE CH2 give $G_{gap} =0.142,~0.191$ for those channels respectively).

Combining these we have for the observed SNR for the as-built system as a function of the idealized SNR for a close-packed system:
$$S_{obs} =F_{\phi } G_{gap} ~S_0  =~0.139~S_0$$
The combined loss factors reduce the observed amplitude by about a factor of 7 compared to an ideal, close-packed system. 

\label{run312sect}
To estimate the effective shower detection threshold, we first must estimate the observed typical signal-to-noise ratio (SNR) per event in our reference channel 3 for a known mean value of $N_e$ and electron energy. To get the observed SNR, we need to average over many events since the individual SNR per event is relatively low due to the loss factors. To ensure that our coherent signal averaging did not introduce additional unmodeled losses, we need to compensate for the trigger jitter, which was of order 1 ns since the trigger was based on the MPPC Cherenkov detector, and the risetime was quite slow compared to the ACE signals. 

\begin{figure}[htb!]
 \includegraphics[width=3.5in]{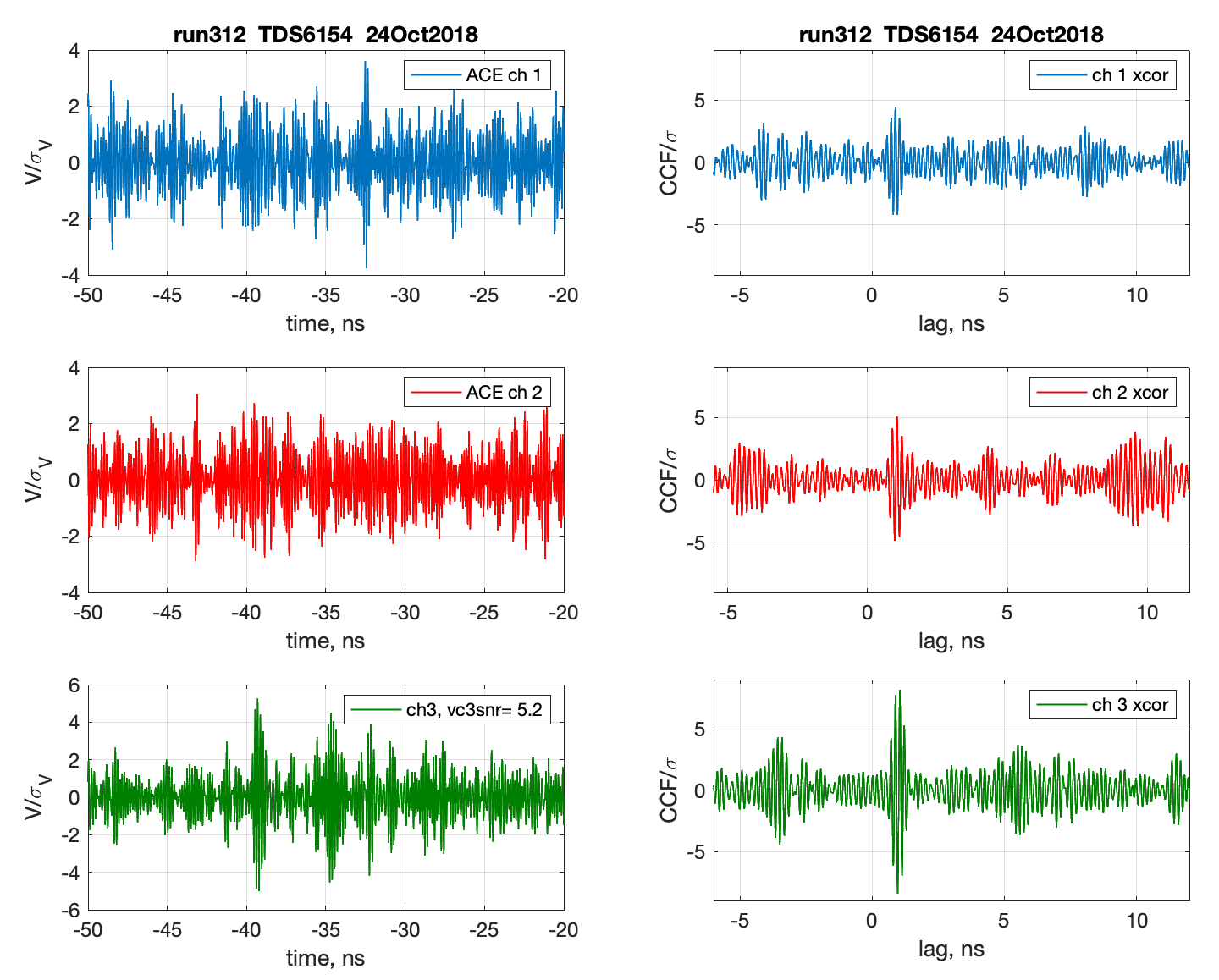}
 \caption{ Left: raw waveforms for a single event in run 312. Right: cross-correlation functions (CCF) with the signal templates. Slight offsets in the template reference time account for the lag differences in the CCF.
 \label{run312event}}
 \end{figure}

We have done this iteratively by first constructing templates with approximate averages, cross-correlating and averaging events to then improve on the templates, and repeating this process until the SNR of the combined average did not increase. To illustrate the importance of the template cross-correlation in ACE we show an example of a single event from run 312, where the waveform SNR in channel 3 is estimated to be $\sim 5.2$, in Fig.~\ref{run312event}. On the left-hand side the raw, thermal-noise-dominated waveforms are shown, with the signal arrival time expected to be around -39 ns on the time scale shown. On the right side the results of the template cross-correlation are shown for each channel, and the signal is detected in each case at a substantially improved SNR, typically 50\% or more. This step is also fundamental to the arrival time estimate, which is provided by the peak of the cross-correlation function.

For the sequence of runs used in this estimate, a total 2475 events were averaged for each waveform -- this was the complete set of waveforms within the run set. No rejection of individual events was done so as to avoid any bias in the resulting SNR, and to facilitate comparison with the results of the ePix camera. 

 The result of the coherent sum for all three ACE channels is shown in Fig.~\ref{ace_avg} above, and the SNR per channel, measured by taking the ratio of  the peak amplitude of the pulse to the RMS voltage of the waveform before the onset of the pulse,  is 39.82, 22.65, and 106.43 for channels 1,2, and 3 respectively. 

For thermal-noise dominated signals, the SNR grows as $\sqrt{N}$ for $N$ events, and we have confirmed this behavior for our data. For the reference channel 3, the average SNR per event is thus ${\bar{S} }_{obs} =(106.43/\sqrt{2475})=2.14$. For this observed SNR the implied idealized SNR $S_0$ for the third ACE element is

$$S_0 =\frac{\bar{S}_{obs} }{F_{\phi } G_{gap} }=\frac{2.14}{0.139}=15.44$$ 

This value represents our estimate of the SNR per event in channel 3 for a close-packed system. Because our minimal timing layer design is a 3-element combination, we also estimate the equivalent SNR for the two upstream elements, if they had retained their full coupling efficiency.

Using the GEANT4 for the double Lorentzian for the two upstream elements in the ideal case, fits, shown in Figures~\ref{ACELfit1} and \ref{noGaps} above,  and recomputing $F_{\phi }$  and $G_{gap}$ for those elements, we find slightly higher expected SNRs for both of these than for the 3rd element, by factors of 1.430 and 2.010 for the 2nd and 1st elements, respectively. For a three-element coherent sum with independent LNAs (and thus additional thermal noise), the expected measurement SNR at a mean shower energy of $E=1754$ GeV is thus  $(1+1.430+2.010=4.44)/\sqrt{3}$ higher than the single-element SNR, and thus 

$$S_0 (3~{\rm element})=\frac{4.44}{\sqrt{3}}(15.44)=39.57$$

Since the signal depends linearly on the excess charge which in turn depends linearly on shower energy, the threshold for a $5\sigma$ amplitude measurement for a timing plane with three ACE elements in the shower is:

$${E_{thr} (5\sigma,~{\rm amplitude} )=5\left(\frac{1754}{39.57}\right)=222~{{\rm GeV}}}$$

This assumes LHe cooling, and a pre-shower similar to what we have modeled here, with the three ACE elements in close proximity to the pre-shower block. The statistical error in this estimate, given the large number of events averaged, is at the several percent level and omitted here. The systematic error is at this point uncertain.

This threshold value is specific to the measurement of the signal {\it amplitude}, not the timing of the shower arrival. For shower arrival timing, we utilize template cross-correlation functions, as described in more detail below (section~\ref{CCFdetails}), and this leads to a significant improvement in the energy threshold for timing. For 25 runs where the average amplitude SNR was $>2$ and the CCF to amplitude ratio could be accurately measured, we find the SNR improvement to be SNR(CCF)/SNR(V)$= 1.53 \pm 0.17$, where the error a convolution of both statistical errors (with 25 samples) and systematics associated with the non-linear CCF process.

Combining this factor with the amplitude threshold results in a $5\sigma$-level timing threshold of 

$$E_{thr} (5\sigma,~{\rm CCF~timing} )=\frac{222~{\rm GeV}}{1.53\pm 0.17} = 145\pm 16~{\rm GeV}$$

where we quote an average standard error between the two asymmetric upper (+18~GeV) and lower (-14~GeV) errors.

We have one additional set of contiguous runs, enumerated 311 to 330, taken 24 Oct. 2018, for which we can repeat the amplitude  analysis to get an independent estimate.
For these runs, the shape of the electron spot as determined by the collimators was unchanged, but the average number of electrons was significantly higher per frame: $\langle N_e \rangle =263$.

\begin{figure}[htb!]
 \includegraphics[width=3.5in]{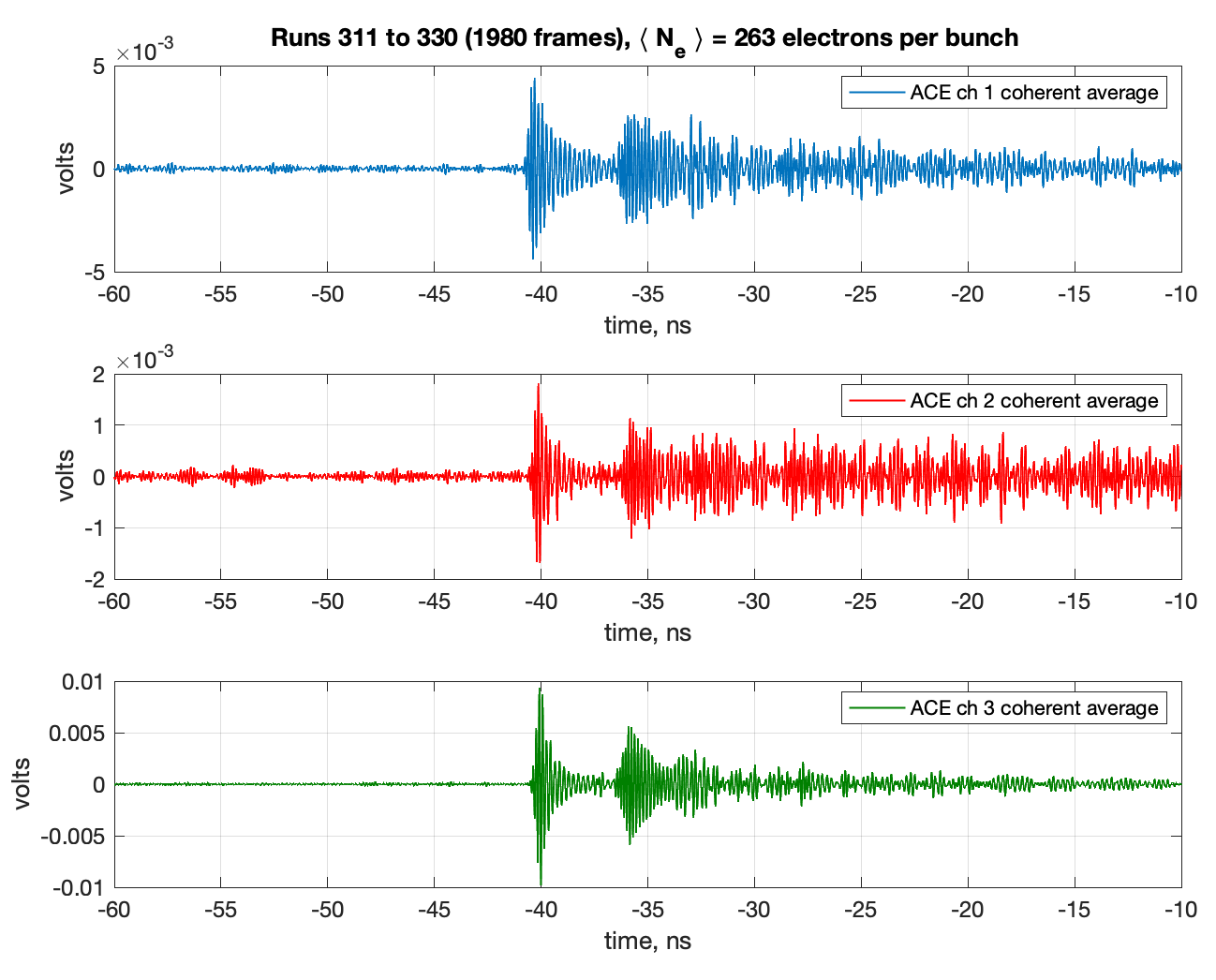}
 \caption{Coherent average of runs 311-330 (1,980 events), similar to Fig.~\ref{ace_avg} above.
 \label{runs311avg}}
 \end{figure}

The averaged waveforms are shown in Fig.~\ref{runs311avg}
and the SNR for each averaged waveform was: ACE CH1: 47.0, ACE CH2: 30.4, ACE CH3: 154.8.
The number of events averaged was 1,980 over the 20 runs. 

For ACE channel 3, since the SNR grows as $\sqrt{N}$ for $N$ events, the average SNR per event is 

$${\bar{S} }_{obs} =(154.8/\sqrt{1980})=3.48$$.

For this observed SNR, since the loss factor terms $F_{\phi}$ and $G_{gap}$ are the same, the implied idealized SNR $S_0$ for the third ACE element is

$$S_0 =\frac{S_{obs} }{F_{\phi } G_{gap} }=\frac{3.48}{0.139}=25.03$$

The shower energy per bunch is now $E_{bunch} =14.5~{\mathrm{G}\mathrm{e}\mathrm{V}})\langle N_e \rangle =14.5\times 263=3814~{\mathrm{G}\mathrm{e}\mathrm{V}}$. Using the same GEANT4 results as the previous section, the idealized 3-element detector SNR per average event is thus:

$$S_0 (3~element)=\frac{4.44}{\sqrt{3}}(25.03)=64.16$$

The implied threshold for a $5\sigma$ amplitude measurement for these data, for a timing plane with three ACE elements in the shower is:

$${E_{thr} (5\sigma~{\rm amplitude} )=5\left(\frac{3814}{64.16}\right)=297~{{\rm GeV}}}$$

and the corresponding CCF timing threshold is 

$$E_{thr} (5\sigma,~{\rm CCF~timing} )=\frac{297~{\rm GeV}}{1.53\pm 0.17} = 194 \pm 22~{\rm GeV}$$


These two results taken together give a first-order estimate of the systematic error in both amplitude , and taking the arithmetic mean, and the variance, we find the ideal $5\sigma$ thresholds, for liquid Helium cooling on the whole detector, are

$${E_{thr} (5\sigma~{\rm amplitude} )\simeq 260\pm 53~{{\rm GeV}}}$$
$${E_{thr} (5\sigma~{\rm CCF~timing} )\simeq 170\pm 39~{{\rm GeV}}}$$

The equivalent ideal 3-element $1\sigma$ thresholds, or {\it least-count} energies $E_{\ell c}$ are

$${E_{\ell c} (1\sigma~{\rm amplitude} ,~{\rm 3~element})\simeq 52\pm 11~{{\rm GeV}}}$$
$${E_{\ell c} (1\sigma~{\rm CCF~timing} ,{\rm~3~element})\simeq 34\pm 8~{{\rm GeV}}}$$

The equivalent ideal single-element $1\sigma$ least-count energies are 

$${E_{\ell c} (1\sigma~{\rm amplitude} ,{\rm~1~element})\simeq 90\pm 18~{{\rm GeV}}}$$
$${E_{\ell c} (1\sigma~{\rm CCF~timing} ,{\rm~1~element})\simeq 59\pm 14~{{\rm GeV}}}$$

While the $1\sigma$ energy threshold values may be of only academic interest from the detection point-of-view, we will later address configurations of the timing planes that can better optimize for lower effective thresholds, and these values will facilitate scaling to these new configurations. In addition, since in practical applications timing planes will be used to estimate transit times for particles or showers with independently established parameters, useful information can be retrieved for SNRs well below the $5\sigma$ level.
\vspace{5mm}

\paragraph*{Liquid Nitrogen or liquid Argon cooling.}

If the system is cooled via liquid nitrogen or liquid argon, the ACE energy threshold will increase as 
$$E_{thr} \propto \sqrt{\frac{T_{sys,{\rm LN2/LAr}}}{T_{sys,{\rm LHe}} }}$$
where the system temperatures in each case depend on the sum of three terms, the LNA noise temperature $T_{LNA}$, the effective temperature $T_{\Omega}$ of the waveguide material due to high-frequency ohmic losses, and the additional emissivity $\eta$ of the $Al_2O_3$ dielectric which depends on the loss tangent and yields and additional thermal noise component of $T_{\epsilon} = \chi T_{amb}$ for ambient temperature $T_{amb}$.

Cryogenic LNA noise temperatures typically decrease as $T_{amb}^{-1}$ down to around 40~K, below which internal blackbody phonon bundles prevent efficient cooling of the device junction temperatures~\cite{Schleeh2014} and lead to a slower reduction of noise figure with temperature. For our LNAs, the room temperature noise temperature is $T_{LNA}(296~K) = 32$~K, implying $T_{LNA}(77-90~K) = 8-10$~K, a factor of 4-5 above the LNA noise temperature at LHe temperatures.

At 77-90~K, copper and silver have conductivities about a factor of 3-5 higher than at room temperature, and these result in additive thermal noise of $\sim 2.2$~K per meter of waveguide length. For an average path length of $\sim 0.5$~m, we expect 1.1~K of increased noise from ohmic losses. (At 4-5K, copper and silver resistivities decrease another factor of 5 or more, implying about 0.2-0.3K of ohmic contributions for LHe temperatures).

Since the loss tangent of high-purity alumina is exceptionally low, generally $\leq 2 \times 10^{-4}$ at room temperature and $\leq 10^{-5}$ at  $T_{amb} \leq 100$~K~\cite{Molla_cryoAlumina}, even at liquid argon temperatures, $\chi \simeq 0.033$ and the alumina contributes at most $T_{\epsilon} \simeq 0.3$~K per meter of pathlength in the waveguide. Use of sapphire, the lowest possible loss tangent $Al_2O_3$ dielectric material would improve this somewhat, but is in this case hardly warranted, and the ready availability and low cost of high-purity alumina is an excellent choice.

\begin{table}[t!]
    \centering\   \caption{Estimated $5\sigma$ particle energy threshold $E_{thr}$ and $1\sigma$ least-count energy $E_{\ell c}$ for CCF timing (based on shower detection) for various ACE timing plane configurations as estimated by our 2018 beam test. }
    \vspace{3mm}
    \begin{threeparttable}
    \begin{tabular}{|l|c|c|}
    \hline \hline
{\it \bf configuration,} & $E_{thr}(5\sigma)$, &  $E_{\ell c}(1\sigma)$, \\ 
 {  ~~cooling}  & GeV &  GeV \\ 
\hline
{\bf 3 $\times$ single-element stack:}\tnote{a} & & \\
~~LHe (immersed), $T_{sys}=2.7$K & $170\pm 39$   & $34\pm 8$   \\
~~LHe (spot)+LN2, $T_{sys}=5$K & $231\pm 53$   & $46\pm 11$  \\
~~LAr (immersed), $T_{sys}\simeq 10.5$K & $335\pm 77$   & $67 \pm 15$   \\ \hline
{\bf 6 $\times$ dual-element stack:}\tnote{b} & & \\
~~LHe (immersed), $T_{sys}=2.7$K & $49\pm 11$   & $9.8\pm 2.3$   \\
~~LHe (spot)+LN2, $T_{sys}=5$K &$67\pm 15$   & $13\pm 3$    \\
~~LAr (immersed), $T_{sys}\simeq10.5$K &   $94\pm 22$   & $19\pm 4.4$   \\ \hline
    \end{tabular}
        \begin{tablenotes}
            \item[a] Based on what was used in our 2018 experiment.
            \item[b] Reference design for FCC-hh, as described in section~\ref{baseline}.
        \end{tablenotes}
     \end{threeparttable}
    \label{table1}
\end{table}


The net effect of these noise contributions leads to $T_{sys,{\rm LN2/LAr}} \simeq 9.5-11.5$~K, a factor of $\sim 4$ above $T_{sys,{\rm LHe}}~\simeq 2.7$~K which is dominated by $T_{LNA}$. The net increase in threshold is thus of order a factor of 2 or slightly more, depending on the cryogen used.

Table~\ref{table1} summarizes these shower energy threshold results for several different configurations of the ACE stackup, and the cooling parameters. Since $\sqrt{10.5/2.7}\simeq 2$, the thresholds and least count energies for LAr immersion are about double those of the LHe-immersed system. Thus for FCC-hh energies, there will be a large fraction of daughter particles that can be taggged to picoseond precision, even for LAr cooling.

In the following sections, where we introduce a reference design, we have not extrapolated LNA noise performance forward to the actual timeframe for FCC-hh deployment. This is likely a quite conservative approach. In fact it is already the case that the desire for quantum-noise-limited performance in microwave LNAs is being pressed by the need for the readout of microwave qubits in the field of quantum information science, and new microwave amplifiers are being developed which have already achieved the quantum limit of 0.1-0.2~K in very cold cryogenic states~\cite{KIT2020}. Such advances, if they can be made to operate at LAr temperatures, could push the usable least-count energy in an ACE system down to several GeV, making it feasible for timing planes for almost all collision products of interest for the FCC-hh.

\subsection{Transverse response function.}

Rectangular waveguide $TE_{10}$ mode electric fields have a characteristic half-cosine amplitude relative to the centerline of the waveguide, that is, if $x$ is the transverse coordinate (along the wider dimension $a$ of the rectangular section) relative to the center, the field strength $E(x) = E(0) \cos(\pi x/(2a))$. Since the shower current and corresponding vector potential is aligned with the $TE_{10}$ field, we may expect that the coupling coefficient would also have a similar dependence on the transverse offset of the current relative to the waveguide axis.

\begin{figure*}[htb!]
 \centerline{\includegraphics[width=5.5in]{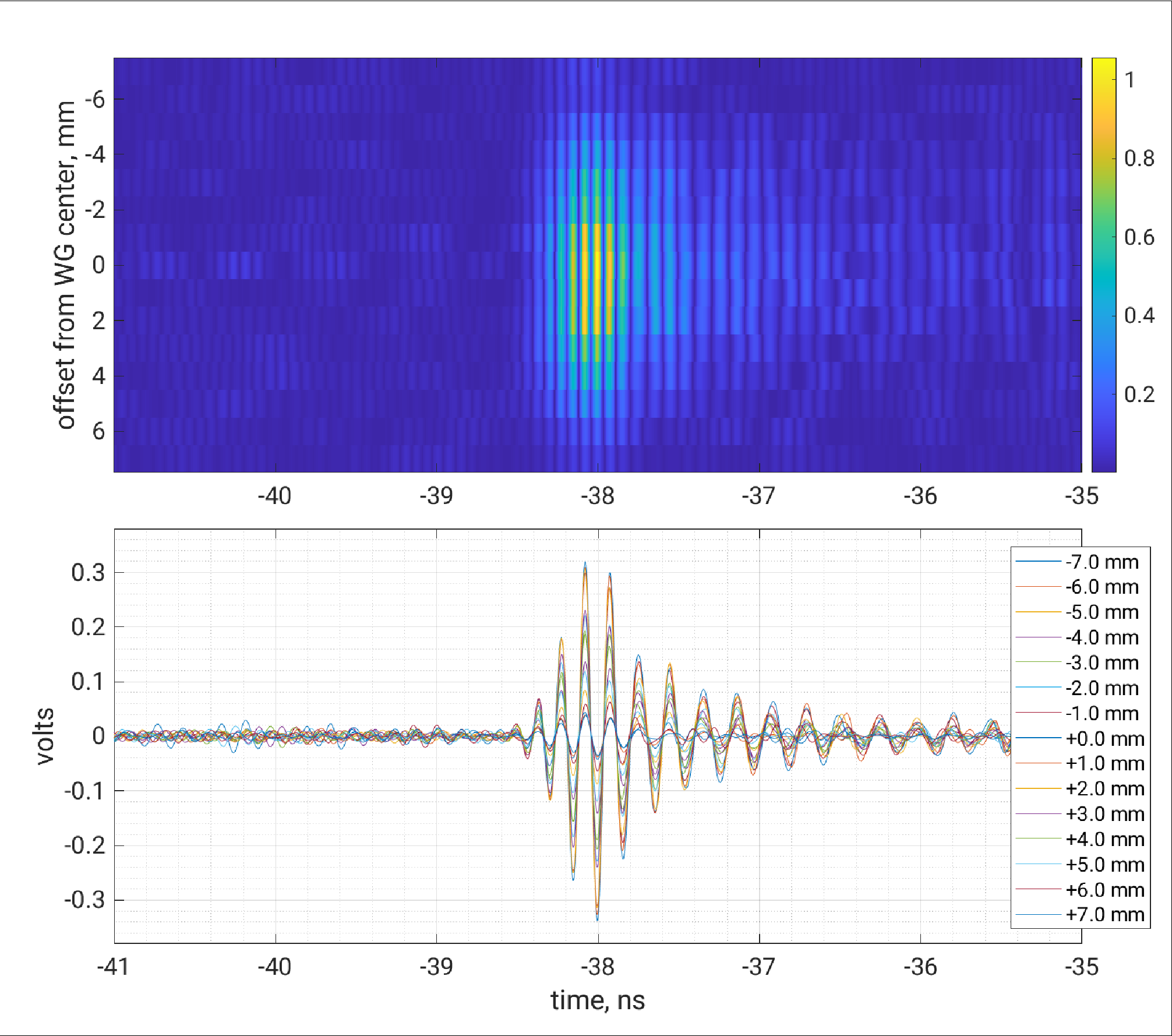}}
\caption{Top: a two-dimensional image of the amplitude of the time-domain response in the vicinity of the primary shower pulse. The vertical axis is offset relative to waveguide center, and the horizontal axis is time. Bottom: overlay of the waveforms for all offsets, labeled by their offset relative to center. 
 \label{offsetimage}}
 \end{figure*}
 
We investigated this response function during a portion of our tests when the beam was relatively stable in both the transverse bunch profile, and in beam current. Fig.~\ref{offsetimage} shows results of these measurements, with the top left pane displaying a two-dimensional image of the amplitude vs. offset and time for the main shower pulse. Waveforms have been aligned via cross-correlation, since they come from separate runs where phase alignment could not be maintained. Thus while the waveform shape is very similar for all of the offsets, we cannot yet confirm that the timing for such offsets is the same as for the center timing.

\begin{figure}[htb!]
 \includegraphics[width=3.5in]{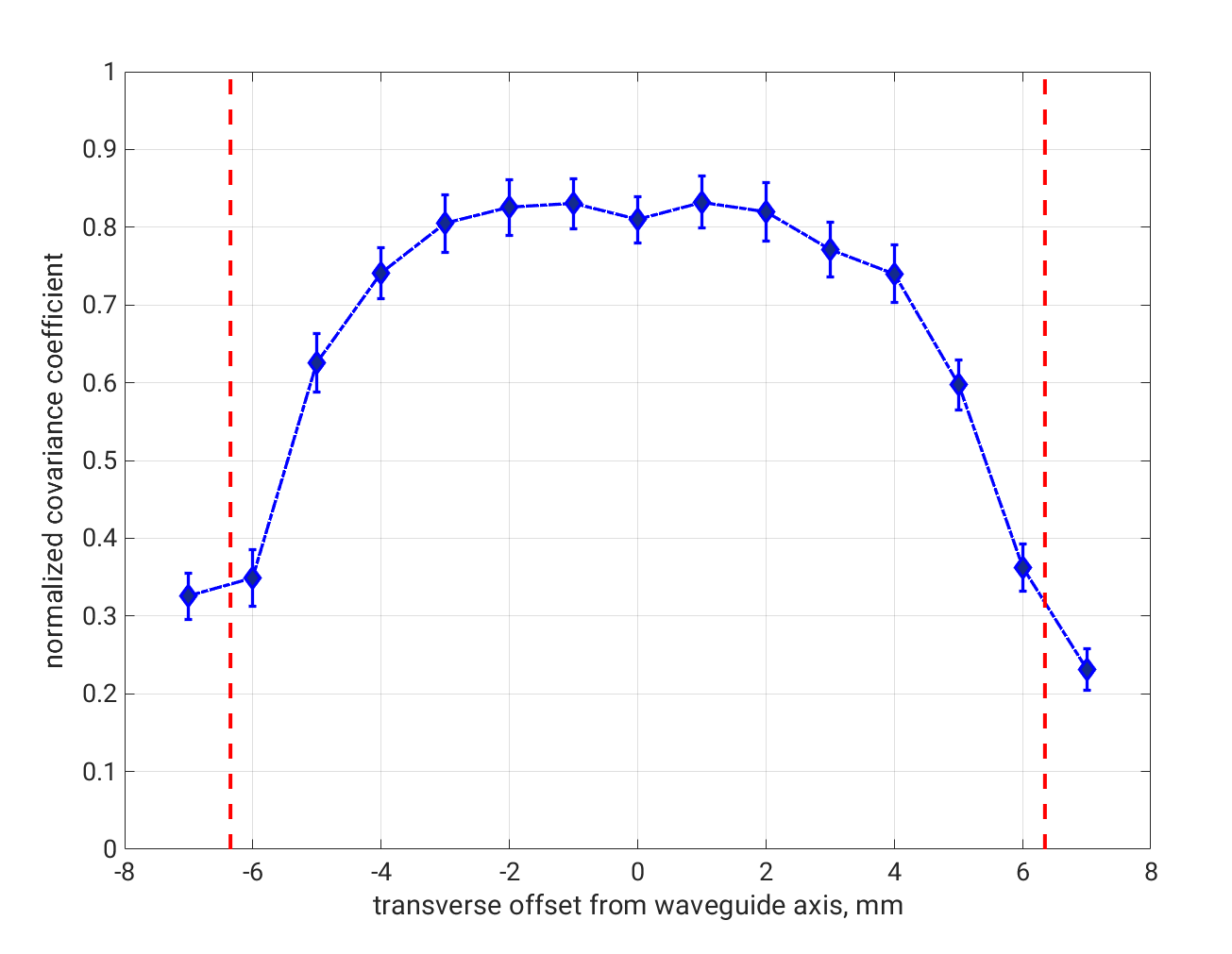}
 \caption{Cross-covariance coefficient for each channel 3 waveform in the offset data sample, vs. the CH3 template waveform.
 \label{xverseCov}}
 \end{figure}
 
 Related to the question of possible timing offsets associated with off-center beam centroids is the question of the degree of correlation between off-center-generated waveforms and the on-axis waveform. This may for example suggest that complete calibration of a waveguide element may require calibration for off-center waveforms as well as on-axis waveforms.  Fig.~\ref{xverseCov} shows the normalized cross-covariance coefficient of all CH3 waveforms from the offset scans with an on-axis template waveform. The covariance coefficient saturates at about 0.83 in the central region, consistent with a slight decrease in covariance due to the residual thermal noise present. The covariance measure maintains this high degree of correlation for about 60\% of the width, and then declines near the edges. Note that this normalized measure is to first order independent of the amplitude of the signals, but some of the decline is likely due to the lower SNR relative to thermal noise of these waveforms. This result suggests that in practice templates for offset positions could be used to advantage in a template library for shower measurements, but it also indicates that centerline templates would perform reasonably well even for showers near the waveguide edges.

\begin{figure}[htb!]
 \includegraphics[width=3.5in]{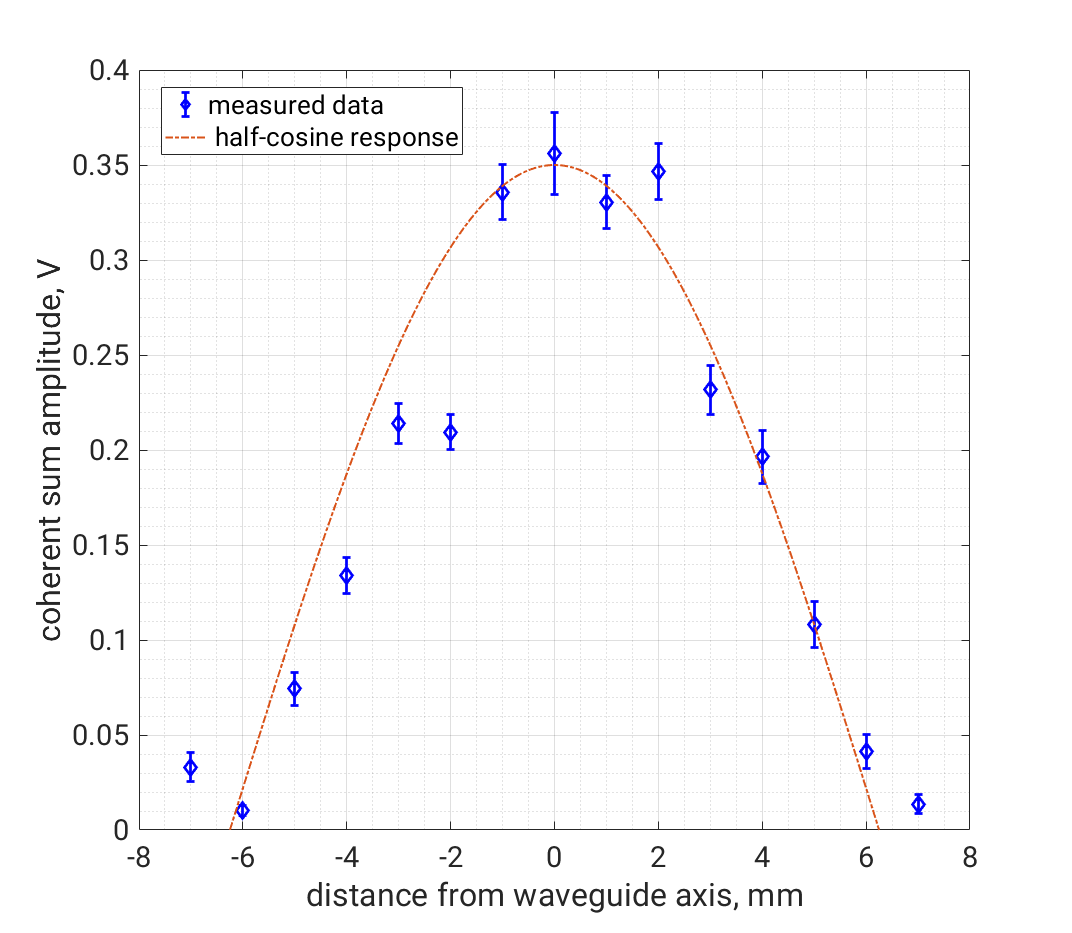}
 \caption{ Response function of the amplitude vs. transverse offset from waveguide axis. The dashed curve shows the expected half-cosine response.
 \label{xresponse}}
 \end{figure}

Fig.~\ref{xresponse} shows a summary of the amplitude variation with offset, along with a half-cosine curve for comparison. We have attempted to deconvolve the asymmetric shape function of the beam spot combined with the GEANT4 shower shape, but the results indicate that the deconvolution was only partially successful, as there remains an asymmetry in the response function. While the expectations are generally confirmed, precise details will likely require tests with a more symmetric and compact beam spot.

The falloff of the amplitude of the response of the waveguide elements near the edges motivates an improved design that includes layers with positions that alternate the centers and edges of the waveguides to ensure more uniform coverage. In a later section we will introduce such an updated design and assess results via simulations which are informed by these beam test results.

 \subsection{Event timing results.}
 \label{CCFdetails}
 
We have addressed the context within which our SLAC timing-plane measurements have been made; each of these amplitude-dependent issues sets scales which will be useful later in applying these results to simulations. For timing, the temporal shape of the waveform is the more critical parameter. To establish that we first use high-beam-current data where the signal is relatively strong, and create a waveform template by averaging many events over a portion of the observations where the beam and experimental configuration are stable. The resulting templates for each channel are then used to generate cross-correlation functions  between the template and an individual event, as shown in the example from run 312 above (Fig.~\ref{run312event} in section~\ref{run312sect} above).
 
To extract timing information from these individual beam shots, the events and templates must be resampled to a finer scale than the original sampling, which is typically 20-40 Gs/second, or 25 to 50 ps/sample. This is done using numerical interpolation methods to sampling rates that will support picosecond precision, and is feasible and numerically stable since the signal is fully band-limited to $<10$~GHz. The other requirement is that the analog-to-digital converter system used to sample the signals must have a jitter level which is below the picosecond level, and a bandwidth that exceeds the $TE_{10}$ single-mode bandwidth. For sampling we have continued to use a Tektronix TDS6154C digital realtime oscilloscope with 15 GHz bandwidth and sub-picosecond jitter specifications. We have also cross-calibrated this scope with two other TDS6804B 8 GHz bandwidth oscilloscopes to verify the timing results.
 
 \begin{figure}[htb!]
 \includegraphics[width=3.65in]{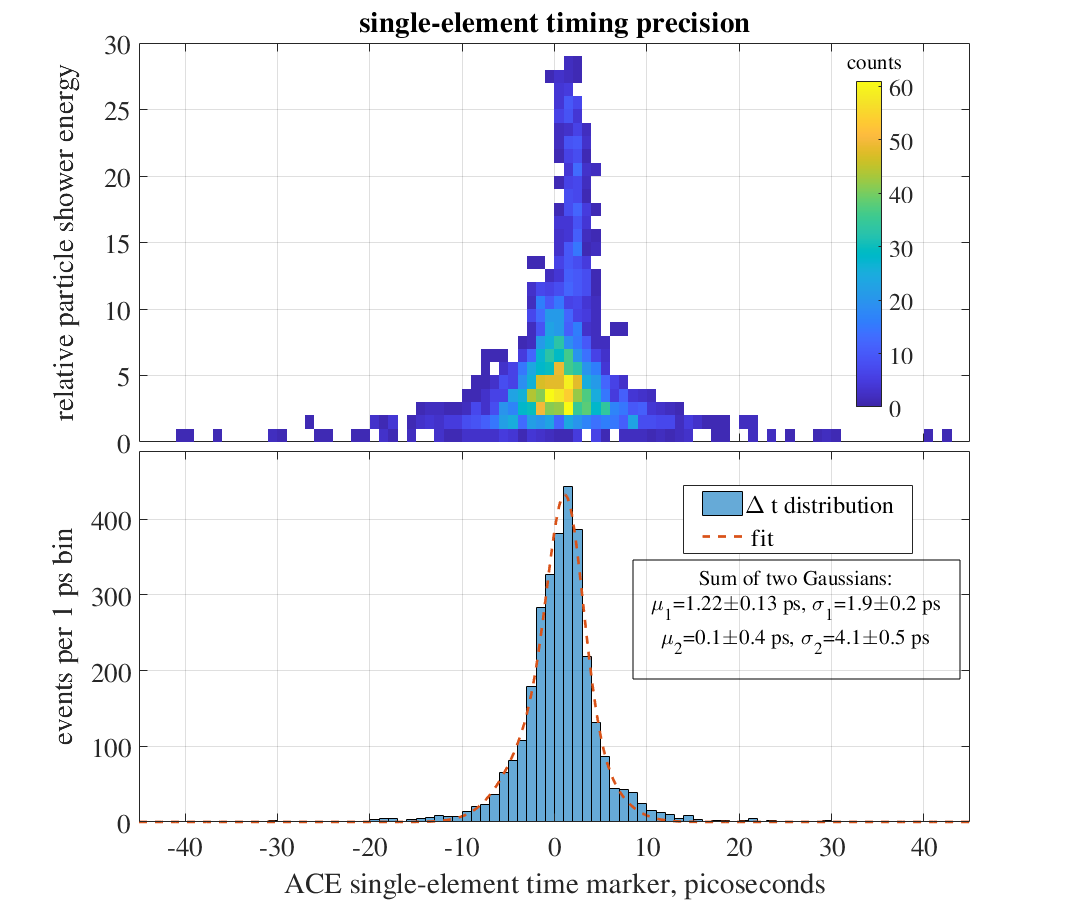}
 \caption{Top: Two-dimensional histogram of relative shower energy vs. measured delay for selected runs. The distribution at higher energies has a slightly different mean delay than the lower energy distribution. Bottom: a sum of two-Gaussian fit to the data, where the higher energies fit to a $\sim 1$~ps delay offset, and a narrower timing distribution than the lower energies.
 \label{timingDist}}
 \end{figure}

The normalized CCF computed between a measured event waveform $w(t)$ and the template $\mathcal{C}(t)$,
evaluated over a window length $T$ is given by
\begin{equation}
 \mathcal{X}(\tau) = \frac{1}{\sqrt{\mathcal{A}_w(0) \mathcal{A}_{\mathcal{C}}(0)}} \int_{-T/2}^{T/2} w(t) \mathcal{C}(t-\tau) dt
\end{equation}
where $\tau$ is called the {\it lag}, or time offset of the product of the functions, and
$\mathcal{A}_w(0)$, $\mathcal{A}_{\mathcal{C}}(0)$ are the autocorrelations of the waveform
and template, respectively, at the zero lag $\tau=0$, which provide the normalization:
\begin{equation}
\mathcal{A}_{w}(\tau)|_{\tau=0} = \int_{-T/2}^{T/2} w(t)^2 dt
\end{equation}
with a similar equation for $\mathcal{A}_{\mathcal{C}}$. In practice these equations are discretized
at the sample period of the data after interpolating to a suitably fine reference grid, typically
1~ps/sample or finer. The peak of the CCF within the window then determines the absolute delay of the waveform. Normalization of the waveforms is not essential to the timing of the peak, but it provides an estimate of the correlation coefficient which can be used to determine a standard error for the delay time estimate.

\begin{figure}[htb!]
 \includegraphics[width=3.65in]{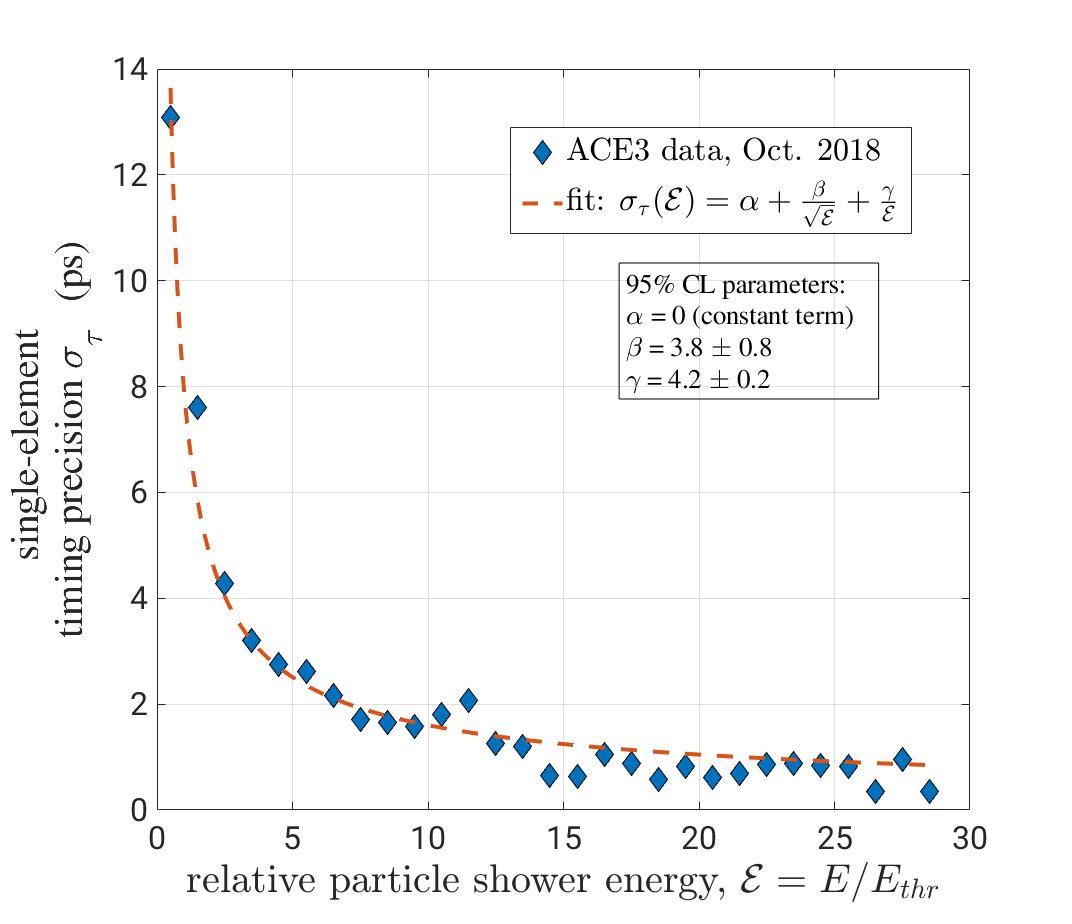}
 \caption{Single ACE element timing precision vs. relative shower energy measured in our experiment. 
 \label{timingCurve}}
 \end{figure}

Fig.~\ref{timingDist} shows the results of single-element timing in the 3rd ACE element for all of the data for which we have good estimates of the  composite bunch energy. The energy is scaled relative to the threshold energy, which for our data is affected by the loss factors detailed above. In the ideal case the single-element CCF timing threshold is $\sim 60$~GeV; in the non-ideal conditions of our experiment geometry the operating threshold was about a factor of 7 higher. 

In the upper panel of Fig.~\ref{timingDist} we show a two-dimensional histogram of the counts vs. relative energy vs. estimated delay, where the relative shower energy in each case is determined by the ePix camera scaling, and the timing is estimated via the CCF process described above. Since the distribution shows evidence for a shift in the mean value as a function of shower energy, the bottom one-dimensional histogram of the data is fit to a sum of two Gaussians, which matches the data well and shows a 1.2 ps offset that appears at higher energies, possibly due to slightly different beam steering for those bunches. The single-element standard deviation in the timing is 1.9~ps at the higher energies and grows to 4.1~ps at the low-energy end. 

Fig.~\ref{timingCurve} shows the dependence of the single-element timing precision on the relative shower energy including a fitted function analogous to what is often used in estimating calorimeter performance:
$$\sigma_{\tau}(E) = \alpha + \frac{\beta}{\sqrt{E}} + \frac{\gamma}{E}$$
where in our case the constant term $\alpha$ is consistent with zero, and the coefficients $\beta,\gamma$ for the square root and linear terms in energy are comparable. Since the timing precision is dominated by thermal noise, we can also expect the timing precision to improve with the number of elements combined as $N^{-1/2}$. Fig.~\ref{timingCurve} may thus be used in combination with the timing threshold energies reporting in Table~\ref{table1} above to estimate timing vs. shower energy. Thus for example, for the 6-element stack (detailed in the next section) shown in Table~\ref{table1}, the least count energy in the ideal case with fill LHe cooling is 9.8~GeV, implying a timing precision of $\sim 4$~ps for $\sim 25$~GeV ($2.5\sigma$) showers, and under 2~ps for 100~GeV ($10\sigma$) showers. This analysis is oversimplified however, since it does not address the detection efficiency for showers, which will drop significantly below the $5\sigma$-level, and will also depend on the type of shower, whether hadronic or electromagnetic, as well as the location and pre-shower column density of the detector. We address these details in the next section with more advanced modeling and simulations.

\section{An updated timing plane design.}

While the single-element results are useful to understand how to scale to a larger instrument with a goal toward a future collider detector, there are several avenues that we have explored, in addition to multiple detector elements, to optimize performance. These include developing couplers to join more than one element to a single low-noise amplifier, capturing a larger radiating pathlength per LNA, and creating adapters with larger bandwith  to capture a larger portion of the emission frequency space. In this section we create a reference detector design incorporating these improvements and a layer depth (in terms of the number of elements) appropriate to a possible deployment scenario in the Future Circular Collider.

\subsection{Other single-element geometries.}

Ultimately it is the single-element sensitivity which limits the energy threshold for our detectors or timing planes, since in any practical deployment, the thickness available to the timing plane will be subject to significant constraints. The single-element sensitivity is in turn constrained by the path length of charged particles transverse to it -- the induced Cherenkov field strength in the waveguide grows linearly with the path length. However, waveguides cannot be of arbitrary dimensions while retaining acceptable modal structure. 

In our previous report, we discussed the possibility of using square waveguides that are the same width (12.7 mm) but twice the height of normal WR51 waveguides. This geometry would support both a $TE_{10}$ and orthogonal $TE_{01}$ mode with twice the pathlength for particles within the waveguide. 

\begin{figure}[htb!]
 \includegraphics[width=3.65in]{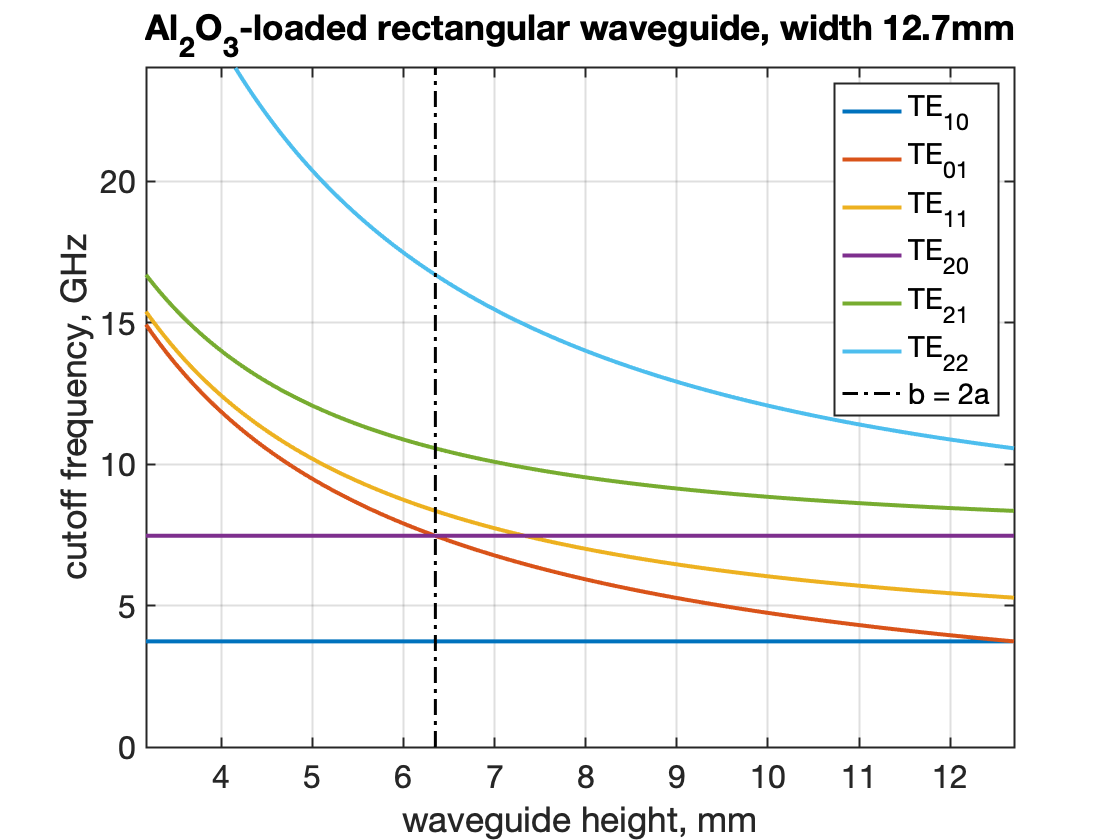}
 \caption{Cutoff frequencies for alumina-loaded waveguides of 12.7 mm width and variable height. 
 \label{wgmodes}}
 \end{figure}
 
We have further explored this option, and while it is feasible from the point of view of the waveguide modes, there are reasons why it is less than optimal in practice. Fig.~\ref{wgmodes} shows a plot of cutoff frequencies as a function of the waveguide height for Alumina-loaded waveguides with width fixed at the 12.7mm inside width of WR51. The cutoff frequency for a $TE_{nm}$ mode is given by
\begin{equation}
f_c = \frac{1}{\sqrt{\mu\epsilon} } \sqrt{ \left ( \frac{n}{2a} \right )^2  \left ( \frac{m}{2b} \right )^2 }
\label{fc}
\end{equation}
where the speed of light in the medium is given in  terms of the absolute permeability $\mu$ and permittivity $\epsilon$,  $a,b$ are the width and height of the waveguide, and we assume $b\leq a$. It is evident from the figure that increasing the height of the waveguide causes the mode frequencies to compress together in their range. At $b=a=12.7$~mm in our case, while we have doubled the pathlength, the single-mode bandwidth is now more than a factor of two smaller, with the $TE_{11}$ mode now infringing on that bandwidth from above. It is unclear whether this will in fact cause problems in practice, since the transverse current element source geometry presented by a shower are in general very inefficient in exciting these other modes.

\begin{figure}[htb!]
 \includegraphics[width=3.5in]{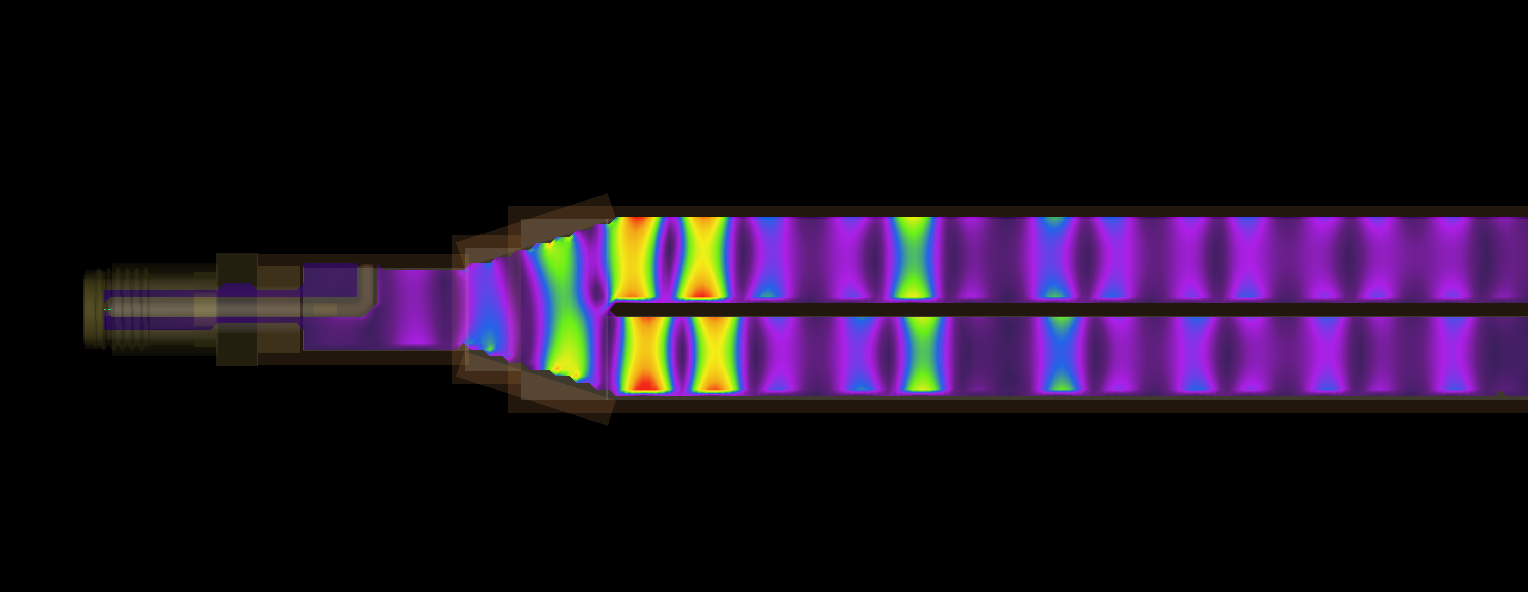}
 \caption{FDTD simulation of dual WR51 waveguide modes arriving into a 2:1 E-plane combiner.
 \label{dualSlice}}
 \end{figure}
 
To avoid this potential bandwidth issue, we have investigated two other ways to increase the shower pathlength in an element: the first is to simply use larger standard waveguide sizes, for example WR62 or WR75, which increase the shower pathlength by 22\% and 47\% respectively. The corresponding center frequencies decrease by a similar amount, as do the single-mode bandwidths available for detection. In addition, the lower center frequency leads to coarser timing precision due to the slower risetime. However, since the gain in sensitivity is linear with increasing shower pathlength (since the signal is coherent), and the loss of sensitivity goes only as the square root of bandwidth (since the noise is incoherent), there is still an improvement in sensitivity to be gained by this approach. Such trades may be acceptable for a given application, if sensitivity is more critical than timing.

\begin{figure}[htb!]
 \includegraphics[width=3.5in]{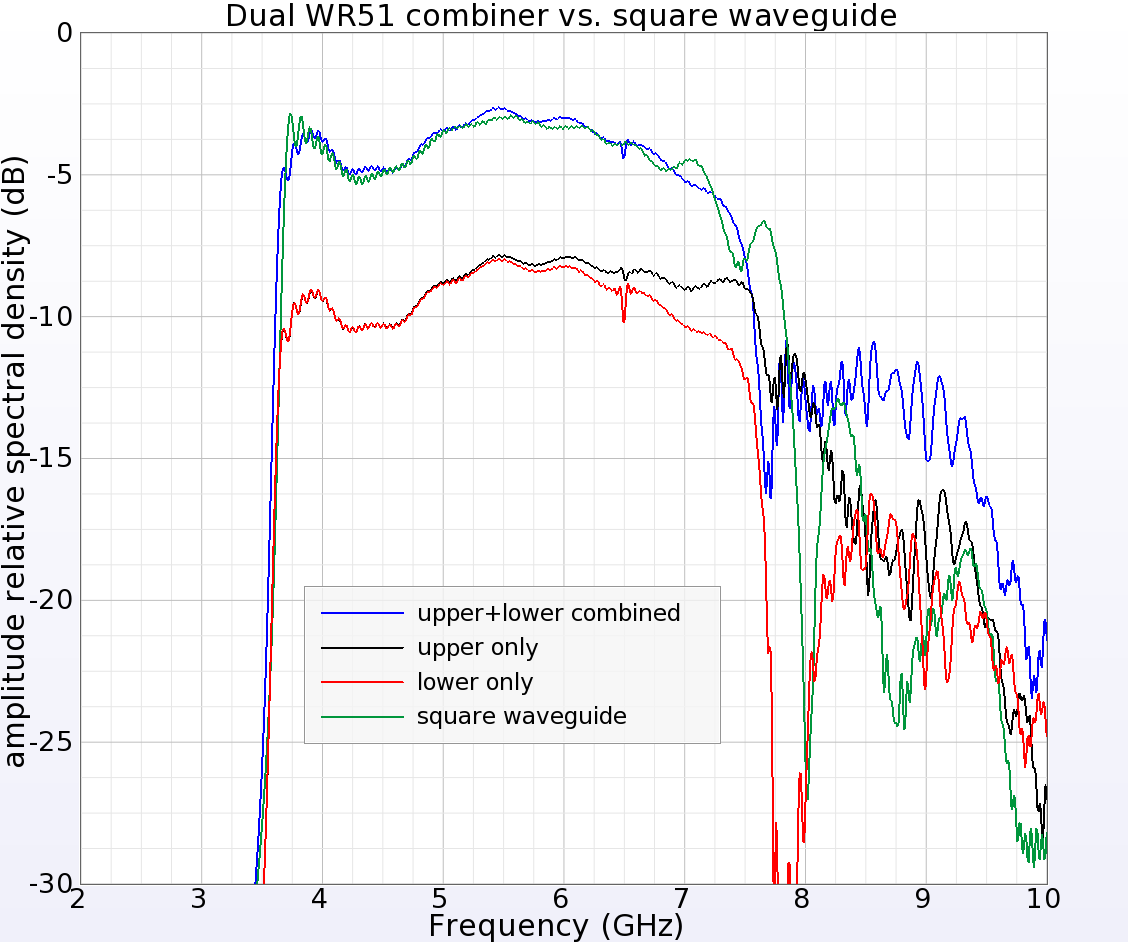}
 \caption{Comparison of amplitude spectra for the individual, combined and square waveguide modes of the FDTD model.
 \label{dualcomp}}
 \end{figure}

\begin{figure*}[htb!]
 \includegraphics[width=6.5in]{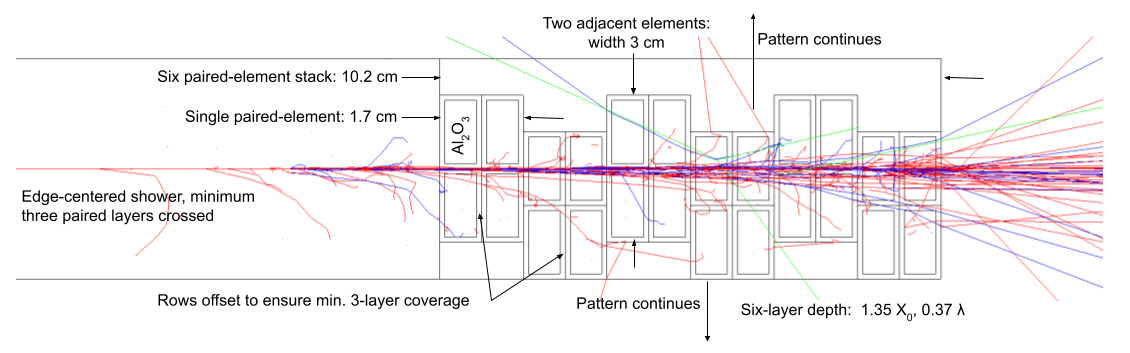}
 \caption{ A 6-layer point design timing-plane detector using the methods we have described here.
 \label{newstack}}
 \end{figure*}

A second option which we found to be more effective is to use a 2:1 E-plane waveguide coupler to combine the signal from a pair of stacked WR51 waveguides into a single LNA receiver at either end. This gains a factor of two in pathlength with no loss of bandwidth or change of center frequency. The two signals generated in the upstream and downstream waveguides are slightly out of phase due to the travel-time delay of the shower bunch as it passes between waveguides. At the center frequency of $\sim 6$~GHz the phase difference for the $\sim 23$~ps shower delay between elements leads to about a 10\% loss in the combined amplitude compared to perfect phasing, an acceptable value relative to the alternative of two independent waveguides, where the signal gain improves only as $\sqrt{2}$. Although efficient combiners can be designed for more than two elements, the phase difference rapidly overcomes any gain in the combiner, and for more than two elements coherent combination is not possible, at least for standard rectangular waveguide elements.

Fig.~\ref{dualSlice} shows the geometry of such a combiner at one end of the stacked waveguide pair, in a section view from the side, with a FDTD simulation overlain showing the magnitude of the upper and lower E-fields as they arrive from the right and merge in the combiner. The combiner gives a good match over the $TE_{10}$ bandwidth of the system, with the output coupling to an SMA microwave coaxial 50$\Omega$ connector at the left. The slight delay of the lower compared to the upper waveguide signal is evident in the figure. This type of tapered combiner is also quite efficient at capturing the $TE_{10}$ mode of a square waveguide, and since it transitions to a normal WR51 section prior to the coaxial adapter, it may also be effective in suppressing unwanted square-waveguide modes.

Fig.~\ref{dualcomp} shows a decibel-scale plot of the relative voltage coupling in this case. In the simulation we were able measure the combined output, or to turn off the stimulus for the upper or lower waveguide respectively, or also to turn off the septum wall that separated the upper and lower waveguides, effectively creating a square waveguide operating in the $TE_{10}$ mode. This plot compares those four cases. The combined signals shows a relative strength that is about 5dB higher than the individual rectangular waveguide sections, consistent with the loss due to imperfect phasing. The amplitude from the square waveguide is quite comparable to the combiner amplitude, showing that the two methods largely arrive at similar outcomes. For now, the wider availability of standard WR51 waveguide leads us to use the dual-waveguide combiner as a reference design.

\subsection{Timing plane geometry.}
\label{baseline}
For purposes of this study we adopt a full timing layer thickness of $\sim10$~cm as our reference design.  This thickness is relatively small compared to either both electromagnetic and hadronic calorimeter designs now under consideration for the FCC-hh. A reference design using six layers of the combined dual-waveguide elements, with a depth of $10.2$~cm is shown in Fig.~\ref{newstack}. Overlain on the figure is an electron-initiated shower, aligned with the interstices of some of the elements, showing how this shower still is fully captured by a minimum of three of the dual-waveguide elements. 

The stackup of this timing layer consists of the paired waveguides as the basic element, with three such pairs aligned in the shower direction, and three more pairs with a half-width offset making up the complete stack of six pairs (12 single elements). This longitudinal base section is tiled and repeated along the transverse coordinate from the beam axis to fill in the region desired. The offsets are required due to the half-cosine response of each waveguide element, which diminishes their efficiency at the edges.

To minimize the column depth of material in the stack, we assume waveguides with 1mm thick aluminum walls. The balance of the material is made up by the alumina loads, and the total depth for electromagnetic showers is thus $1.35~X_0$, and for nuclear showers, $0.37~\lambda_{N}$. Copper waveguide walls would double the radiation length depth to $2.75~X_0$, and the nuclear interaction depth to $0.47~\lambda_{N}$. For the $\sim 13\%$  of tracks that align with the interstitial gaps, the depths would be significantly larger with copper walls; for aluminum, the depths actually decrease compared to those paths through the detectors because of the higher density of $Al_2O_3$.

The overlapping sub-layers guarantee measurement by at least three of the paired dual waveguides for showers aligned with the waveguide walls of one set, and better sampling for other showers at intermediate locations. A minimum configuration with four paired-dual elements is also possible with some increase in turn-on threshold, but for less than four, the sampling will be incomplete and the threshold higher as well. However, a configuration with several four element stacks arranged at longitudinal intervals to capture separate portions of a shower is also a viable possibility and could yield useful partial calorimetry which might supplement a nearby calorimeter, especially at the high-energy end where saturation of other detectors may be an issue.

\begin{figure*}[htb!]
\centerline{ \includegraphics[width=4.75in, trim=0.5cm 0 12cm 0, clip]{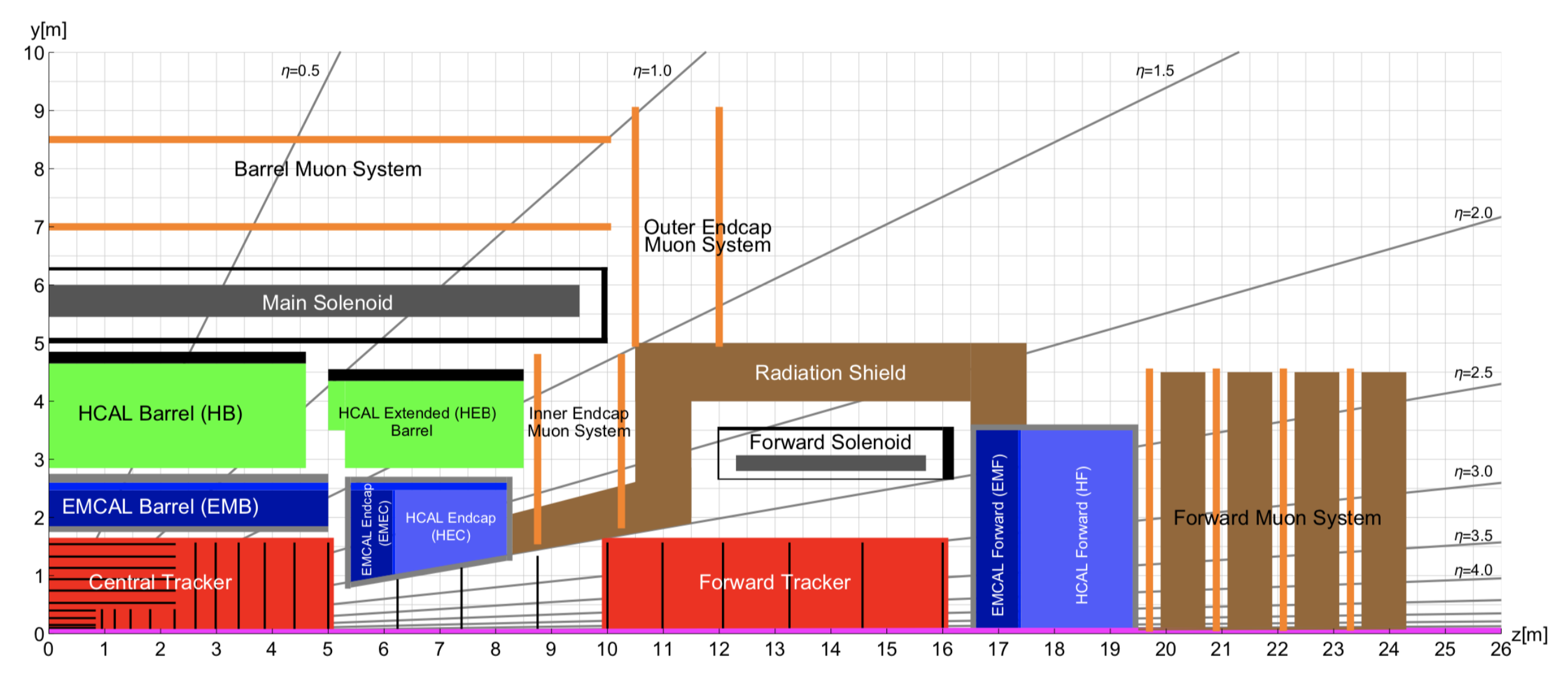}
 \includegraphics[width=2.25in, trim=0 0 0 0, clip]{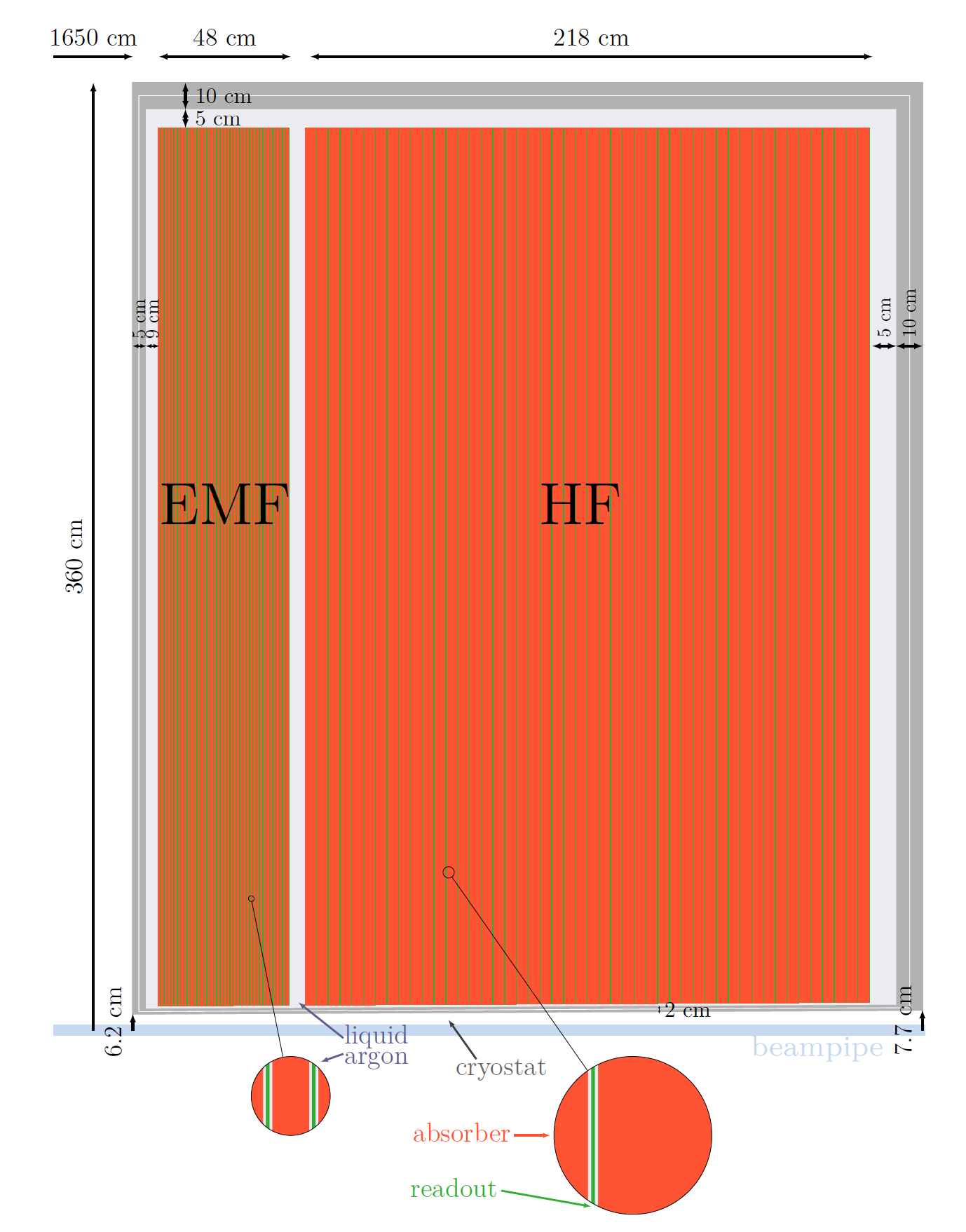}}
 \caption{Left: FCC-hh reference detector section view schematic, the forward EMCAL+HCAL is in blue in the 17-19m region. Right: a zoom of the forward calorimeters. ACE timing planes could be located just inside the dewar ahead of the EMCAL, and just after the EMCAL before the HCAL.
 \label{fcchhref}}
 \end{figure*}
 
From our cross-correlation analysis above, we find the effective $5\sigma$ CCF timing threshold for this stackup is 
$$E_{thr}(5\sigma {~\rm CCF~timing},~ {\rm 3~of~6~elements}) = 56 \pm 12~{\rm GeV}$$
for events along the waveguide edges, and for the average event which showers within all six paired elements, $$E_{thr}(5\sigma~ {\rm CCF~ timing},~{\rm typical}) \simeq 49 \pm 10~{\rm GeV}$$ 
For particles above this energy showering in the timing plane, we expect to measure the shower arrival time to $\sigma_{\tau} \leq 3$~ps. At liquid nitrogen or liquid argon temperatures, the $5\sigma$ threshold will increase by about a factor of two. 

Lower energy showers will be detected with some efficiency down to around twice the thermal noise level, thus even down to 20~GeV for this detector configuration, yielding timing resolution still better than 10~ps. These thresholds are still high compared to the sub-GeV least-count values for most collider detector instruments. For the FCC-hh, even at 100~TeV COM energies, there will be many particles of interest that will fall below detectability for these ACE timing planes, especially in the barrel region at low rapidity values. 

However, in the forward, higher-rapidity portions of the detector, particles are far more likely to carry much of the primary beam energy, and thus we believe that these timing layers will be most effective as augmentations to forward physics measurements. The forward region is also that portion of the detector system where radiation exposure is most extreme. ACE timing planes, consisting almost completely of passive, radiation-hard materials, are also very well-suited to deployment in these regions.

\section{Simulations for a future collider.}

Based on suggestions from Chekanov {\it et al.}~\cite{Chekanov20} that timing planes for the FCC-hh would be best co-located with the EMCAL and HCAL detectors, and our conclusions that the most promising parameter space for our relatively high-threshold systems is in the forward region, we focus a reference design on two locations: first, a layer just in front of the forward FCC-hh EMCAL, as described in the reference design~\cite{FCChh}, and an additional layer between the EMCAL and HCAL in the same region. 
 
Fig.~\ref{fcchhref}~(Left) shows a section view of the FCC-hh reference detector. The forward EMCAL and HCAL elements, covering $\eta = 2.5-6$ can be seen within the 16.5-19.5m longitudinal region. Both elements consist of liquid argon + copper sandwich systems, as shown in the zoomed view on the right side of the figure. We assume that $\sim 10$~cm spaces can be made available for timing planes in both locations.

To simulate the timing of shower generated by several different types of secondaries, we first create a GEANT4 model of each timing layer including any additional column depth upstream of it. In the case of the preshower detector, the total depth of the forward tracker is small, and we neglect it, but we include the estimated mass budget for a reference design for a preshower detector, about $3.5~X_0$, just upstream of the timing plane. In the case of the second timing layer after the forward EMCAL and just in front of the HCAL, we include a block mass upstream of the timing layer with comparable depth and average $Z, A$ similar the liquid argon + copper EMCAL. 

\subsection{Time delays of the shower centroid.}

\label{timedelays}

One of the critical issues for our methodology is the variance of the delay of the effective electromagnetic centroid of the excess charge distribution of the shower relative to the arrival time of the initiating particle. We term this as the ``electromagnetic centroid'' since the observed microwave pulse is proportional to a sum of the field amplitude from each particle weighted by its phase factor, which itself is determined primarily by its arrival time delay as it passes through a given waveguide element.

\begin{figure*}[htb!]
 \includegraphics[width=7in, trim=40mm 0 40mm 0, clip]{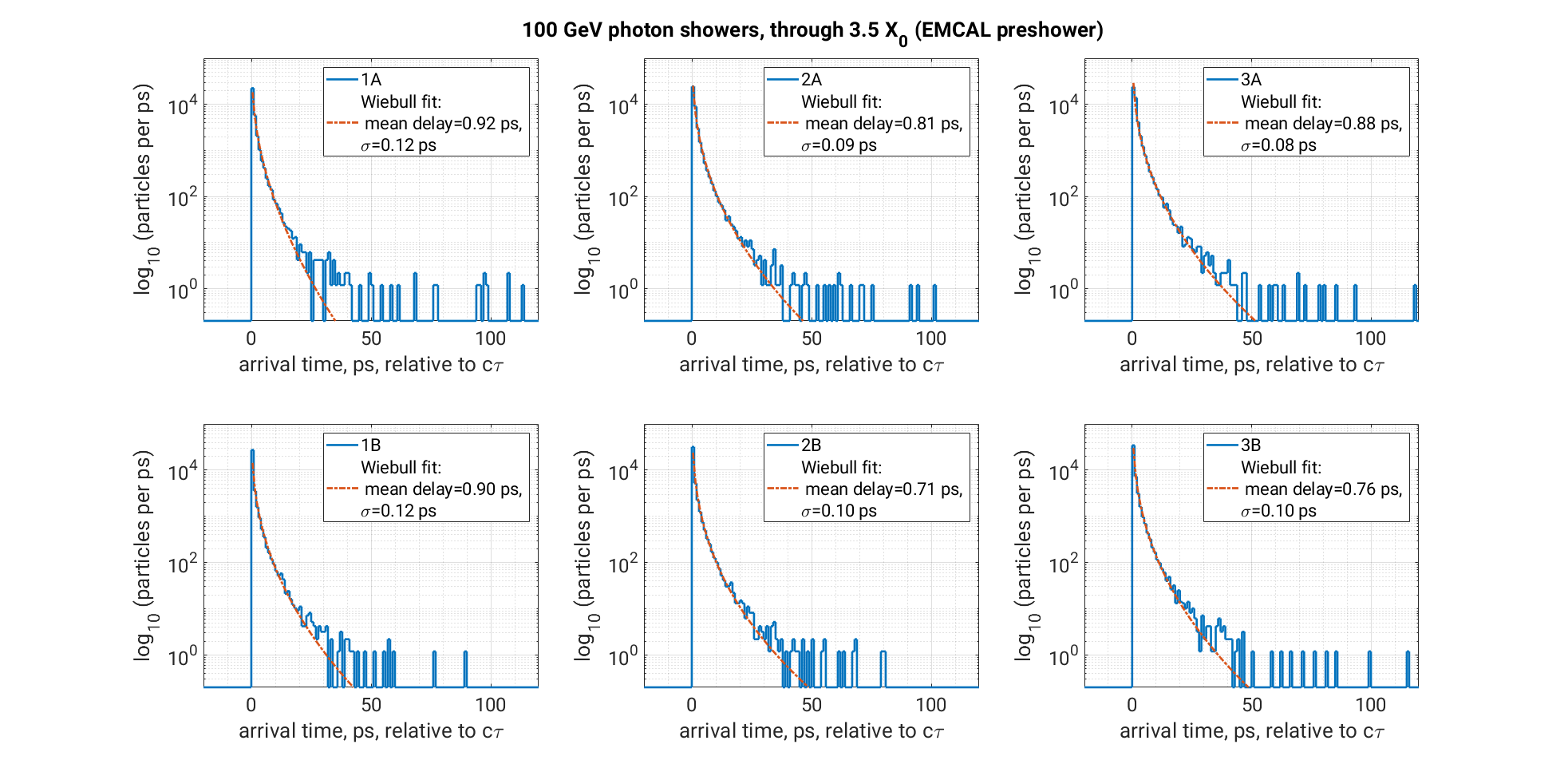}
 \caption{GEANT4-simulated time delay distributions of 100 GeV photon-initiated shower particles relative to the light travel time from the injection point, along with fits of a Wiebull probability density.
 \label{timeoffsetGamma}}
 \end{figure*}
 
In future beam test experiments the delay between the bunch EM centroid and the resulting shower EM centroid is observable if we use an upstream waveguide element to capture the bunch transit prior to the shower onset; this was however not the case for our 2018 experiment. For now we have done detailed GEANT4 simulations of the timing profile of the showers to measure the particle time of passage in simulation. We take this for now as a proxy for the EM centroid delay and its variance.

Fig.~\ref{timeoffsetGamma} shows the results of simulated shower particle exit times for 100 GeV photon showers in the pre-shower configuration described above. Results are shown for each of the six elements, indicated alphanumerically as 1A, 1B,...,3A,3B, with the number indicating the order in the 3-paired-element stack, and the letter indicating whether the element is first or second in the pair. In each case the light-travel-time offset has been removed to set the origin of the plot. We find the resulting arrival times are described reasonably well by a Wiebull distribution, which is fitted and shown as an overlain curve, with the fitted mean of the distribution indicated in the legend.

The probability density function for the Wiebull distribution~\cite{Wiebull}, here given as a function of time $t$  is given by:
\begin{align}
 f(t) &=    ~0~&t<0,\\ \nonumber
 f(t) &= ~\frac{k}{\lambda} \left ( \frac{t}{\lambda}\right )^{k-1} e^{-(t/\lambda)^{k}}~&t\geq 0
\end{align}
where $\lambda>0$ is the scale parameter and $k>0$ is the shape parameter.
The mean of the distribution, which gives the offset delay of the shower relative to the primary particle, is given by
\begin{equation}
    \mu = \lambda \Gamma \left ( 1 + \frac{1}{k} \right )
    \label{Wiebullmean}
\end{equation}
where $\Gamma$ is the gamma function.

\begin{figure*}[htb!]
 \includegraphics[width=7in, trim=40mm 0 40mm 0, clip]{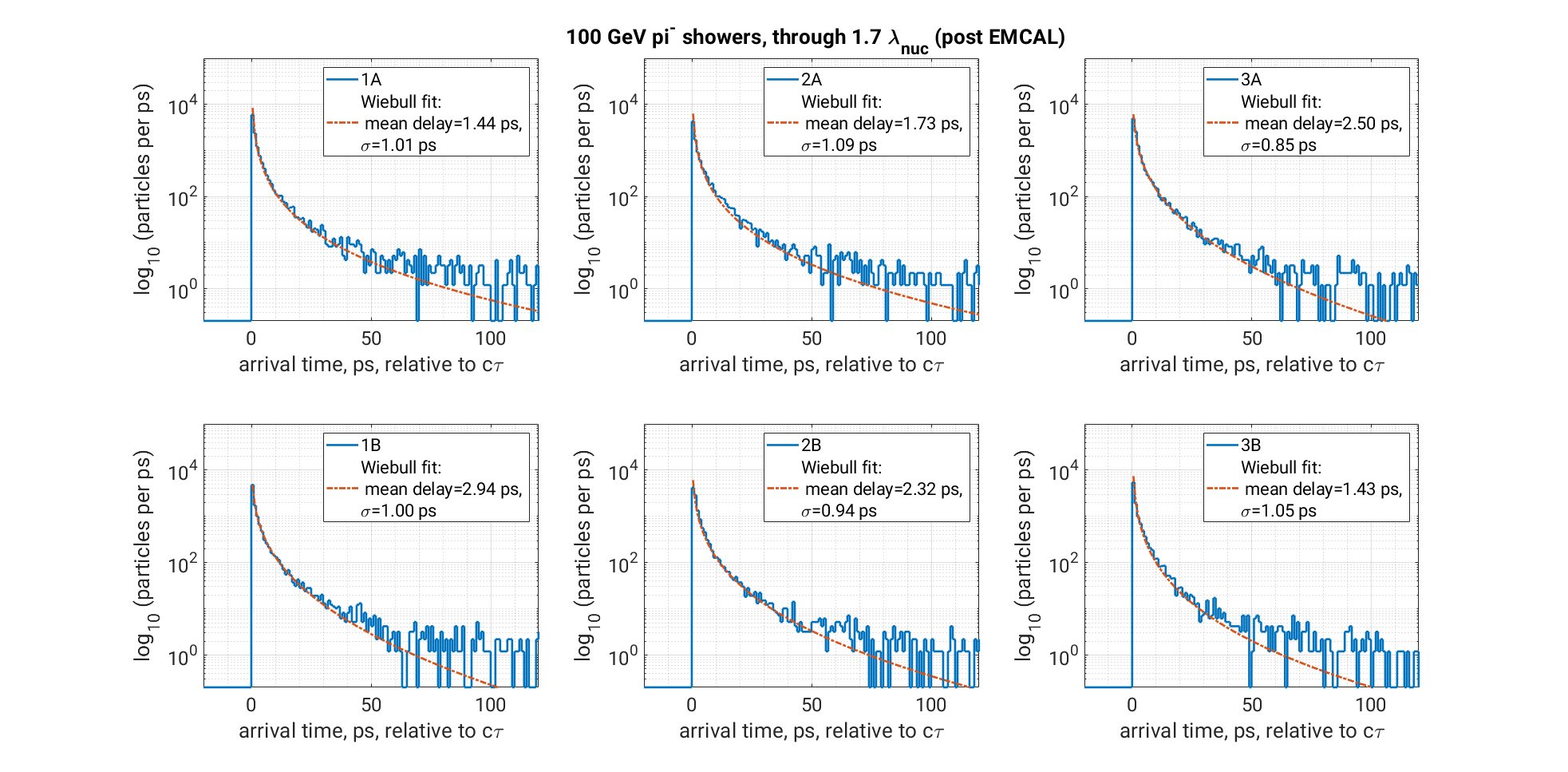}
 \caption{GEANT4-simulated time delay distributions of 100 GeV pion-initiated shower particles relative to the light travel time from the injection point, along with fits of a Wiebull probability density.
 \label{timeoffsetPion}}
 \end{figure*}

While the fitted mean value for the offset delay may be removed from the timing as a nuisance parameter, the standard error on the mean of these distributions is irreducible and represents the noise floor on timing measurements using the shower centroid as a proxy for the primary particle. To determine the mean and its standard deviation, we have used a Monte Carlo method which generates Wiebull random deviates based on the fits to $\lambda,~k$, and then using  equation~\ref{Wiebullmean} to estimate the corresponding mean and its variance. For the photon showers, the typical standard deviation is around 0.1 ps, small compared to the picosecond timing goals here.

Fig.~\ref{timeoffsetPion} shows the results of simulated shower particle exit times for 100 GeV pion showers, now measured in a timing plane between the EMCAL and HCAL detectors, thus with $1.7$ nuclear interaction lengths upstream of it. While the shower initiation is governed by the nuclear cross sections, the shower development rapidly becomes dominated by electromagnetic cross sections, and the resulting showers are much broader in time delays than for the photon-initiated showers with a relatively small material burden. The resulting fits show standard deviations of around 1 ps, which is comparable to our timing goals, and thus will appear in quadrature with it, and defines the timing noise floor for a layer at this location in the system.

\begin{figure*}[htb!]
 \centerline{\includegraphics[width=3.25in, trim=0mm 0 0mm 0, clip]{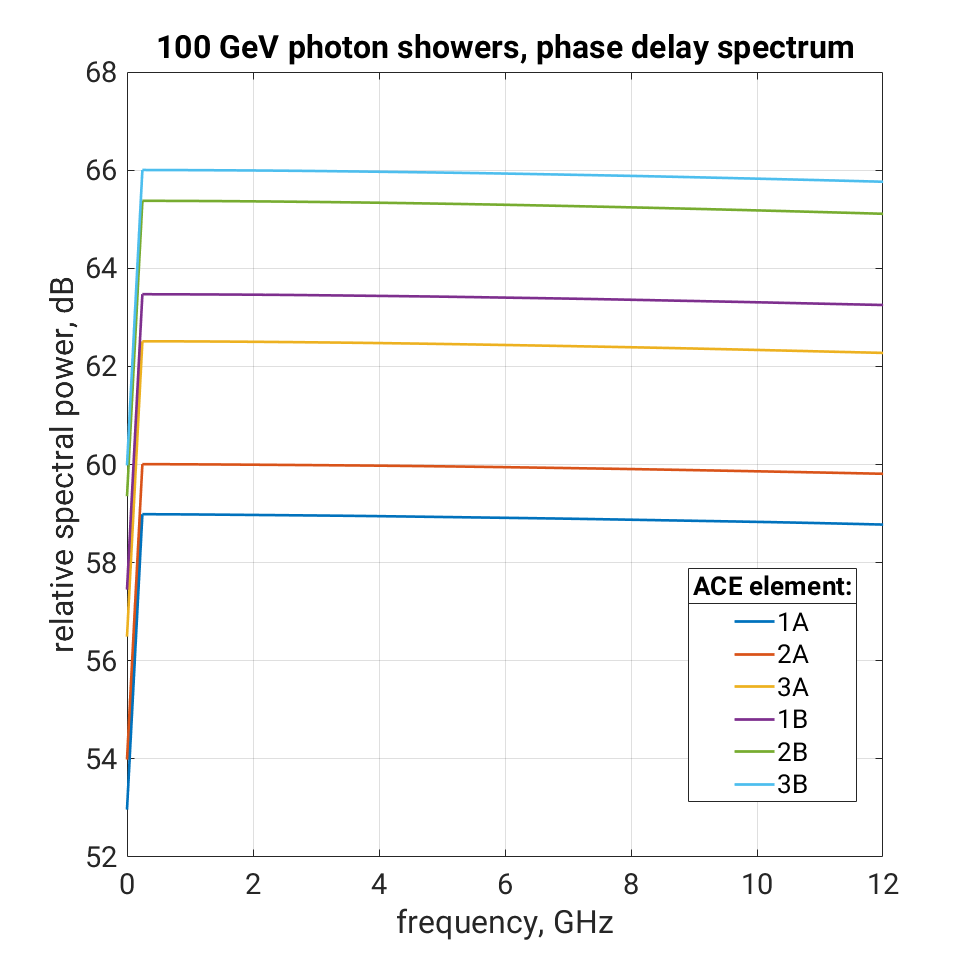}
 \includegraphics[width=3.25in, trim=0mm 0 0mm 0, clip]{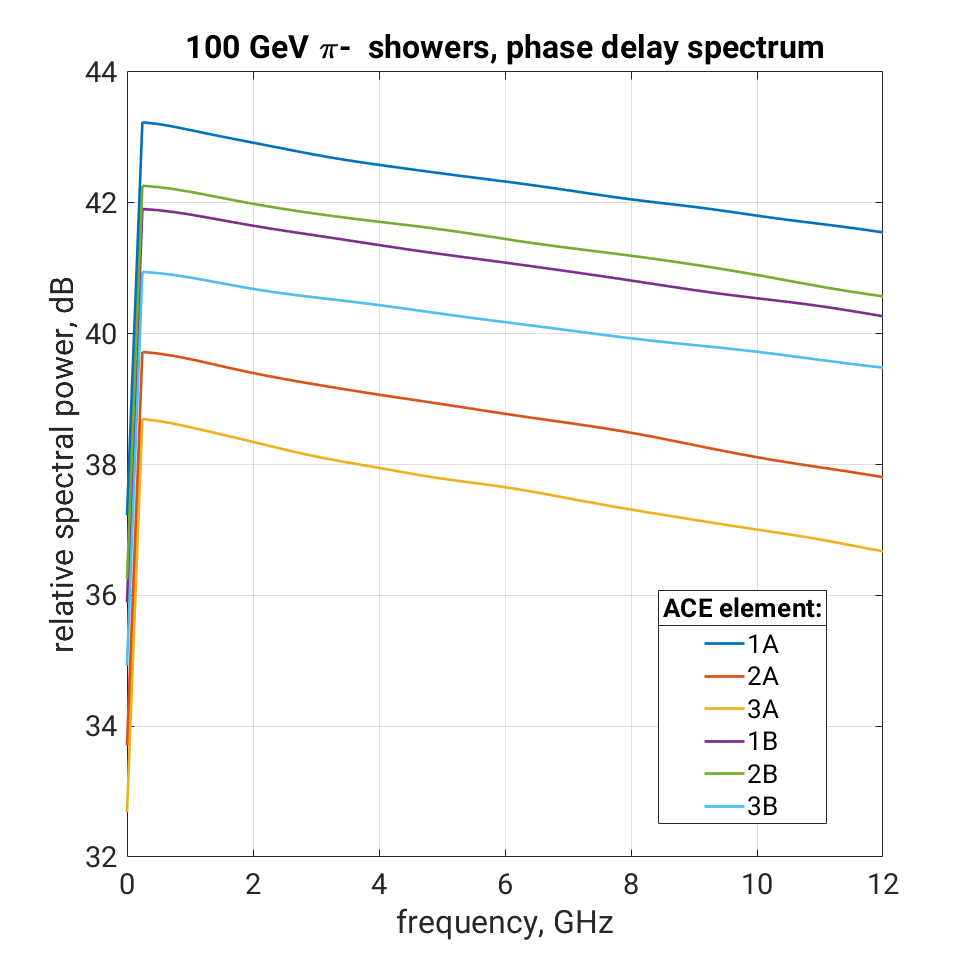}}
 \caption{Relative frequency spectral power for the time delay spectra shown previously. Left, for photon showers, and right, for pion showers. 
 \label{delayspectra}}
 \end{figure*}

\begin{figure*}[htb!]
\centerline{ \includegraphics[width=1.3in, trim=0mm 0 10mm 0, clip]{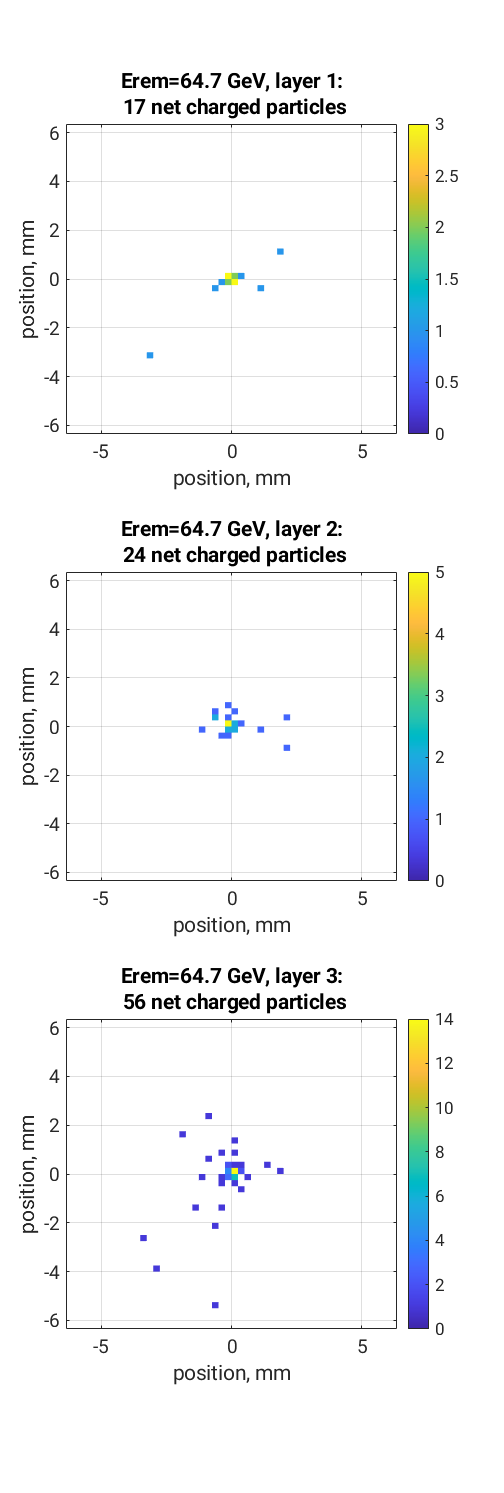}
 \includegraphics[width=6in, trim=40mm 0 30mm 0, clip]{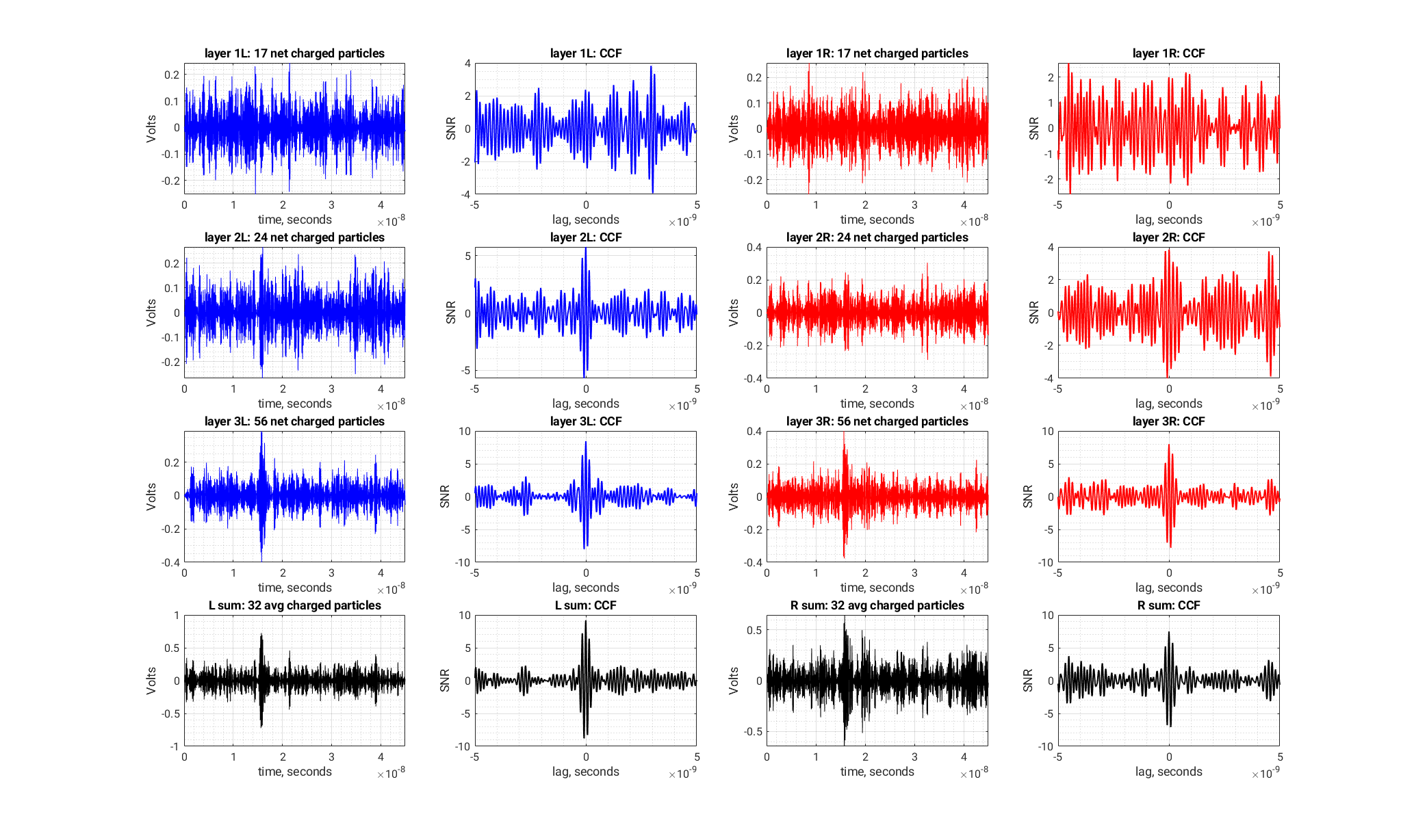}}
 \caption{Left column: distribution of throughgoing charged particles for a single realization of a photon shower transiting three layers of our timing plane. The remaining four columns show, from left to right, (1) the microwave signal at the left end of the ACE waveguide, (2) its cross-correlation with the template response function for the signal; (3) and (4), similar waveforms and CCFs for the microwave signal at the right end of the waveguides. From top to bottom, we show the signals and CCFs for each of the three layers, followed by the coherent sum in the bottom row.
 \label{simwfms}}
 \end{figure*}

 One other issue regarding these particle delays must be addressed: the phase delay distribution of the secondaries leads to a partial loss of coherence in the summation of the individual particle electric fields. We can estimate this effect by looking at the Fourier spectrum of the delay distribution.
 Fig.~~\ref{delayspectra} shows the effect of these delay distributions on the resulting field amplitude as a function of frequency,  in the microwave range of interest. For the photon-initiated showers in the pre-shower region, the effect is completely negligible, due to the fact that all particles are quite prompt on the scale of 8~GHz (a cycle time of 125 ps), the upper frequency band limit for our detector. 
 
 For the pion showers, with a significantly longer tail, the effects are not completely negligible: a slope of about 0.5-0.7 dB across the 5-8 GHz band is evident, corresponding to an amplitude loss of 5-8\% across the band. This effect is still quite moderate given the large material burden ahead of this timing plane, and does not significantly degrade the efficiency of the timing plane in this configuration.

\subsection{Timing distribution results for the FCC-hh.}

The possible range of particle showers to be studied for timing purposes is very large, both in particle type and particle energy. We limit ourselves in this study to a small number of variations which are indicative of the expected results. For the FCC-hh, the dynamic range of desire measurements of outgoing particle momenta for a typical event is enormous: from below a GeV/c up several tens of TeV/c, four orders of magnitude or more. It is improbable that any single detector will be able to accommodate this entire dynamic range without several different operating modes. 

For ACE, the limiting element in dynamic range is likely to be first-stage low-noise amplifier compression, but it is possible to achieve LNA dynamic range of order $10^4$ in amplitude ($\gtrsim 80$~dB) with careful design. For example the LNAs used in our 2018 tests (Low Noise Factory LNF-LNC4\_8C models) have a noise temperature of 2.2~K at LHe ambient temperatures, and an output compression point of -12 dBm. The thermal noise of the LNA is equivalent to -100.4~dBm, which implies a nominal dynamic range of $\sim 88$~dB. As per our estimates above, the $1\sigma$ thermal noise level corresponds to 15-30~GeV signal amplitude depending on $T_{sys}$, and thus the ratio of kinematic limit of 50~TeV to our least-count energy is as high as 3300, a 70 dB range in microwave amplitude. It is evident that even current LNA technology easily covers the full dynamic range of events expected for the FCC-hh, with significant margin.

\begin{figure*}[htb!]
\centerline{ \includegraphics[width=2.75in, trim=10mm 0 10mm 0, clip]{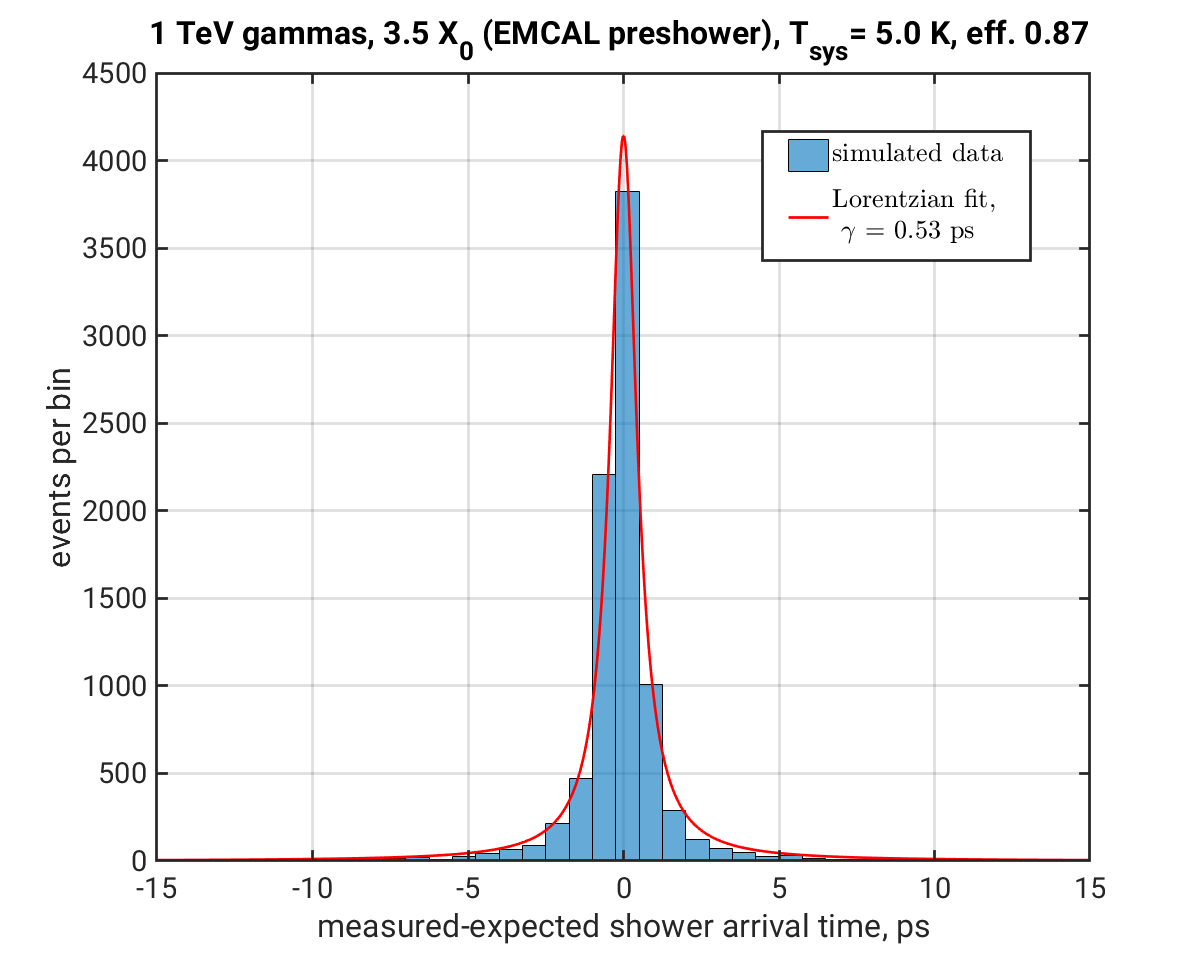}
 \includegraphics[width=4in, trim=10mm 0 10mm 0, clip]{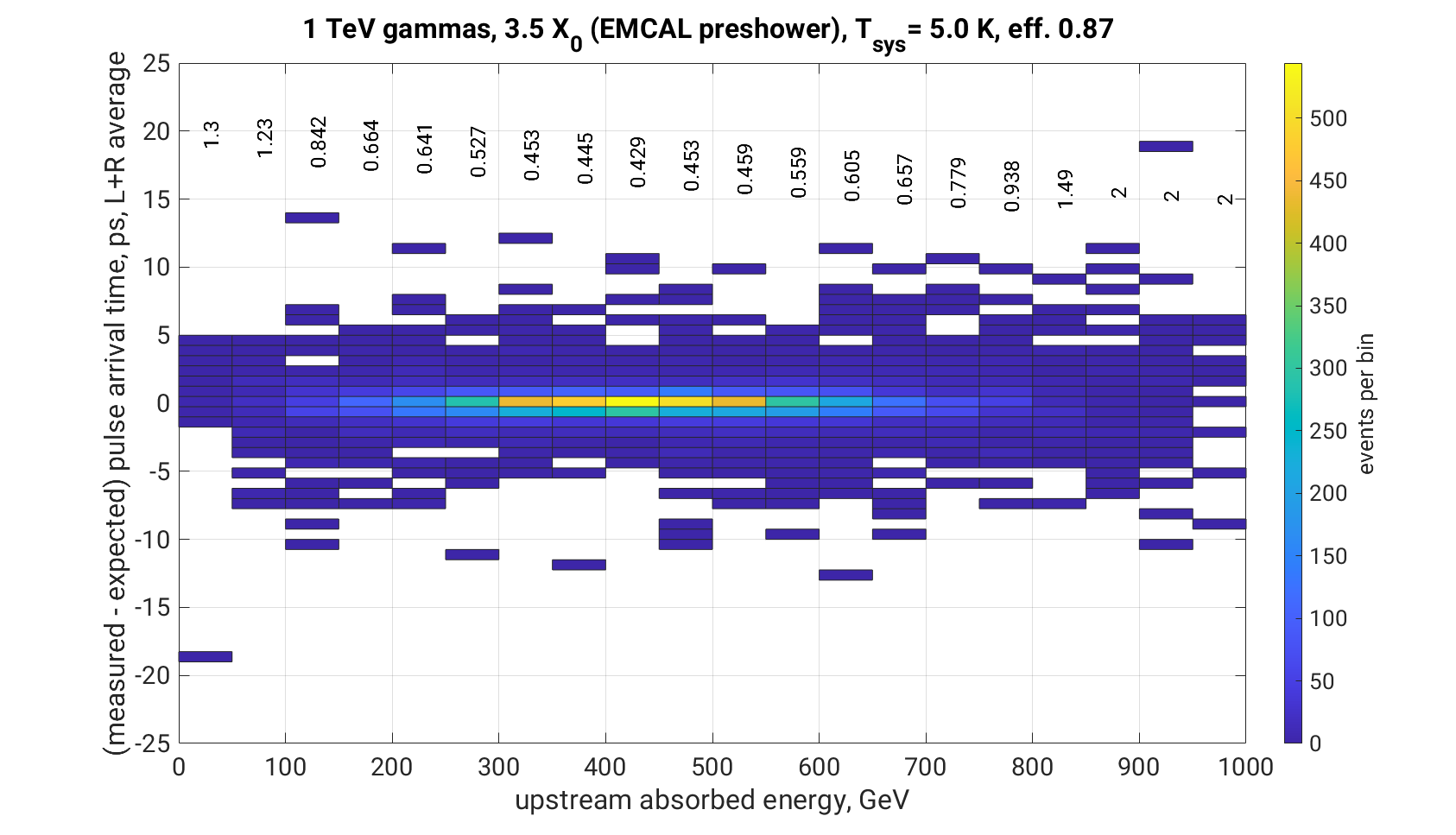}}
 \centerline{ \includegraphics[width=2.75in, trim=10mm 0 10mm 0, clip]{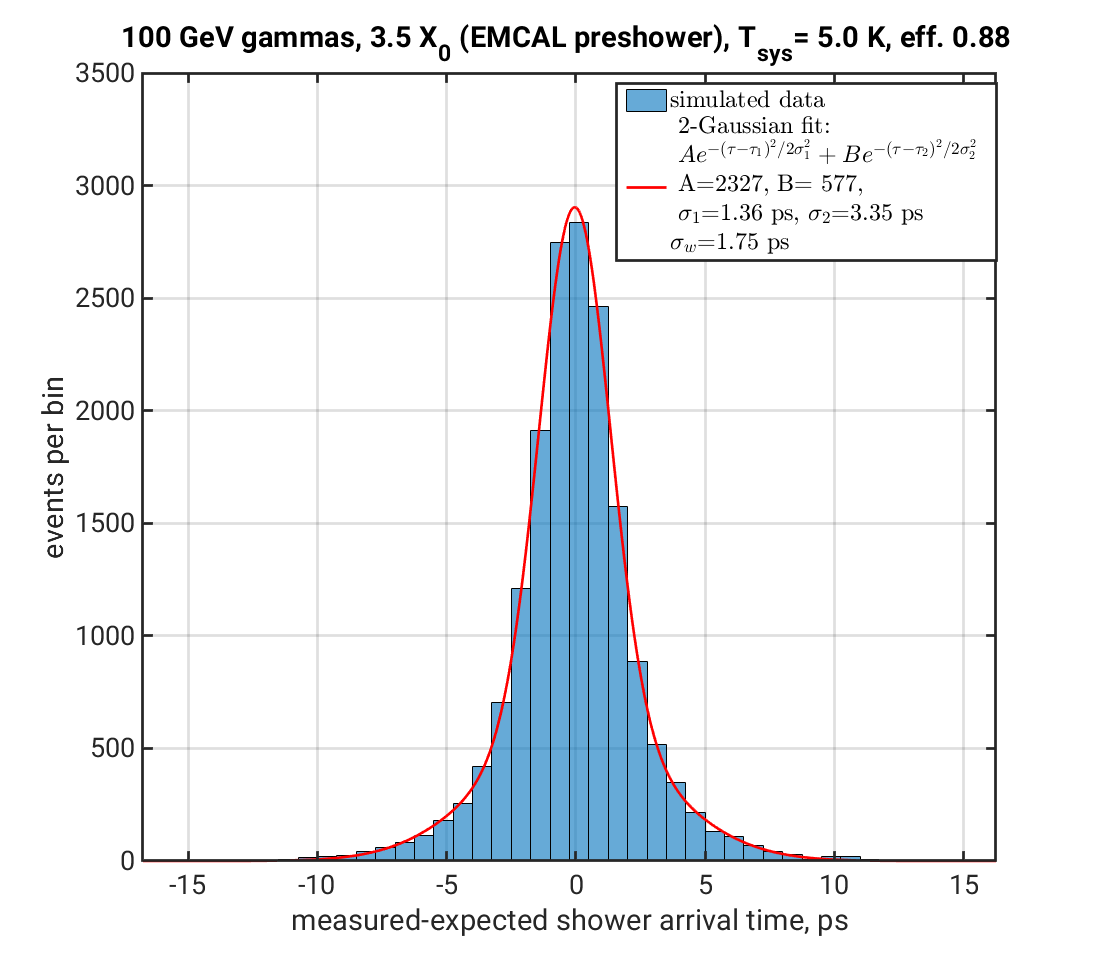}
 \includegraphics[width=4.2in, trim=10mm 0 10mm 0, clip]{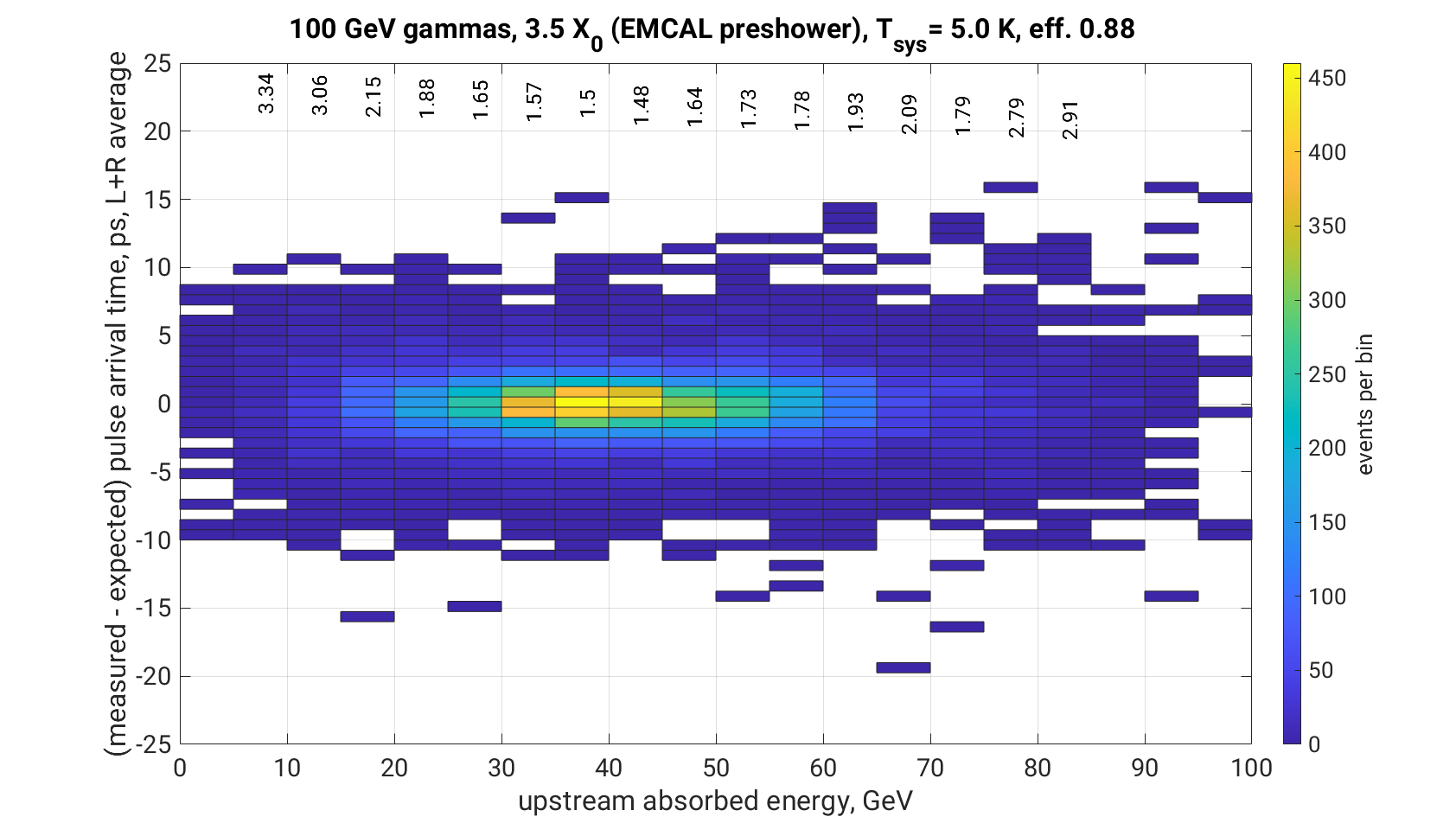}}
 \caption{Distributions of all times and time vs. absorbed energy prior to the timing plane, for photons at 1 TeV and 100 GeV. Here our timing plane is between a preshower block and the forward EMCAL, with about 3.5 radiation lengths of material ahead of it.
 \label{gammaTiming}}
 \end{figure*}

For our current study, the range of $\eta = 2.5-6$ for the forward calorimeters where we consider our timing planes, results in significant boosts of the particles and the resulting showers observed in those regions. For a given transverse momentum $p_T$, the energy deposited in the calorimeters will be of order $E(\eta)\simeq p_T/\cosh{\eta}$, corresponding to factors of between 6$^{-1}$ and 200$^{-1}$. For our $5\sigma$ shower energy threshold of 50~GeV, the equivalent threshold in particle transverse momentum varies between $\sim 8$~GeV ($\eta=2.5$), to a fraction of a GeV for the highest pseudorapidities. In practice this means that our efficiency for yielding timing results will be much higher in the forward direction that it would in the barrel region.

We consider two different incident particles: photons and pions, at two different total energies, 100~GeV, and 1~TeV. Photons are of interest because they are not observed in the trackers, but will be efficiently observed via a preshower layer, and pions are ubiquitous in virtually every collision, creating hadronic showers, which will be observed in our second timing layer. 

Fig.~\ref{simwfms} displays a complete set of simulated data for a single shower event, for the case of a 100 GeV photon-initiation shower, in a timing plane placed after the preshower block and just prior to the EMCAL. For each event the simulation uses measured GEANT4 particle distributions in the center of a 1 meter long waveguide stack, as shown on the left figure column for the three dual layers. The simulation then convolves in the electromagnetic response based on our FDTD results as validated by our test beam data. The expected thermal noise floor is imposed on the data, and the signal is received at both ends (denoted left and right) of the waveguide.  We then perform a cross-correlation using the response template waveform on each of the left and right received signals. The received waveforms and corresponding CCF are shown in rows for each of the three dual layers, followed by a coherently summed version for the left and right signals, and the corresponding CCF, at the bottom row. 

For this event, in which the photon shower had deposited about 35 GeV in the preshower block, the individual element timing was above threshold for the second and third layers as the shower grew, but the signal in the first layer fluctuated below threshold. The coherently combined data showed a significant boost in SNR as expected, and the CCF shows another SNR increase as well. 

\subsubsection{Photon showers at 1 TeV and 100 GeV.}

\begin{figure*}[htb!]
\centerline{ \includegraphics[width=2.85in, trim=10mm 0 10mm 0, clip]{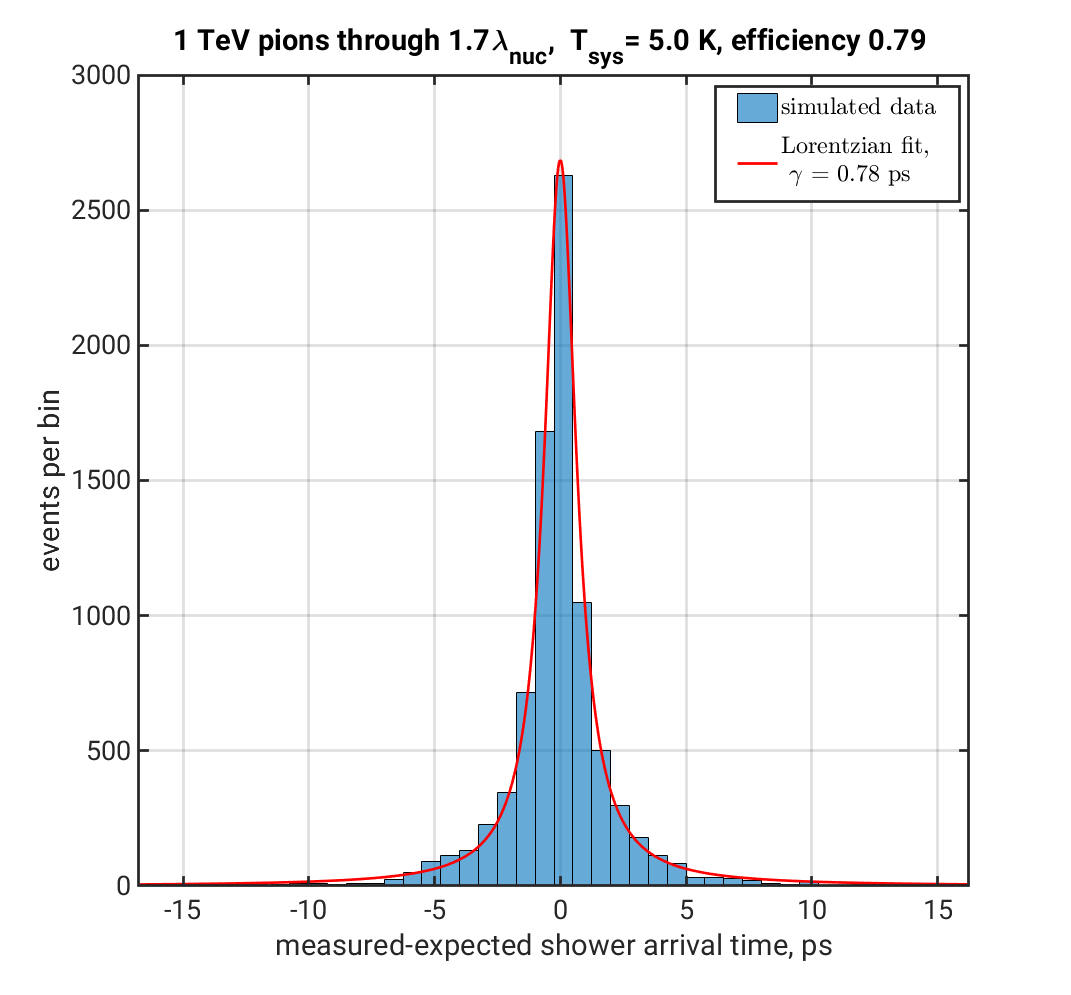}
 \includegraphics[width=4.15in, trim=10mm 0 10mm 0, clip]{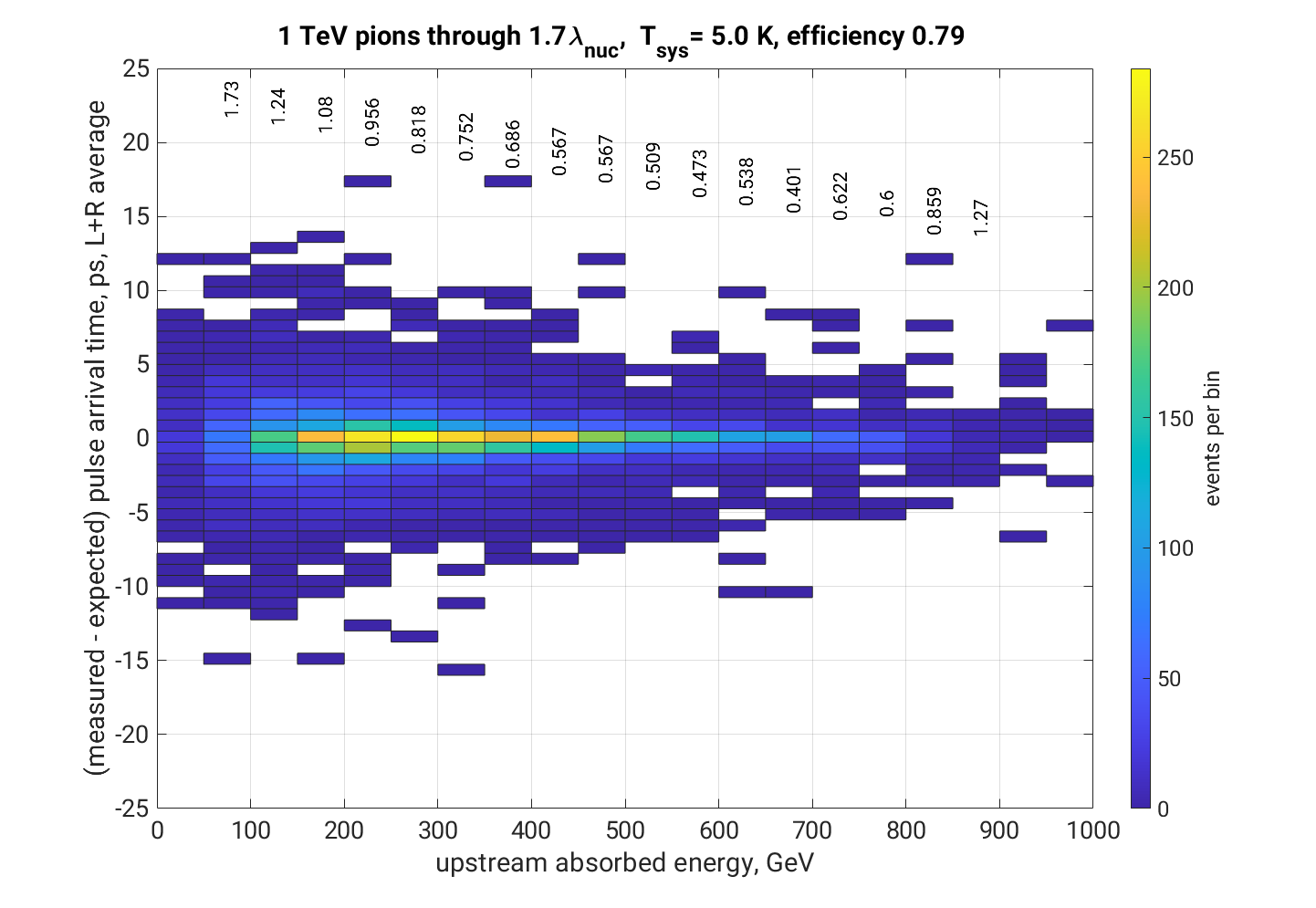}}
 \centerline{ \includegraphics[width=2.95in, trim=10mm 0 10mm 0, clip]{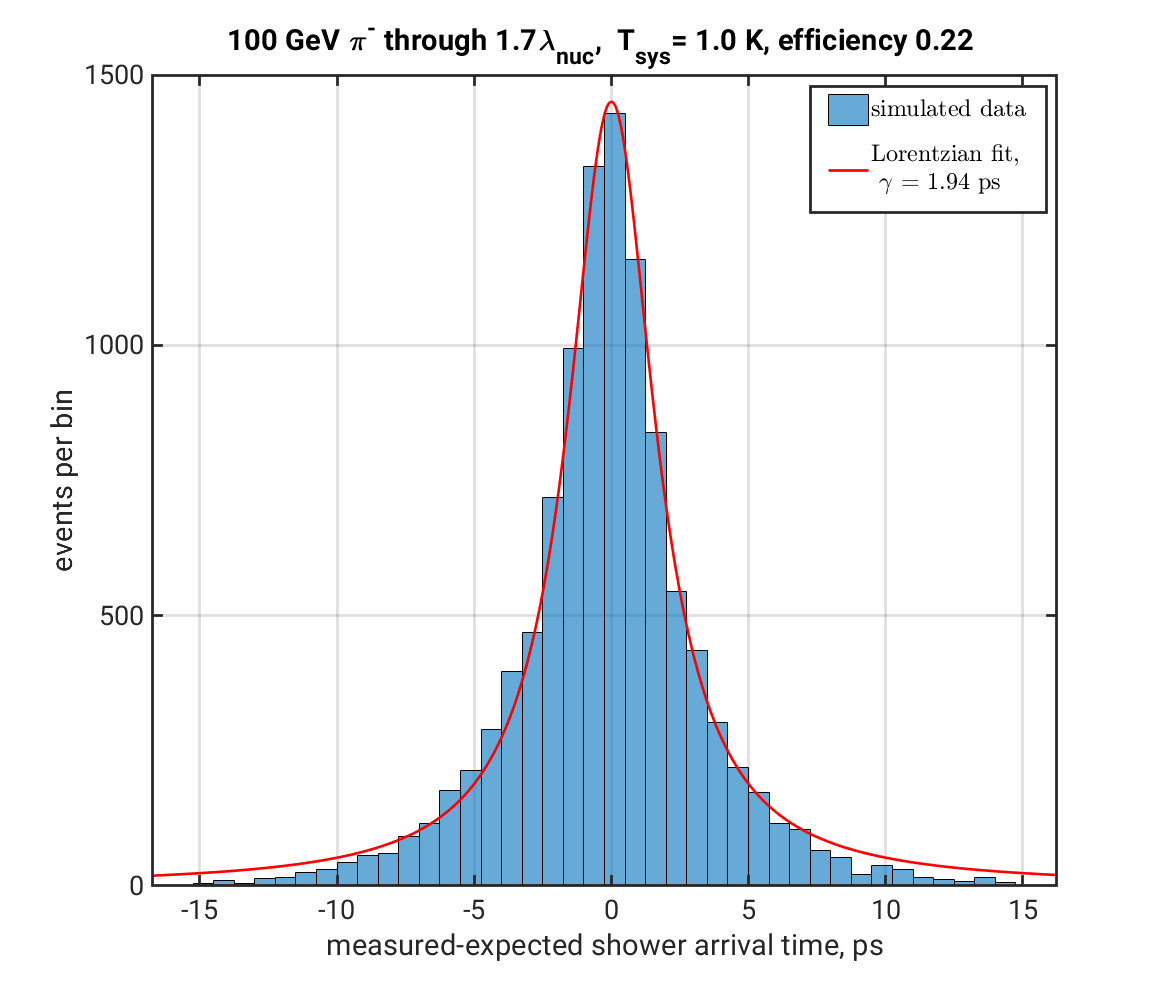}
 \includegraphics[width=4.1in, trim=10mm 0 10mm 0, clip]{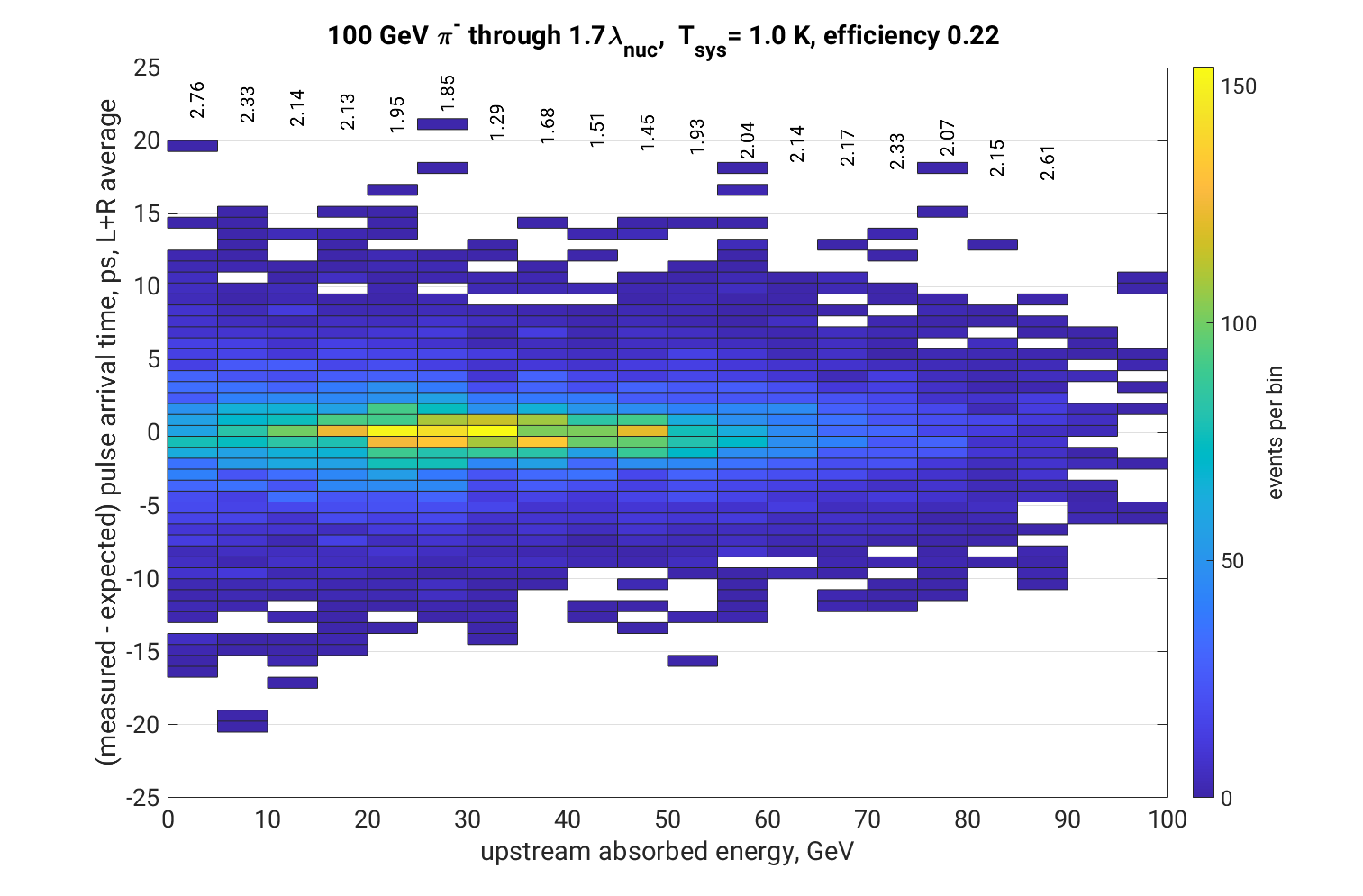}}
 \caption{Distributions of all times and time vs. absorbed energy prior to the timing plane, for charged pions at 1 TeV and 100 GeV. Here our timing plane is between the forward EMCAL and HCAL, with about 1.7 nuclear interactions lengths of material ahead of it.
 \label{pionTiming}}
 \end{figure*}

Fig.~\ref{gammaTiming} shows the timing vs. remaining energy results of the timing plane between the preshower element and the forward EMCAL. Here we have assumed that the timing plane is within the EMCAL cryostat and it thus at liquid argon temperatures, but with additional cold-head cooling of the LNA, giving an overall $T_{sys} \simeq 5$~K, about half of which is produced by the high-frequency ohmic losses in the waveguide. 

The left column of figures shows the full timing histogram the entire distribution of events relative to the expected timing.
For the 1 TeV photons, the copious secondaries in the shower result in sub-picosecond timing over a wide range of shower development. For the 1D histogram at left, the best fit is for a Lorentz distribution, giving about 0.5 ps as a typical timing error. Efficiency here is a measure of the number of showers that give an average combined amplitude SNR>3.0 for left and right, for the coherent combination; this level represents the minimum that gives a reliable timing solution. Most of the lost events are showers that are largely absorbed in the preshower, or photons that punch through with almost no energy deposited. 

At 100 GeV, the efficiency is still high, but the timing distribution is no longer well described by either a Lorentzian profile, or a single Gaussian, but a sum of two Gaussians fits well. The weighted average standard error is now more than a factor of 3 larger than the 1 TeV results, but still provides overall timing errors below 2~ps.

\subsubsection{Charged Pion showers at 1 TeV and 100 GeV.}

For the second timing layer, we simulated charged pions as the hadronic reference particle, since at these energies interactions dominate completely over decay. Simulations were done at 1 TeV and 100 GeV, as for the photon simulations.
The current baseline design for the forward EMCAL imposes a material burden of about $1.7$ nuclear attenuation lengths in front of the HCAL. 

For a timing plane at this location, the variability of showers will be much larger than for the preshower timing layer. Our GEANT4 simulations indicate about 80\% efficiency for 1~TeV pion shower detection using the same detector parameters as for the preshower timing layer, but we find that the detection efficiency for 100 GeV pions was only of order 10\% using the $T_{sys}=5$~K design. For this case, we therefore assumed advances in affordable cryocooler technology that would allow us to achieve $T_{sys}\sim 1$~K, which boosts the efficiency to over 20\%.

Fig.~\ref{pionTiming} shows the results of the timing layer just in front of the HCAL for charged pion showers. These do not include the $\sim 1$~ps additional uncertainty due to the absolute variability of the shower electromagnetic time centroid with respect to the parent pion, as determined in section~\ref{timedelays} above, so the actual timing should include the 1~ps as a root-sum-squared increase in the timing uncertainty.

At 1~TeV, the pion shower timing distribution is best fit by a Lorentzian, giving $\sim 0.8$~ps intrinsic uncertainty with $\sim 80\%$ efficiency, which translates to an overall uncertainty of $\sim 1.3$~ps including the uncertainty in the pion-to-shower offset. The corresponding transverse momentum of pions at this energy for pseudorapidities of 2.5, 4.0 and 6.0 within the HCAL range are $p_T = 160$~GeV/c,37~GeV/c, and 5~GeV/c, respectively. 

At pion energies of 100~GeV, the efficiency has dropped to 22\%, and the timing uncertainty, including the RSS with the 1~ps pion-to-shower uncertainty, is 2.2~ps for the subset of showers that are above threshold. These showers in the forward regions now correspond to transverse momenta which are a factor of 10 lower that at 1~TeV.

\begin{figure*}[htb!]
\centerline{ \includegraphics[width=5.5in, trim=10mm 0 10mm 0, clip]{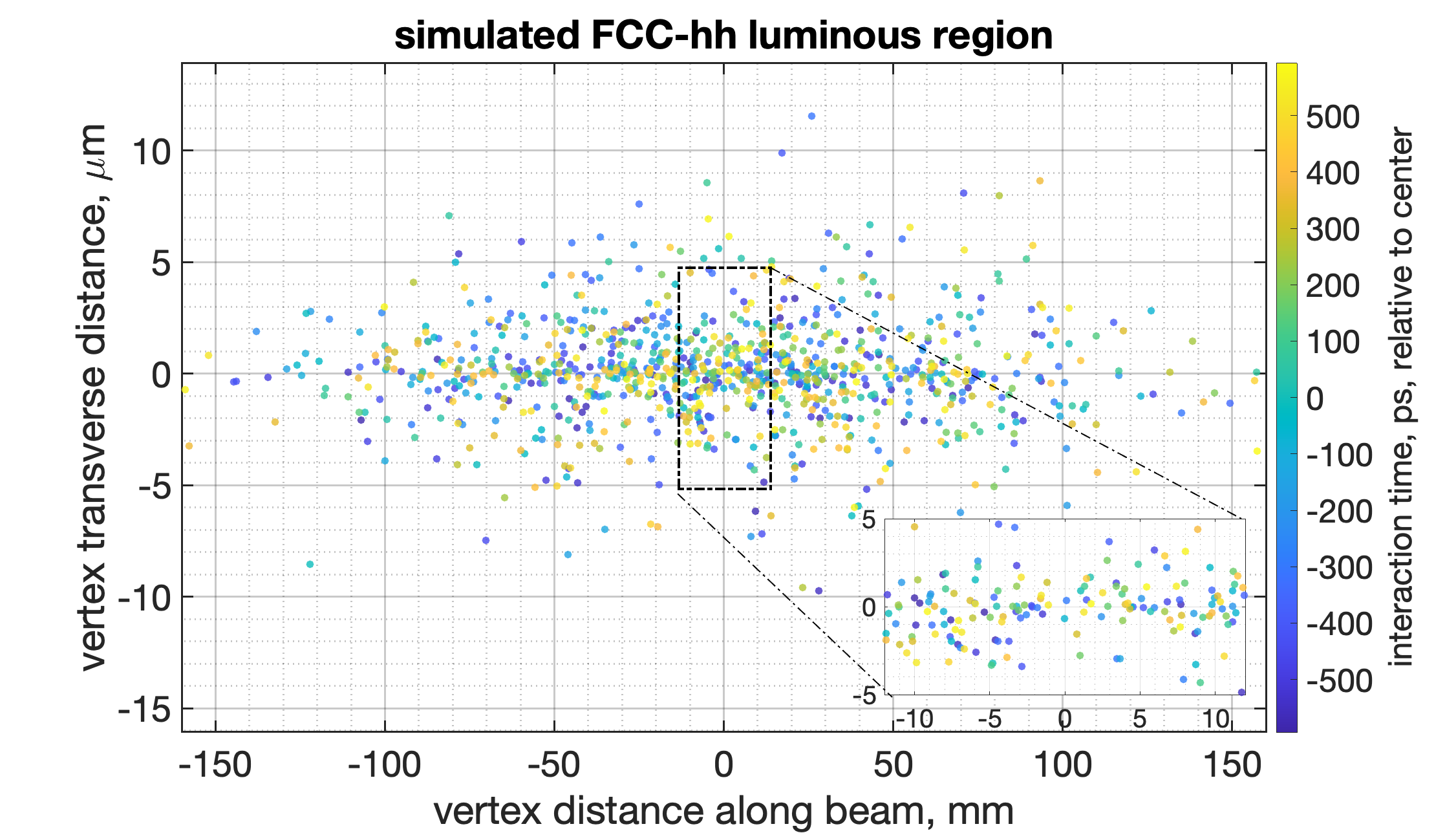}}
 \caption{ A simulation of  the FCC-hh luminous region for a single event, based on the expected bunch sizes and interaction rate. The inset shows the core region, and the colors encode the time of the interaction at each vertex relative to the center time.
 \label{pileup}}
 \end{figure*}

\subsection{Applications.}

We have demonstrated that picosecond timing of collider secondary particles, via a $\sim 10$~cm longitudinal sample of their electromagnetic or hadronic showers, is possible above a threshold energy -- at least several tens of GeV or more -- which is relatively high compared to the least-count energy of current collider detectors. While the FCC-hh will present an entirely new domain of extreme energy particles and showers, there will be many collision products that fall below our least-count energy. Optimizing the design for timing layers in the forward region mitigates this issue to a large degree, but it is still important to explore the strengths and limitations of our methods using examples that are tied to the physics goals of the FCC-hh. We will thus look briefly at several applications here.

\subsubsection{Higgs to two photon decays.}

Because secondary photons are not easily tagged in tracker instrumentation, EM calorimeters take on the role of identifying such photons, measuring their energy, and providing constraints on their geometry. For the two photon Higgs decay, other diphoton processes produce an irreducible background, but neutral pion decays can add to this if the photons cannot be separated or the pion vertex cannot be distinguished from the Higgs vertex. Particularly in the forward direction, the higher energies and large diffractive components involved can lead to more difficult pileup background rejection. Timing can be used to significant advantage in this situation.

\begin{figure}[htb!]
\centerline{ \includegraphics[width=3.5in, trim=10mm 0 10mm 0, clip]{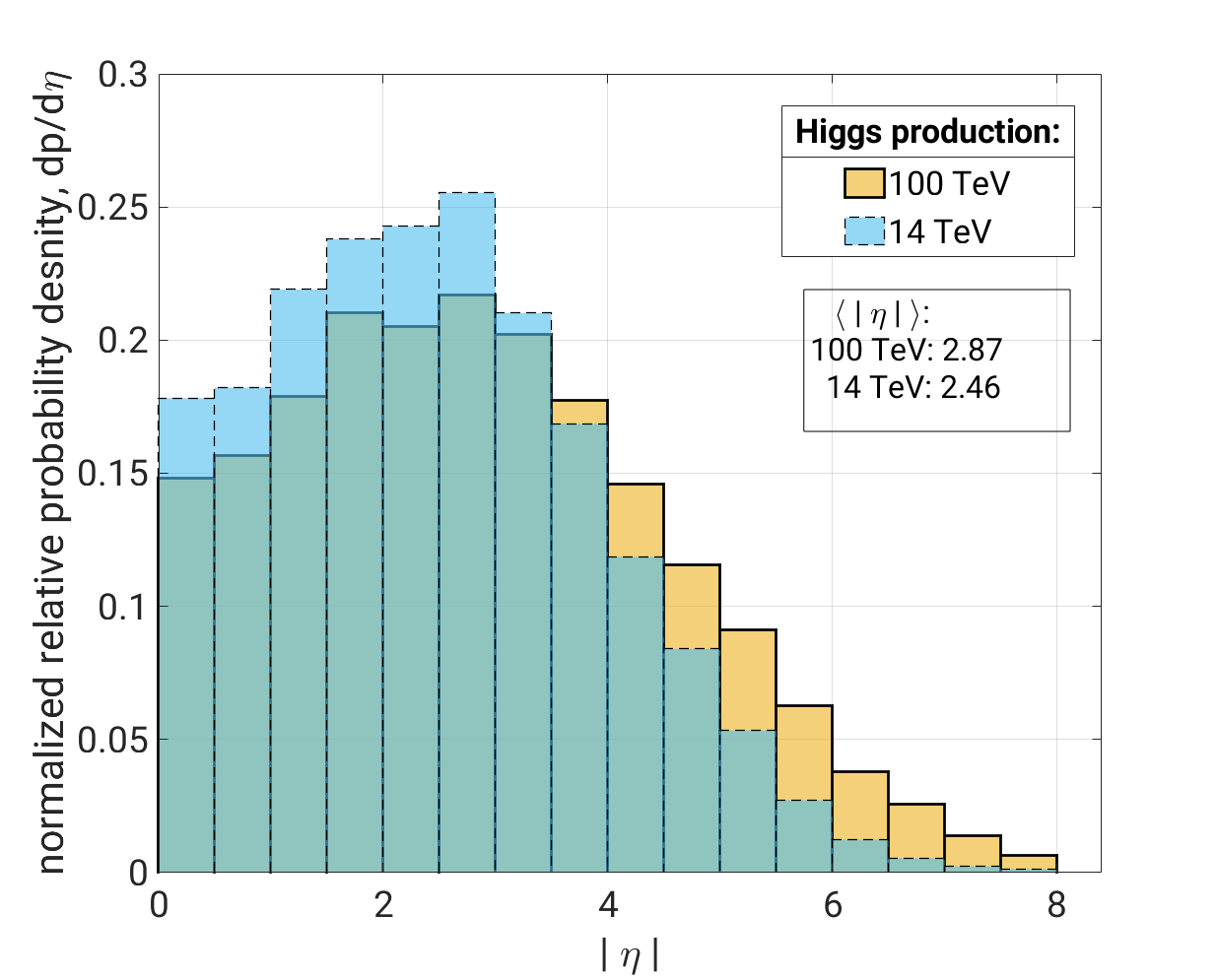}}
 \caption{ Distribution of pseudorapidity for $H\rightarrow \gamma\gamma$ at 14 and 100 TeV. Mean values of $|\eta|$ are also shown.
 \label{Higgs1}}
 \end{figure}
 
 \begin{figure}[htb!]
\centerline{ \includegraphics[width=3.5in, trim=10mm 0 10mm 0, clip]{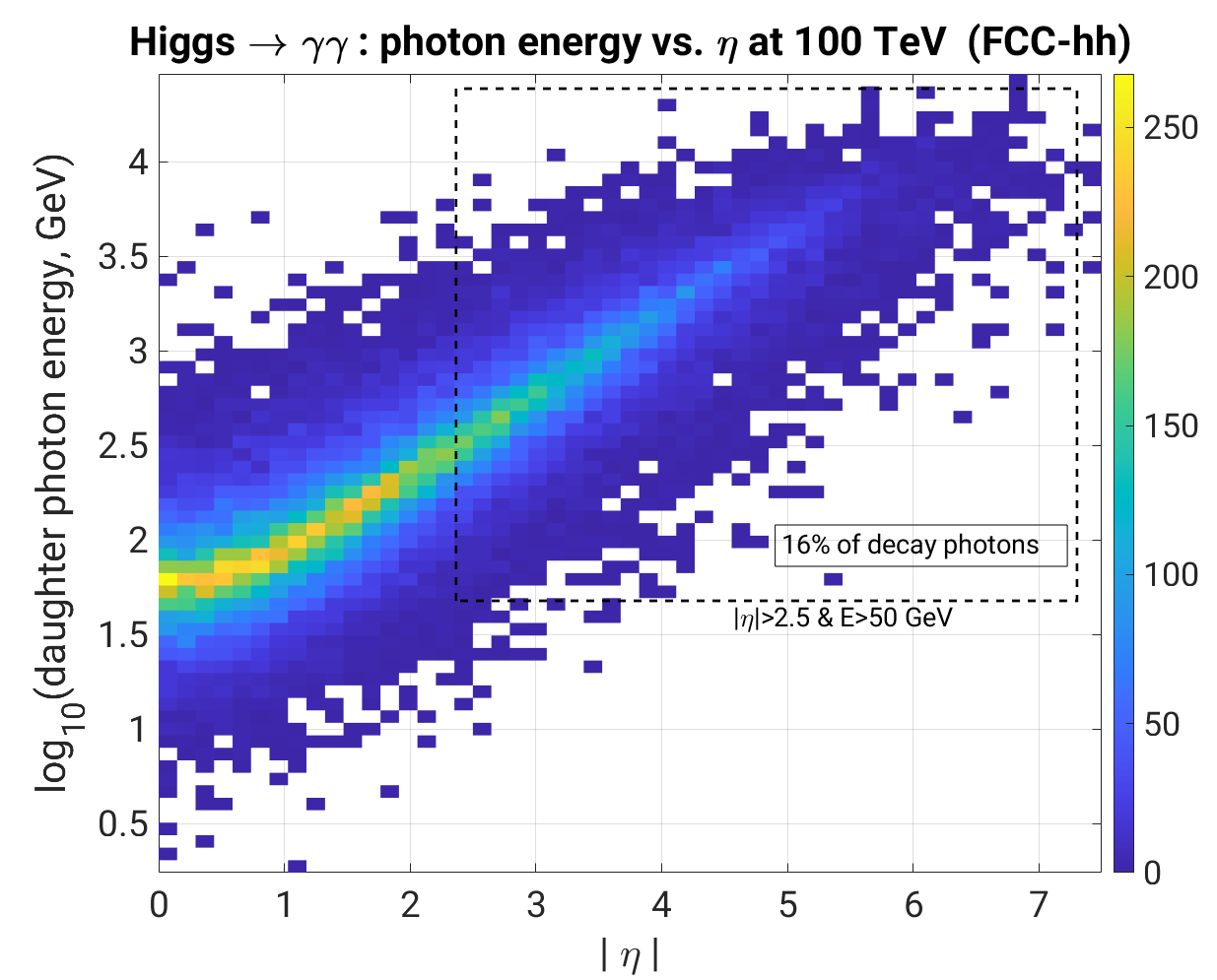}}
 \caption{ Daughter photon energy vs. pseudorapidity for $H\rightarrow\gamma\gamma$ at 100~TeV. The box indicates the region covered by our baseline preshower timing layer.
 \label{Higgs2}}
 \end{figure}

To illustrate the scale of the problem, Fig.~\ref{pileup} shows a plane projection of a simulated sample of the luminous region for a single FCC-hh collision, involving $\sim 1000$ interaction vertices. These are color-coded by time relative to the time at the center of the bunch crossing in the plot. The inset shows the inner $\sim 20$~mm region. Tracker systems may be able to resolve the longitudinal positions of most, if not all, vertices which produce a charged particle. However, in the core of the luminous region there are several vertices per mm. While timing resolution of order several tens of ps should resolve all ambiguities in the barrel region of the detector, the problem becomes more acute in the forward direction, and the geometric foreshortening along the beam axis in those regions exacerbates the problem. The foreshortening of the spatial $z$-direction is fortunately compensated by the improved resolution in the time axis along $z$, and while timing along does not yield the $z$ coordinate directly, at the picosecond level it separates virtually all vertices. If a particular vertex is a candidate for a $H\rightarrow \gamma\gamma$ interaction, the two photons must share a vertex and a picosecond time tag for both photons will eliminate any other vertex with nearly 100\% efficiency.

In Fig.~\ref{Higgs1} we show pseudorapidity distributions of Higgs production for both 14 and 100~TeV center-of-momentum energies, illustrating the importance of the forward region for this benchmark standard model process. These results are derived from PYTHIA-8 simulations created as part of the HEPSIM archive~\cite{Hepsim}. The mean of $|\eta|$ shifts from under 2.5 to almost 2.9, well into the forward EMCAL acceptance,  and a much larger fraction of Higgs are produced at $|\eta|>4$.

 \begin{figure*}[htb!]
 \centerline{ \includegraphics[width=6.5in, trim=10mm 0 10mm 0, clip]{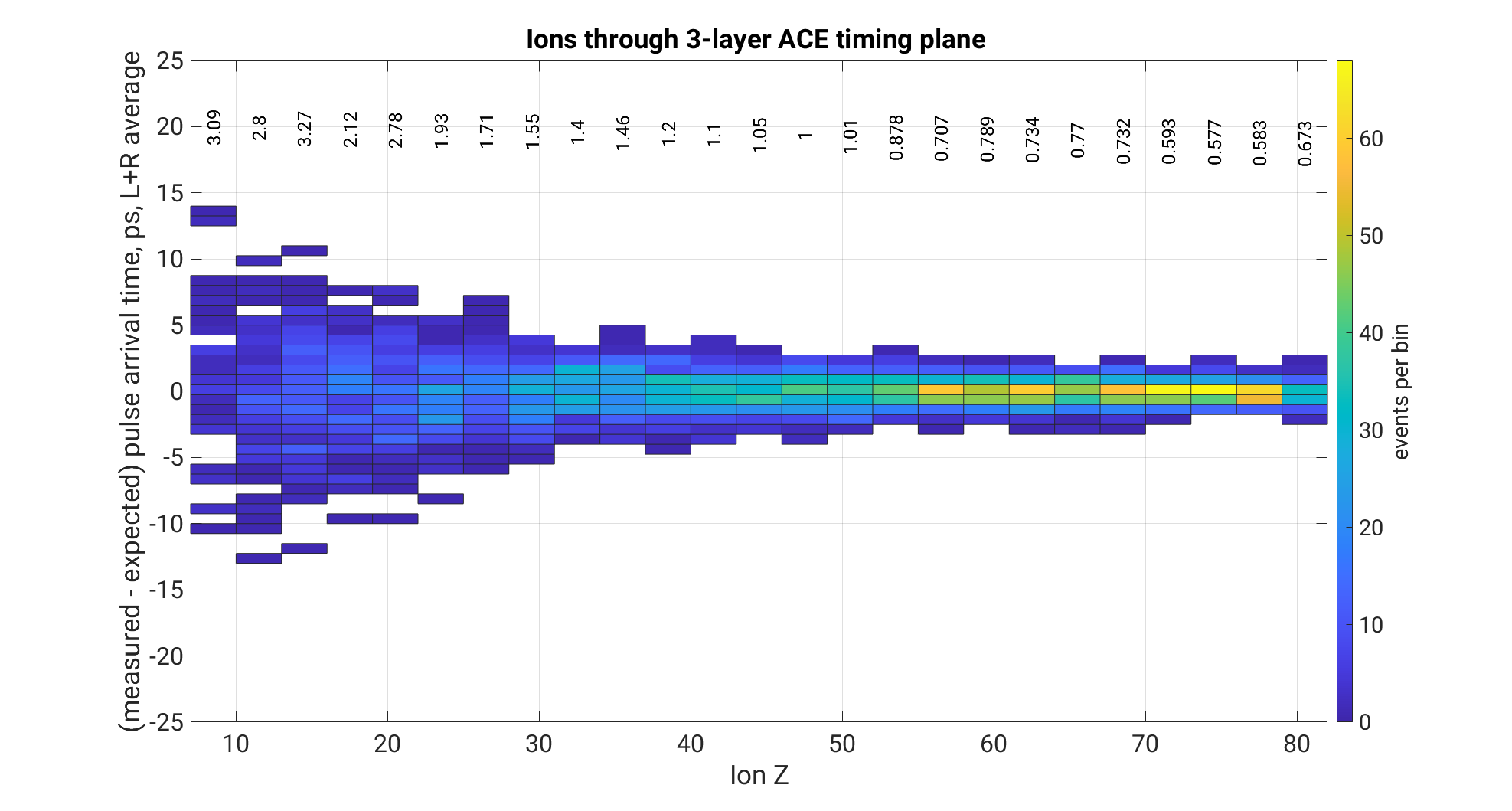}}
 \caption{Timing results vs. nuclear $Z$ for ions in our timing planes, where the signal is produced from microwave Cherenkov via the nuclear charge rather than the excess charge of a shower.
 \label{ionTiming}}
 \end{figure*}

In Fig.~\ref{Higgs1} we show HEPSIM PYTHIA-8 simulation results for the distribution of $\eta$ for the $H\rightarrow\gamma\gamma$ process at 100~TeV. The boxed region indicates the coverage of our baseline timing plane positioned just after a preshower block in the leading portion of the forward EMCAL. Virtually all daughter photons within the forward EMCAL acceptance are above our threshold and will thus receive a picosecond time tag, although this does not guarantee both photons in the event will be tagged since the pairs can have significantly different $\eta$. For this process, a forward timing layer will provide time tags for about 16\% of all photons. Adding similar timing layers to the endcap EMCALs ($1.5<|\eta|<2.5$) would increase this fraction to $\sim 28\%$, and even at $|\eta|\simeq 1.5$, 99\% of daughter photons are above the timing layer energy threshold.

Lacking a full-scale detector simulation, a detailed estimate of the background reduction afforded by our timing resolution is beyond our scope, but it is evident that any vertex ambiguities for this process should be completely resolved over all of the forward region. 

\subsubsection{Time-of-flight measurements.}

Time-of-flight (TOF) measurements are an obvious application of picosecond timing, since they are sensitive to the momentum of the particle, and thus the mass can be derived in the energy is also measured. Since in our application a shower is required, and the timing layers are colocated with calorimeters, all of the elements are present to provide direct measures of particle mass via the timing and calorimetry. In this case however, our relatively high energy threshold works against the TOF application, since the momentum resolution at a given timing precision decreases linearly as the particle energy increases, at least for particles that are highly relativistic above our $\sim 50$~GeV threshold. For light particles at the highest energies, time delays relative to $c$ are largely unobservable with any current technology. For light and heavy ions however, we may expect that picosecond timing will produce precision momentum measurements over a wide range of energies.

\begin{figure}[htb!]
\centerline{ \includegraphics[width=3.65in, trim=10mm 0 10mm 0, clip]{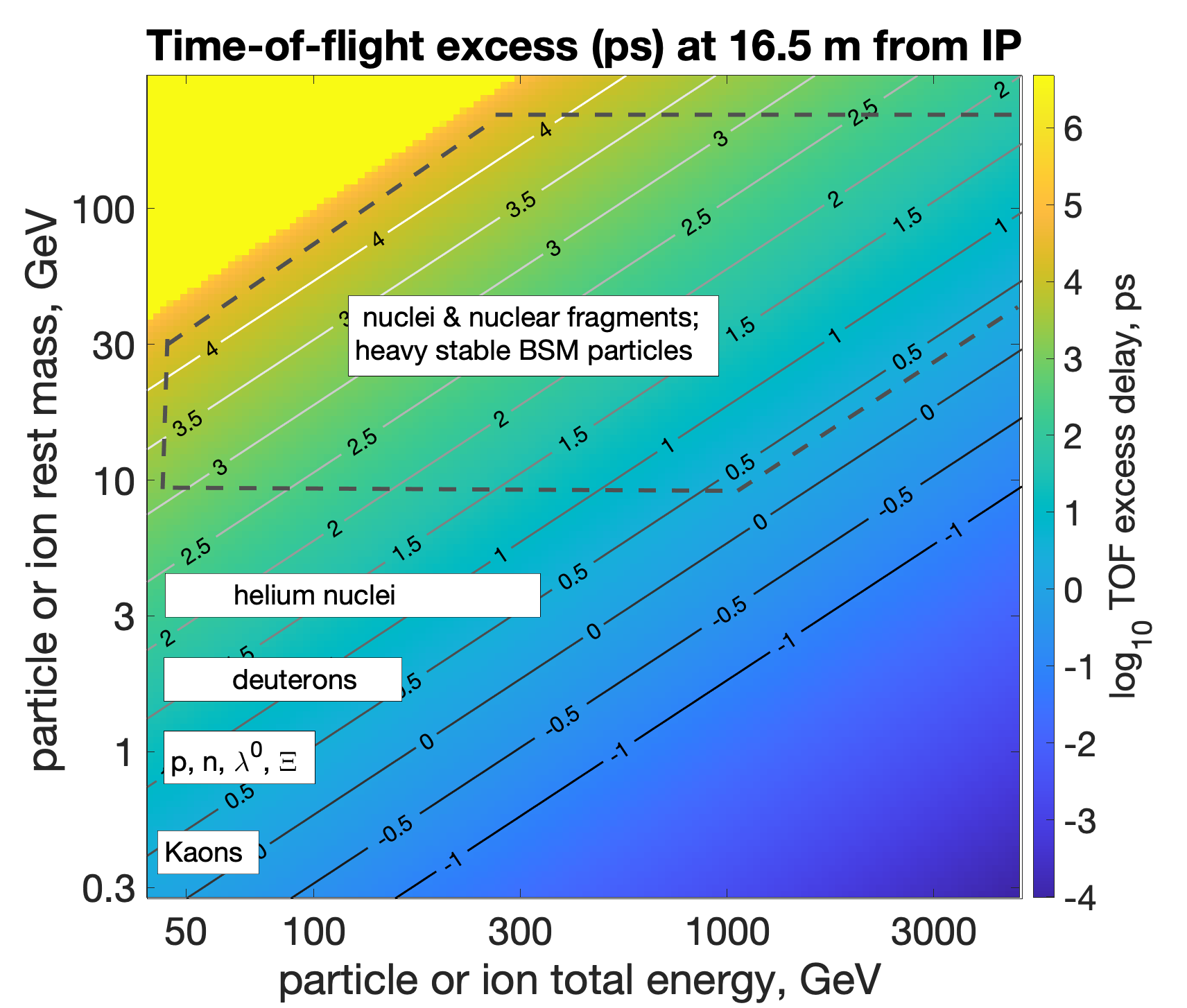}}
 \caption{ Excess time-of-flight for particles or ions at 16.5 m relative to the interaction point, for  various masses and energies over the range of the FCC-hh. The overlain text boxes show some specific examples of particles subject to TOF measurements. The energy scale starts at roughly the limiting ACE threshold $\sim 50$~GeV, and timing limits can achieve 1-3~ps with current technology. Thus the region to the lower right is currently inaccessible. For timing of heavy ions, nuclear fragments, or long-lived stable BSM particles, the energy is total energy, not energy per nucleon, and the timing range extends from $\sim 1$~ps to the limit where a new bunch arrives at 25 ns. Thus for very heavy ions the momentum precision will be parts in $10^4$ or better.
 \label{TOF}}
 \end{figure}
 
We have not previously discussed simulations of ions in our timing planes. For baryons and light ions, detection is via hadronic showers in a manner very similar to the charged pion simulations above, with some improvement in sensitivity and efficiency due to the reduction of large fluctuations in showers initiated by nuclei. For heavier nuclei, above $Z\simeq 10$ a new detection opportunity obtains: direct detection of the microwave Cherenkov produced by the charge of the ion itself rather than the secondary shower. 

Fig.~\ref{ionTiming} shows a simulation of the timing resolution for ions as a function of $Z$, for a $T_{sys}= 1.65$K, which is currently achievable for commercial off-the-shelf LNAs. Because there is essentially no loss due to the form factor of the transiting charge, the sensitivity extends down to $Z\simeq 10$, where $\sim 3$~ps resolution is possible, and that resolution rapidly improves to sub-picosecond levels  at high $Z$. In addition to the timing, the amplitude of the observed microwave signal is also directly proportional to $Z$, providing a resolution that we estimate to be
$$ \frac{\Delta Z}{Z} \simeq \frac{30\%}{(Z/10)} $$
which may be useful to supplement other measures.

Fig.~\ref{TOF} shows TOF results in the mass vs. total energy projection for a timing plane located at the front of the forward EMCAL for the FCC-hh, at about 16.5 m from the interaction region. The color scale and associated diagonal labeled contours indicate the logarithm of the TOF excess for a given particle compared to light speed in picoseconds. Since the ACE threshold energy is (in our current design) of order 50~GeV, the energy axis begins there. Several regions are highlighted with text boxes according to the particle or ion mass, and the boxes indicate a region that extends over that portion of the space where our timing layers can yield a TOF excess that is measurably different from $c$. 

For example,  deuteron showers should be detectable above $\sim 50$~GeV (or $\sim 25$~GeV/nucleon, with 30-50 ps delays expected near threshold, measurable with 10\% precision or better, and yielding measurable excess delays up to 150-200~GeV. Helium nuclei at 50-100~GeV will yield 30-300 ps excess delays, and measurable excess delays up to 300~GeV. For heavier ions, the energy range for measurable delays extends well into TeV energies, or hundreds of GeV per nucleon. One issue that may lead to confusion: for very heavy ions with energies below 100~GeV, delays may be long enough so that arrival at the forward detectors begins to encroach on the event from the subsequent collision 25~ns later. But at higher energies this is not an issue.

The reach of the system includes some parameter space for kaons (as we will discuss in the next section) although the time delay would provide very limited resolution. For stable and relatively long-lived baryons in the O(1 GeV) mass range, 20-50\% resolution is possible, which may be useful for tagging neutrals. For light nuclei, the reach is somewhat better, providing 5-10\% or better mass resolution over the 50-100~GeV range for deuterons and even better resolution for helium. For nuclei above He, TOF data with picosecond resolution, combined with calorimetry to determine energy, will provide precision mass measurements over a wide range of high energies, up to the kinematic cutoff.

The FCC-hh may also produce beyond-standard-model (BSM) particles that are sufficiently stable to reach the calorimeters prior to decay. For these BSM particles at the distance of the forward EMCAL, they will be observed via their showers, which may be either electromagnetic or hadronic. For a 100 GeV BSM particle, its TOF will exceed 30 ps out to 3~TeV, where it can be measured to $\leq5\% $ precision; below 1 TeV, the TOF measurement precision with be a fraction of a percent, and the mass measurement below this energy will thus be limited by the precision of the calorimetry rather than the TOF. Similar results have been discussed for hypothetical timing precision of 10~ps or better~\cite{Chekanov20}; in our case $\leq3$~ps timing can already be achieved currently with our methodology.

\subsubsection{Jet physics.}

The utility of picosecond timing in studies of jets and their underlying physics is still a developing topic. To determine in detail the effects of improved timing on the many observables in jets requires integration of timing elements within a full detector simulation, which is beyond the scope of our current study. In lieu of this we can look at distributions of jet components, and determine which of these are within the reach of our proposed detectors within our current design point for the forward region of the FCC-hh. 

\begin{figure}[htb!]
\centerline{ \includegraphics[width=3.65in, trim=10mm 0 10mm 0, clip]{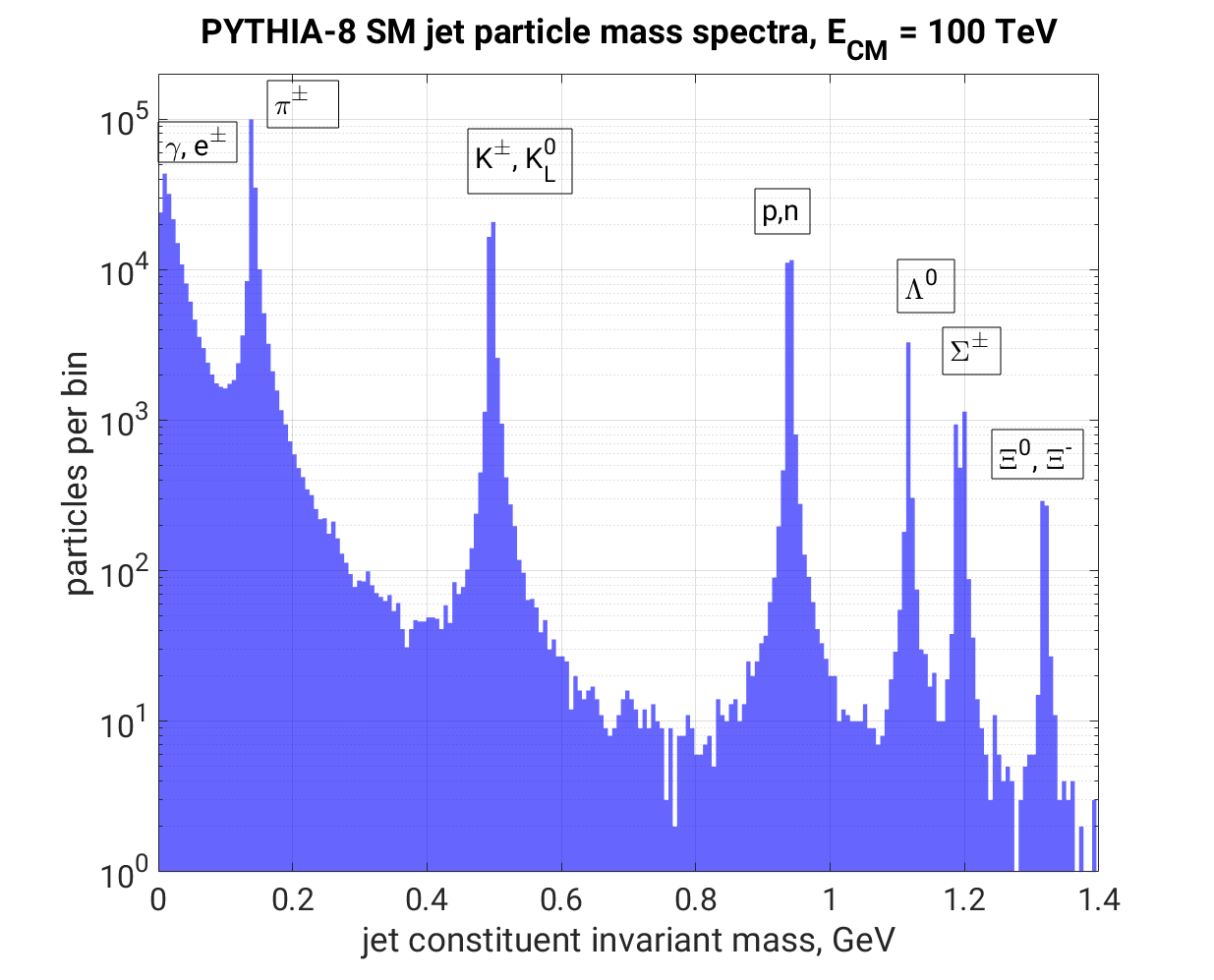}}
 \caption{ Jet invariant mass spectrum from the HEPSIM simulations for 100 TeV collisions.
 \label{invmass}}
 \end{figure}

To do these we use the same 100~TeV PYTHIA-8 simulations as detailed above, now focusing on a simulation data of standard model jets at the quasi-stable particle level, with jet total transverse momenta $p_t > 300$~GeV/c, again from the HEPSIM archive. The jets are analyzed and the components detected using an anti-kt algorithm with $R=0.5$,  and for each simulated 100~TeV center-of-momentum event, only the leading jet is used for our analysis, with approximately $10^4$ simulated events used.

\begin{figure}[htb!]
\centerline{ \includegraphics[width=3.65in, trim=10mm 0 10mm 0, clip]{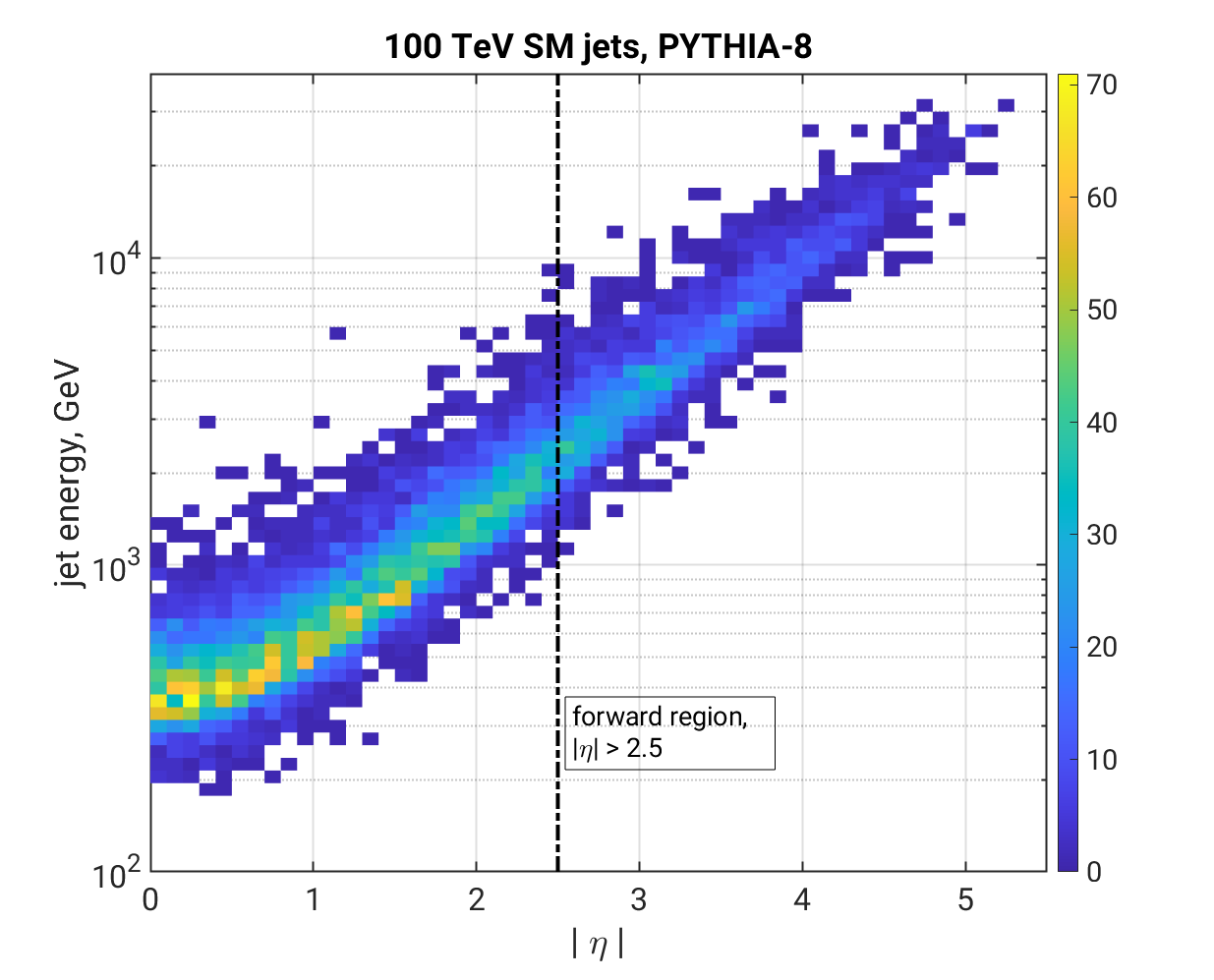}}
\centerline{ \includegraphics[width=3.65in, trim=10mm 0 10mm 0, clip]{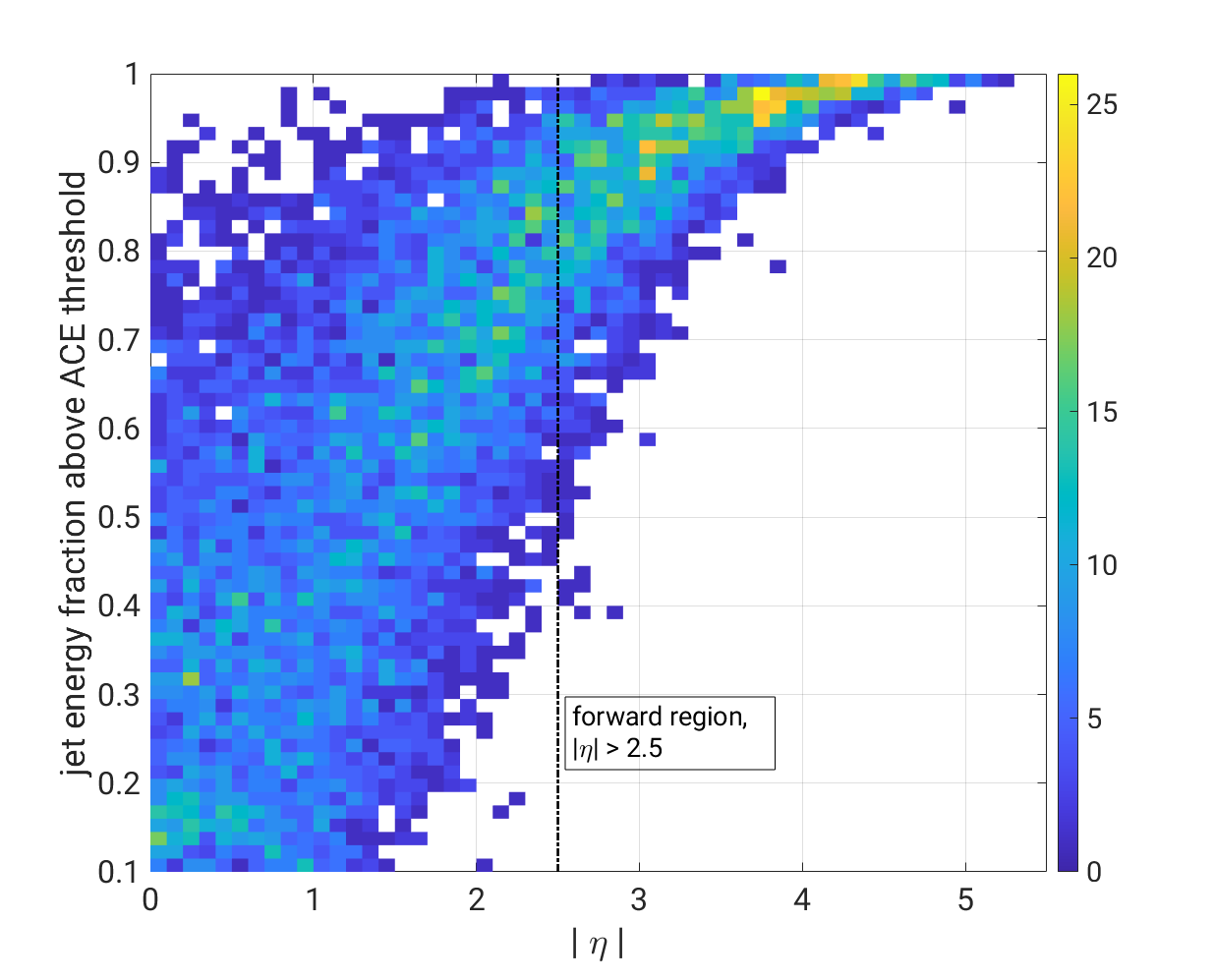}}
 \caption{ Top: total jet energy vs. pseudorapidity. The vertical line indicates the forward region where our study is focused. Bottom: fraction of jet component particles that are expected to be above the ACE timing threshold as a function of pseudorapidity, again with the vertical division indicated the forward region.
 \label{jetEnergy}}
 \end{figure}

The role of TOF measurements of jet components requires an understanding of the typical jet mass spectrum. To this end, Fig.~\ref{invmass} shows the invariant mass spectrum of all of the $10^4$ leading jets analyzed in our study, with the corresponding particles indicated. The population of jet components is dominated by charged pions and photons, but with subdominant numbers of kaons, stable baryons, and light hyperons. While the $\pi^{\pm}$ and $K^{\pm}, K^0_L$ lifetimes are such that decays are irrelevant at primary energies greater than a few GeV, this is not the case for the $K^0_S$ and hyperons, which have mean lifetimes in the subnanosecond range, or $\langle c\tau \rangle \simeq 2-8$~cm. 

Fig.~\ref{jetEnergy} shows two views of the energy distribution of forward SM jets at 100~TeV. Fig.~\ref{jetEnergy}(Top) shows the total energy vs. pseudorapidity, with the forward region indicated. In the forward region at $|\eta|\geq 2.5$, which contains 23\% of all jets, total jet energy exceeds 1~TeV in all cases, with the mean value well above 2~TeV. Particle multiplicity is high, distributing this energy over many components, and thus in Fig.~\ref{jetEnergy}(Bottom) we show the fraction of jet component particles that are above our expected energy threshold, also as a function of pseudorapidity. 

\begin{figure}[htb!]
\centerline{ \includegraphics[width=3.65in, trim=10mm 0 10mm 0, clip]{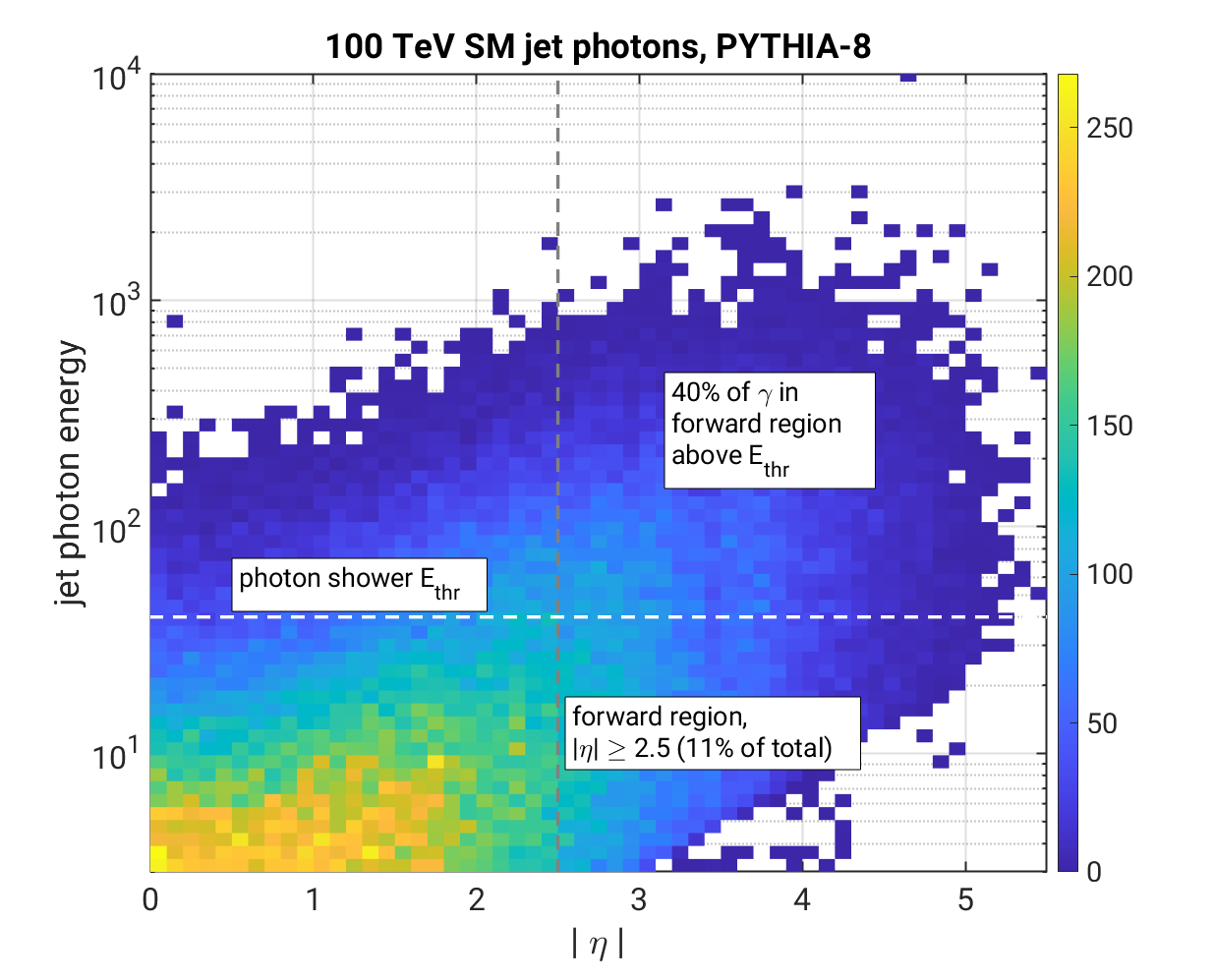}}
\centerline{ \includegraphics[width=3.65in, trim=10mm 0 10mm 0, clip]{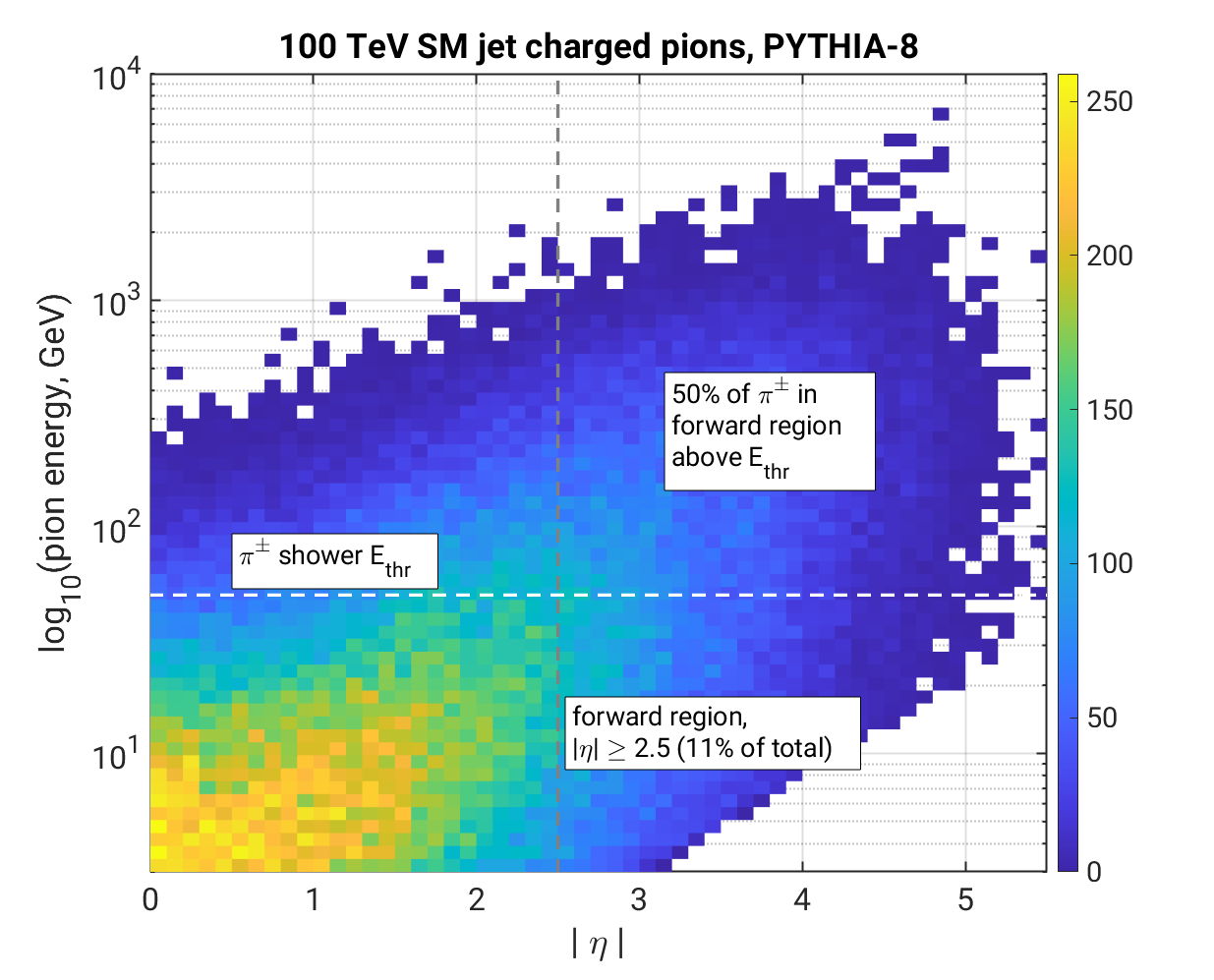}}
 \caption{ Top: Jet photon energy distribution vs. $| \eta |$ with the forward region and energy threshold for ACE detection marked. Bottom: similar distribution for jet $\pi^{\pm}$.
 \label{pigamma}}
 \end{figure}

While this distribution is somewhat oversimplified since we have not accounted for components that decay in flight before reaching the timing layer, the efficiency for timing of jet components is still quite high, and since photons and pions dominate the jet population, these will also dominate the timing solutions for the jet components. The distribution also does not account for fluctuations in the hadronic showers, which reduce the efficiency of our assumed single timing layer, but this could be improved with additional timing layers stationed within the HCAL stack.

As indicated in Fig.~\ref{TOF} above, time-of-flight excesses for pions over photons at these energies are not detectable even at the picosecond level, but the timing constraints will be useful for mitigating pileup, and correct association of jet components with vertices. ACE elements with detectable pulses at both ends will yield both $x$ and $y$ positions for the shower core. Because of the high index of refraction ($n\simeq 3.2$) of the dielectric load within the waveguides, the spatial precision given a time difference from end-to-end in the waveguide is $\sim 100~\mu m$ per picosecond. In the orthogonal direction, a fit of the shower amplitudes in the $\sim 6$ waveguide stack will determine position to $1-2$mm.

Given that $\pi^{\pm}$ and $\gamma$ constitute a major fraction of jet constituents, we separately histogram their distribution in energy and pseudorapidity in Fig.~\ref{pigamma}. In the top pane of the figure, we show the photon distribution, and in the bottom the pion distribution. The forward regions and the achievable energy thresholds are also marked. These thresholds are based on detection levels guided by the external shower information from the calorimeters, allowing us to make useful measurements of the particle timing down to roughly the $\sim 3\sigma$ level in amplitude by constraining the search space of the waveform data. 

In the forward region, 40-50\% of jet photons and pions can be effectively timed at the several ps level or better. These results do not include the shower fluctuation efficiency losses due to having only one timing plane each for electromagnetic and hadronic showers; the shower efficiency can be improved with additional timing planes within the calorimeters if such timing is deemed critical. While understanding the ultimate utility of picosecond timing for jet parameter estimation will require a full detector simulation, it is evident that ACE timing planes can already provide orders-of-magnitude improvement in calorimeter timing for a large fraction of jet constituents.

\begin{figure}[htb!]
\centerline{ \includegraphics[width=3.65in, trim=10mm 0 10mm 0, clip]{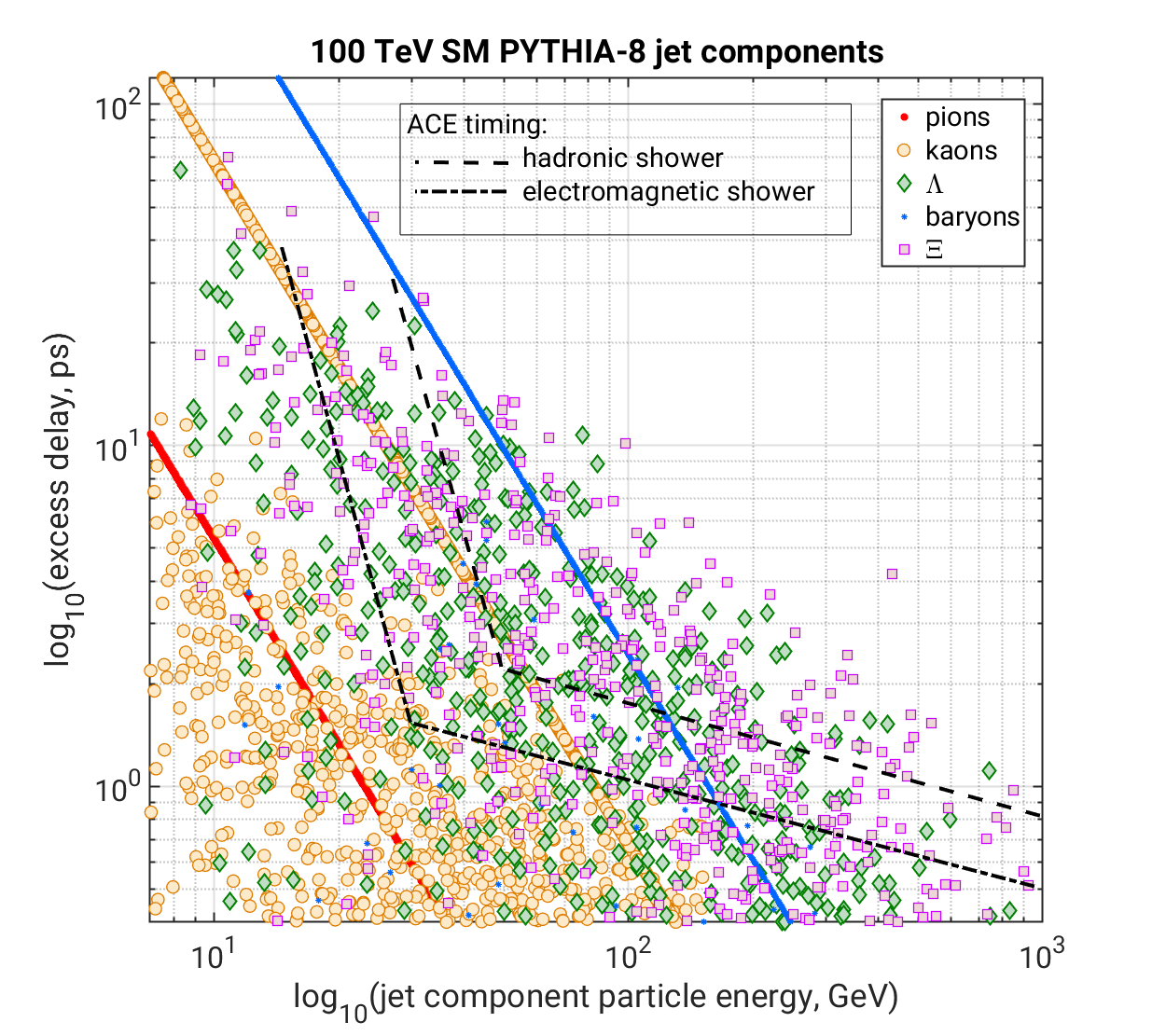}}
 \caption{ TOF simulation for jet components with symbols denoting the particle type in each case, and including the decays of the shorter-lived mesons and hyperons. The dashed and dot-dash lines provide lower bounds on the regions over which ACE can provide picosecond timing for either electromagnetic or hadronic showers for particles to the right and above the given lines. Quasi-stable particles appear to stack closely along diagonal lines corresponding to their momenta.
 \label{jetTiming}}
 \end{figure}

The TOF of photons from the jet vertex to the preshower ECAL timing plane set a reference time, and for practical purposes pions are too light to resolve delays at the picosecond level, so they provide the same reference for the HCAL timing plane in our reference design.
Using the same time-of-flight estimation done in the previous section, we can Monte Carlo the particle decays, and compute TOF directly for each jet component to a forward timing-plane location at a distance of $\sim 16.5$~m as above.  The results are shown in Fig.~\ref{jetTiming} where each jet particle is color coded and represented by a single data point, and we have only included jets with $|\eta|>2.5$. Stable or quasi-stable particles appear as diagonal lines in the log-log scaled excess time vs. primary particle energy space. For decaying particles, we assume that the daughter particles from the decay effectively transit the remaining distance after the decay at light speed, which is a good approximation for the predominant decay modes. 

Stable baryons and long-lived kaons produce well-defined loci of delay vs. energy values. In contrast,
the shorter-lived hyperons and $K^0_S$ produce a continuum of TOF values as the plot shows. The decay products are dominated by a mix of photons and pions and thus can be detected in either the forward preshower timing plane or the second plane after the EMCAL, and we show estimated thresholds for each type of shower detected, denoted by photon and pion timing limits, respectively. For these continuous distributions of TOF from intermediate decays, unique particle and momentum identification may not be possible event-by-event, but statistically these components will be distinguishable from light mesons up to several hundred GeV in many cases.

While this simulation glosses over the details of the decay daughter particle momenta, it does illustrate that picosecond timing, even with the relatively high threshold for ACE, will give useful constraints on a significant fraction of jet components. The TOF measurement yields total momentum,  the shower core is localized to a few mm or better in both the $\eta$ and $\phi$ directions, and the calorimeters yield total energy. Thus for a subset of jet components, including neutrals, full five-dimensional constraints will be determined. 

\section{Conclusions}

We have reported on results of beam tests intended to further explore the utility and sensitivity of $Al_2O_3$-loaded microwave waveguides with fields excited by the charge excess in an electromagnetic shower, initiated by 14.5~GeV electron bunches at the SLAC National Accelerator Laboratory. These tests were able to establish the absolute sensitivity of the method, calibrated with a particle camera of known response, and timing studies confirm earlier results that such showers can be used to establish several picosecond or better arrival time tags for the electron bunch, which acts as a proxy for a primary particle with the composite bunch energy. We have extended these measurements to also establish the amplitude sensitivity along the transverse dimension of each waveguide elements, using these results to inform further design parameters with the goal of developing compact timing-plane geometries for particle collider detectors.

Using the results validated by the beam tests, we have developed detailed electrodynamic simulations which are also validated by vector potential analysis, and have used these simulations to design improved waveguide systems, with paired elements coupled together to yield lower particle energy thresholds and thus improved sensitivity.  

We have chosen a timing-plane point design for the planned Future Circular Hadron Collider, and using GEANT4 results, have developed microwave simulation tools to estimate the response of the timing plane to both electromagnetic and hadronic showers, for locations within the baseline forward calorimeters for the FCC-hh, for $| \eta | \geq 2.5$. We study photon and charged pion showers in detail to estimate timing precision and detection efficiencies for these events, and utilize these simulation tools for several physics examples, showing the utility of picosecond timing in practical cases, including the canonical $H\rightarrow \gamma\gamma$ signal, and for time-of-flight measurements of ions and jet components. 

While the relatively high least-count energy of our method currently limits its usefulness as a general-purpose timing layer, at least in our current realizations, we have achieved timing precision at least an order of magnitude better than any other technology now available, using a system that is inherently radiation hard and with extremely large dynamic range. For future colliders we offer the possibility of finally achieving particle timing precision commensurate with the spatial precision of other detector elements.

We thank the excellent staff at the SLAC National Accelerator Laboratory for their
support of this project. This material is based upon work supported by the Department
of Energy under Award Numbers DE-SC0009937, DE-SC0010504, and DE-AC02-76SF0051.
We thanks S. Chekanov and Argonne National Laboratory for developing and hosting the 
HEPSIM archive.

\end{document}